\documentclass[10pt]{article}
\usepackage{amsmath, amsthm, amssymb, epsfig, overpic, subfig, graphicx}

\newcommand{\E}{\ensuremath{\mathbb E}}

\newcommand{\Z}{\ensuremath{\mathbb Z}}
\newcommand{\cM}{\mathcal{M}}
\newcommand{\cT}{\mathcal{T}}

\newcommand{\stsp}{\Psi}

\renewcommand{\bar}{\overline}
\newcommand{\prob}{\mathcal{P}}
\newcommand{\m}{\cM}

\newcommand{\head}{r}
\newcommand{\mfix}{\cM_{TR}}
\newcommand{\mvar}{\cM_{EF}}
\newcommand{\mdyck}{\cM_{DK}}

\newcommand{\mtower}{\cM_{{CR}}}

\newcommand{\arc}{\overrightarrow}
\newcommand{\comment}[1]{}
\usepackage{ifthen} 
\newboolean{includefigs} 
\setboolean{includefigs}{true} 
\newcommand{\condcomment}[2]{\ifthenelse{#1}{#2}{}}
\newtheorem{Theorem}{Theorem}

\newtheorem{Definition}{Definition}
\newtheorem{Remark}{Remark}
\newtheorem{Lemma}[Theorem]{Lemma}

\comment{
      {\bigpar{\bf Proof of #1.}\ %
            \setlength{\saveindent}{\parindent}
                       \ignorespaces}%
      {\stopproof\ignorespaces\bigbreak \setlength{\parindent}{\saveindent}}

}
\title{Algorithms for Sampling 3-Orientations of Planar Triangulations}

\begin{document}
\date{}

\author{Sarah Miracle\thanks{College of Computing, Georgia Institute of Technology, Atlanta, GA 30332-0765.  Supported in part by a DOE Office of Science Graduate Fellowship, NSF CCF-0830367 and an ARCS Scholar Award.}\and Dana Randall\thanks{College of Computing, Georgia Institute of Technology, 
Atlanta, GA 30332-0765.  Supported in part by NSF CCF-0830367 and CCF-0910584.} \and Amanda Pascoe Streib\thanks{School of Mathematics, Georgia Institute of Technology,
Atlanta, GA 30332-0280.  Supported in part by a NPSC Fellowship and NSF CCF-0910584.} \and Prasad Tetali\thanks{School of Mathematics and School of Computer Science, Georgia Institute of Technology, Atlanta, GA 30332-0765.  Supported in part by  NSF DMS-1101447 and CCR-0910584.}}

\maketitle

\begin{abstract}

Given a planar triangulation, a 3-orientation is an orientation of the internal edges so all
internal vertices have out-degree three.  Each 3-orientation gives rise to a unique edge coloring 
known as a {\it Schnyder wood} that has proven powerful for various computing and combinatorics applications.
We consider natural Markov chains for sampling uniformly from the set of 3-orientations.  First, we study a  
``triangle-reversing'' chain on the space of 3-orientations of a fixed triangulation that reverses the orientation 
of the edges around a triangle in each move.  It was shown previously that this chain connects the state space and we show that
(i) when restricted to planar triangulations of maximum degree six, the Markov chain is rapidly mixing,
and (ii) there exists a triangulation with high degree on which this Markov chain mixes slowly. 
Next, we consider an ``edge-flipping''  chain on the larger state
space consisting of 3-orientations of all planar triangulations on a fixed number of vertices.   It was also shown previously that
this chain connects the state space and we prove that the chain is always rapidly mixing.
The triangle-reversing and edge-flipping Markov chains both arise in the context of sampling other combinatorial structures,
such as Eulerian orientations and  triangulations of planar pointsets,
so our results here may shed light on the mixing rate of these related chains as well.
\end{abstract}

\section{Introduction}\label{intro}
The $3$-orientations of a graph have given rise to beautiful combinatorics and computational applications.  A 3-orientation of a planar triangulation is an orientation of the internal edges of the triangulation such that every internal vertex has out-degree three. In this paper we study natural Markov chains for sampling 3-orientations in two contexts, when the triangulation is fixed and when we consider the
union of all planar triangulations on a fixed number of vertices.  In the case that the triangulation is fixed, we consider moves that reverse the
orientation of edges around a triangle if they form a directed cycle.  We show that the chain is rapidly mixing (converging 
in polynomial time to equilibrium) if the maximum degree of the triangulation is six, but can be slowly mixing (requiring exponential time)
if the degrees are unbounded.  To sample from the set of all 3-orientations of triangulations
with $n$ vertices we use a simple ``edge-flipping'' chain and show it is always rapidly mixing.  
These chains arise in other contexts such as sampling Eulerian orientations and triangulations of fixed planar point sets, so there is 
additional motivation for understanding their convergence rates.

\subsection{3-orientations}
Given an undirected graph $G=(V, E)$ and a function $\alpha:V \to \Z^+$, an $\alpha$-orientation is an orientation of $E$ where each vertex $v$ has outdegree $\alpha(v)$. Several fundamental combinatorial structures -- spanning trees, bipartite perfect matchings,  Eulerian orientations, etc. -- can be seen as special instances of $\alpha$-orientations of planar graphs. We refer the reader to \cite{f1,f2,fz} and references therein for extensive literature on the subject.
Not surprisingly, counting $\alpha$-orientations is $\#P$-complete.  Namely, consider an undirected Eulerian graph $G$ (with all even degrees); the $\alpha$-orientations of $G$, where $\alpha(v)=d(v)/2$, correspond precisely to Eulerian orientations of $G$. The latter problem has been shown to be $\#P$-complete by Mihail and Winkler \cite{mw}, and more recently Creed~\cite{c} showed that it remains $\#P$-complete even when restricted to the class of planar graphs.

The term \emph{$3$-orientation} refers to an $\alpha$-orientation of a planar triangulation where all internal vertices (vertices not bounding the infinite face) have $\alpha(v) = 3$ and all external vertices have $\alpha(v) = 0$.  
Each 3-orientation gives rise to a unique edge coloring, known as a \emph{Schnyder wood}, whose many combinatorial applications include graph drawing \cite{sc1,ccl} and poset dimension theory \cite{sc2}.  Several intriguing enumeration problems remain open, such as the complexity of enumerating 3-orientations of a planar triangulation (see e.g., \cite{fz}.)  We study the problem of sampling 3-orientations of a fixed (planar) triangulation, as well as sampling 3-orientations of all triangulations with $n$ internal vertices.  In particular, we analyze the mixing times of two natural Markov chains for these problems, which were introduced previously but had thus far resisted analysis.  

\subsection{Results}

First, we study the problem of sampling 3-orientations of a fixed triangulation, which was stated as an open problem by Felsner
and Zickfeld \cite{fz}.  Although there is no known efficient method for counting exactly, there are polynomial-time algorithms for approximately counting and sampling 3-orientations due to a bijection with perfect matchings of a particular bipartite graph (see Section 6.2 in \cite{fz}).  This bijection allows us to sample 3-orientations in time $O^*(n^7)$ using an algorithm due to Bez\'akov\'a et al.~\cite{bsvv} (improving on the celebrated results of Jerrum, Sinclair and Vigoda~\cite{jsv}), but this approach is indirect and intricate.

We consider instead a natural ``triangle-reversing'' Markov chain, $\mfix$, that reverses the orientation of a directed triangle in each step, thus maintaining the outdegree at each vertex.   Brehm~\cite{br} showed that for any fixed triangulation $T$, $\mfix$ connects the state space $\stsp(T)$ of all 3-orientations of $T$.  
We also consider a related ``cycle-reversing'' chain, $\mtower$, that can also reverse directed cycles containing more than one
triangle.   The chain $\mtower$ is a non-local version of $\mfix$ based on ``tower moves'' reminiscent of those in~\cite{lrs}.
We show that both of these chains are rapidly mixing.  
Let $\Delta_I(T)$ denote the maximum degree of any internal vertex of $T$.  We show:

\begin{Theorem}\label{AllTriangulationsTower}
If $T$ is a planar triangulation with $\Delta_I(T)\leq 6$, then the mixing time of $\mtower$ on the state space $\stsp(T)$ satisfies $$\tau(\epsilon) = O(n^5\ln\epsilon^{-1}).$$
\end{Theorem}

\noindent 
We can use a standard comparison argument together with Theorem~\ref{AllTriangulationsTower}
to infer a bound on the mixing time of the triangle-reversing chain $\mfix.$  
Thus we prove:

\begin{Theorem}\label{AllTriangulations}
If $T$ is a planar triangulation with $\Delta_I(T)\leq 6$, then the mixing time of $\mfix$ on the state space $\stsp(T)$ satisfies
$$\tau(\epsilon) =O\left(n^8\ln\epsilon^{-1}\right).$$
\end{Theorem}

\noindent Note that the class of planar triangulations with $\Delta_I \leq 6$ is exponentially large in $n$, the number of vertices.
An interesting special case is when the fixed triangulation is a finite region $\Lambda$ of  the triangular lattice, since 
sampling 3-orientations on $\Lambda$ corresponds to sampling Eulerian orientations.    Creed \cite{c} independently solved the sampling problem in
this special case using an approach similar to ours based on towers;  he shows that for certain subsets of the triangular lattice
the tower chain can be shown to mix in time $O(n^4)$.  In addition, it was previously shown that similar cycle-reversing chains are rapidly mixing in the context of sampling Eulerian orientations on the Cartesian lattice \cite{lrs} and the 8-vertex model \cite{gr}.
Our analysis here bounding the mixing rate of $\mtower$ in the  general setting of arbitrary planar graphs with maximum degree 6 requires additional combinatorial insights because we no longer have the regular lattice structure.  In particular, we make use of a combinatorial structure outlined by Brehm \cite{br}.  In fact, this structure allows us to extend our analysis to certain non-4-connected triangulations that can have vertices of degree greater than six.  

Next, we prove that when the maximum degree is unbounded, the chain $\mfix$ may require exponential time.  Specifically, we prove:

\begin{Theorem}\label{slowmixthm} For any (large) $n$, there exists a triangulation $T$ of size $n$ for which the mixing time of $\mfix$ on the state space $\stsp(T)$ satisfies $$\tau(\epsilon) =\Omega( 2^{n/4}).$$
\end{Theorem}
\noindent Based on the construction we give here, Felsner and Heldt~\cite{fh} recently constructed another, somewhat simpler, family of graphs for which the mixing rate of $\mfix$ and $\mtower$ is exponentially large.  However, we note that their family also has maximum degree
that grows with $n$.   It would be interesting to know if there is such a family with bounded degree.

The second problem we study is sampling from $\stsp_n$, the set of all 3-orientations arising from all possible triangulations on $n$ internal vertices.  The set $\stsp_n$ is known to be in 1-1 correspondence with all pairs of non-crossing Dyck paths, and as such has size $C_{n+2}C_n - C_{n+1}^2$, where $C_n$ is the $n$th Catalan number.  Since exact enumeration is possible, we can sample using the reduction to counting; this was explicitly worked out by Bonichon and Mosbah \cite{bm}.  We consider a natural Markov chain approach for sampling that in each step selects a quadrangle at random, removes the 
interior edge, 
and replaces it with the other diagonal in such a way as to restore the out degree at each vertex.  
Bonichon, Le Sa\"ec and Mosbah~\cite{blm} showed the the chain $\mvar$ connects the
state space $\stsp_n$ and we present the first bounds showing that the chain is rapidly mixing.  
Although the exact counting approach already yields a fast approach to sampling, the chain $\mvar$ is compelling because it arises in other contexts where we do not have methods to count exactly.  For example, it has been proposed as a method for sampling  triangulations of a fixed planar point set, a 
problem that has been open for over twenty years.  In addition, there is additional interest in the mixing rate of this chain precisely because the number is related to the
Catalan numbers;  there has been extensive work trying to bound mixing rates of natural Markov chains for various families
of Catalan structures (see, e.g., \cite{mt}).  Specifically, we prove:

\begin{Theorem}\label{mixing_mvar} The mixing time of $\mvar$ on the state space $\stsp_n$ satisfies $$\tau(\epsilon) = O(n^{10}\log(n/\epsilon)).$$
\end{Theorem}

\subsection{Techniques}
The primary challenge behind the proofs of these results is extracting the right combinatorial insights to understand the dynamics in the context of Schnyder woods and 3-orientations.  Fortunately, there is a long history examining the rich structure of Schnyder woods.   We
were able to extend these results in several new ways, thus allowing us to bound the mixing rates of these chains.  
The proof of Theorem \ref{AllTriangulationsTower} for $\mtower$ involves a complex coupling argument that is straightforward if $T$ is the triangular lattice, but requires more work to generalize to all triangulations with $\Delta_I\leq 6$.  To prove Theorem~\ref{slowmixthm}, we produce an intricate triangulation $T$ which is carefully constructed to reveal an exponentially small cut in the state space $\stsp(T)$.  Although our choice of $T$ may seem complicated, it was carefully architected using properties of 3-orientations to show that the Markov chain may be slow.  
Our proof of Theorem~\ref{mixing_mvar} involves a detailed application of the comparison method to bound the mixing time of $\mvar$ by relating it to a local Markov chain on Dyck paths, $\mdyck$, whose mixing time is known (see \cite{lrs,wil}).  The key obstacle here is decomposing moves of $\mdyck$ into moves of $\mvar$ while avoiding congestion.  This is especially challenging because although $\mdyck$ is local in the setting of Dyck paths, in the context of 3-orientations it can make global changes to a 3-orientation in a single step.  

\section{Preliminaries}\label{prelim}
\begin{figure}[!htb]
\centering
\subfloat[]{\begin{overpic}[scale=.1]{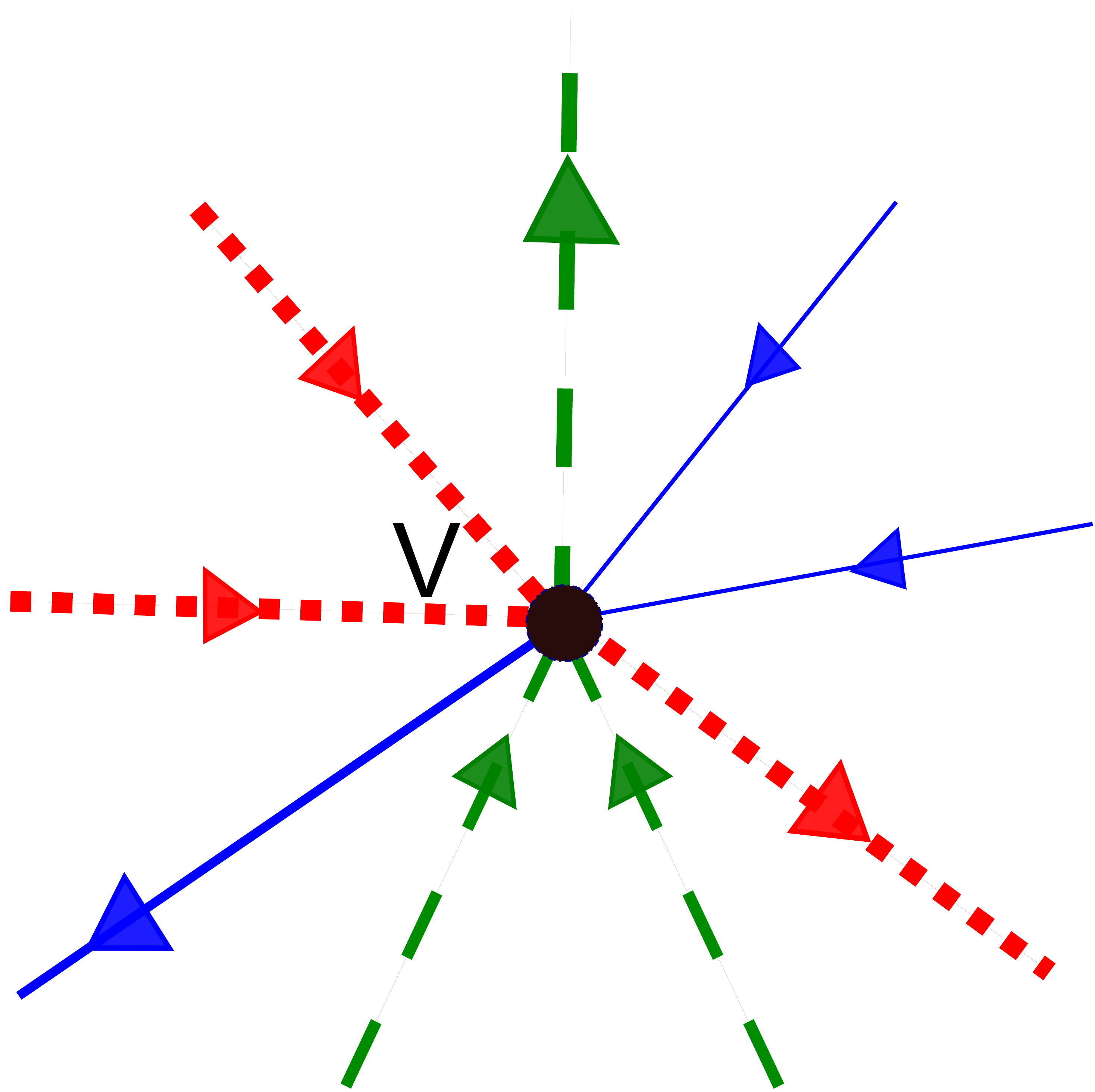}
\end{overpic}\label{vertexconditionpic}}
\hspace{1in}
\subfloat[]{\begin{overpic}[scale=.04]{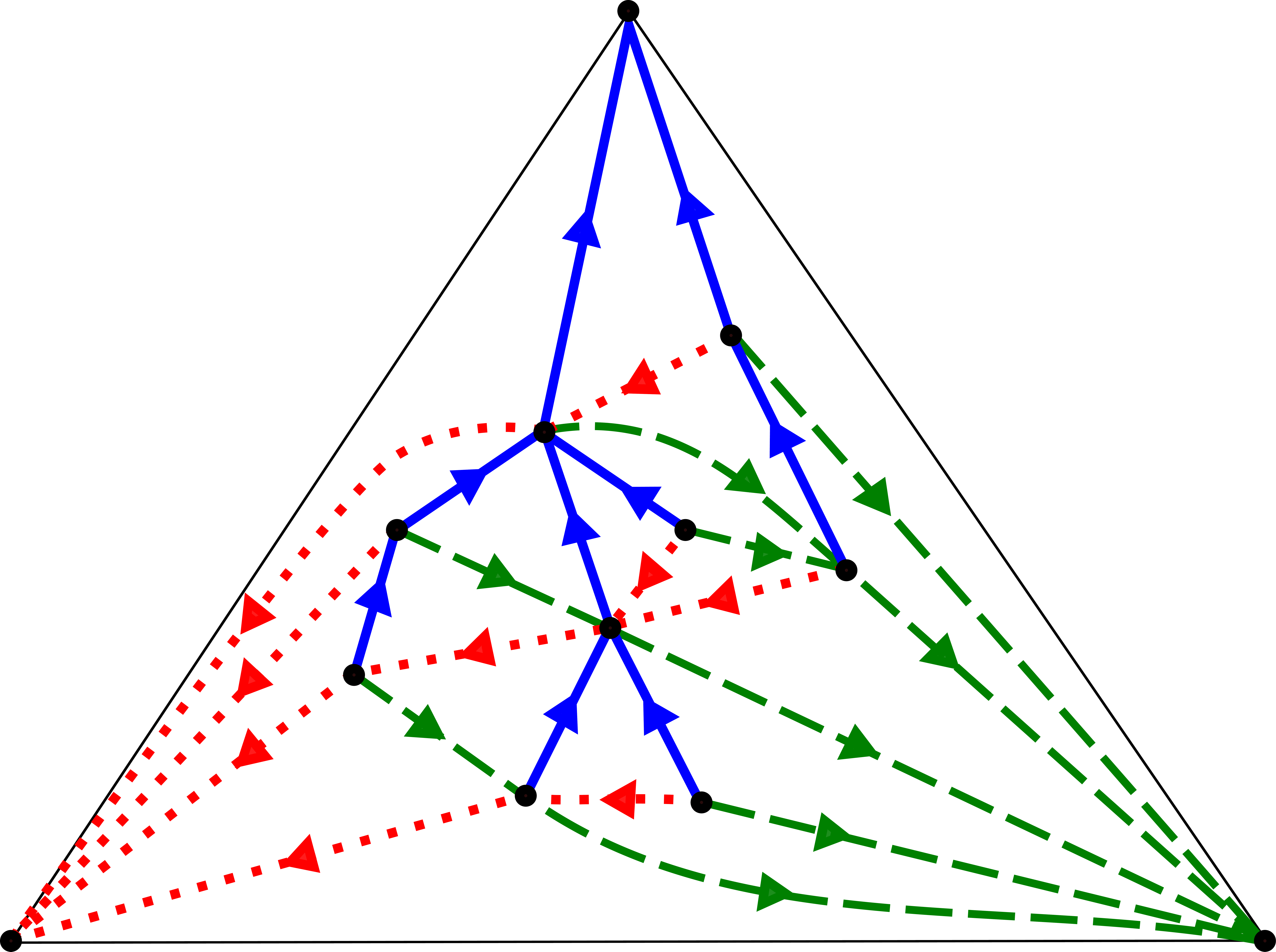}\put(100,0){$s_{green}$}  \put(-5,-5){$s_{red}$}\put(50,72){$s_{blue}$}
\end{overpic}\label{schnyderexample}}
\caption{(a) The vertex condition.  (b) An example of a 3-orientation of a triangulation on 9 internal vertices.}
\end{figure}

We begin with background on 3-orientations, Schnyder woods, and Markov chains.  The \emph{external vertices} and \emph{edges} are those on the infinite face of a planar triangulation.  All other vertices and edges are called \emph{internal}.  Let $\cT_n$ be the set of planar triangulations with $n$ internal vertices.  Given a triangulation $T\in \cT_n$, let $\stsp(T)$ be the set of 3-orientations of $T$, where a 3-orientation is an orientation of the internal edges of $T$ such that every internal vertex has out-degree 3 and every external vertex has out-degree 0.  Let $\stsp_n=\cup_{T\in \cT_n} \stsp(T)$ denote the set of all 3-orientations over all triangulations on $n$ internal vertices.  There are several interesting bijections between 3-orientations of triangulations and other combinatorial objects; for example, in Section~\ref{dyck} we present a bijection between $\stsp_n$ and pairs of Dyck paths.  Fraysseix and Ossona de Mendez defined a bijection between $\stsp(T)$ and the Schnyder woods of $T$ \cite{fm}.  A {\it Schnyder wood} (see Figure \ref{bijection}) is a 3-coloring and orientation of the edges of $T$ such that for every internal vertex $v$, 

\begin{itemize}
\item $v$ has out-degree exactly 1 in each of the 3 colors blue, red and green, and

\item the clockwise order of the edges incident to $v$ is: outgoing green, incoming blue, outgoing red, incoming green, outgoing blue and incoming red (see Figure \ref{vertexconditionpic}).
\end{itemize}

\
\noindent {\it In our figures, we differentiate the colors of edges in the Schnyder woods by dashed lines (green), dotted lines (red), and solid lines (blue).}  See Figure \ref{schnyderexample} for an example of a Schnyder wood.  The orientation of the edges of a Schnyder wood is a 3-orientation and each of the colors forms a directed tree which spans the internal vertices and is rooted at one of the external vertices.  See \cite{fm} for a proof of the bijection.  

It is often convenient to consider the Schnyder wood associated with a given 3-orientation, because the vertex condition allows us to infer some of the global information about the 3-orientation by looking locally at each vertex.  Throughout the paper, when we refer to the colors of the edges of a 3-orientation, we mean the colors of the Schnyder wood associated with that 3-orientation.  We will use the additional information provided by the bijection with Schnyder woods extensively throughout the proofs.  
 
Next, we present some background on Markov chains.  The time a Markov chain takes to converge to its stationary distribution $\pi$ is measured in terms of the distance between $\pi$ and $\prob^t$, the distribution at time~$t$. The \emph{total variation distance} at time~$t$ is
$$\|\prob^t,\pi\| _{tv} = \max_{x\in\stsp}\frac{1}{2}\sum_{y\in\stsp} |\prob^t(x,y)-\pi(y)|,$$
where $\prob^t(x,y)$ is the $t$-step transition probability and $\stsp$ is the state space.  For all $\epsilon>0$, the \emph{mixing time} $\tau$ of $\m$ is defined as
$$\tau (\epsilon)=\min \{t: \|\prob^{t'},\pi \|_{tv}\leq \epsilon, \forall t' \geq t\}.$$
We say that a Markov chain is \emph{rapidly mixing} if the mixing time is bounded above by a polynomial in $n$, where in this case, $n$ is the number of internal vertices of the triangulations.

%%%%%%%%%%%%%%%%%%%%%%%%%%%%%%%%%%%%%%%%%%%%%%%%%%%%%%%%%%%%%%%%%%%%%%%%%%%%%%%%%%%%%%%%%%%%%%%%%%%%%%%%%%%%%%%%%%%%%%%%%%%%%%%%%%%%%%%%%%%%%%%%%%%%%%%%%%

\section{Sampling 3-orientations of a fixed triangulation}\label{fixed}
\begin{figure} [!htb]
\centering
\subfloat[]{\begin{overpic}[scale=.12]{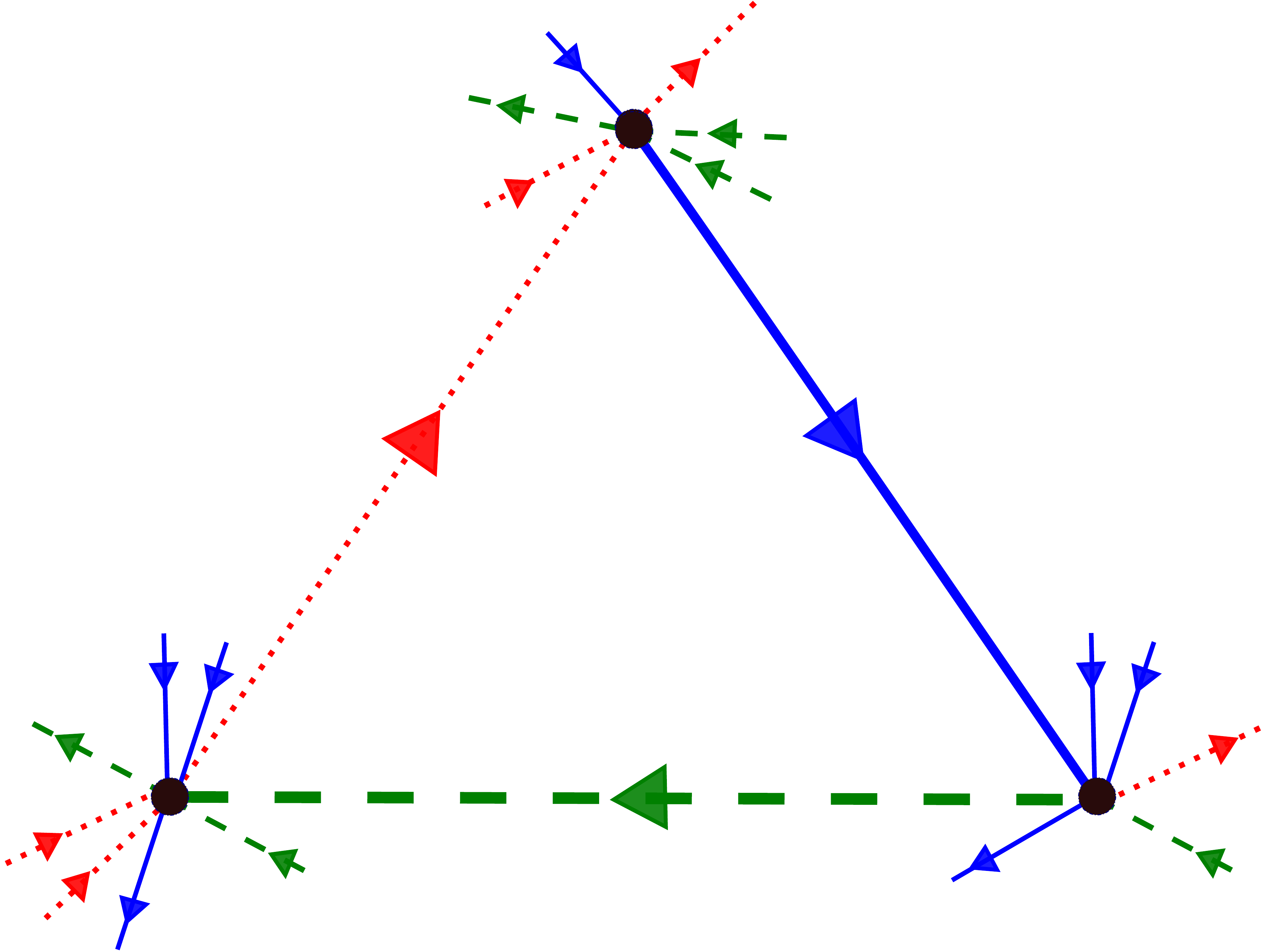}
 \put(49,70){$b$}
 \put(13,5){$a$}
 \put(85,5){$c$}
\end{overpic}\label{trianglemove1}}
\hspace{.5cm}
\subfloat[]{\begin{overpic}[scale=.12]{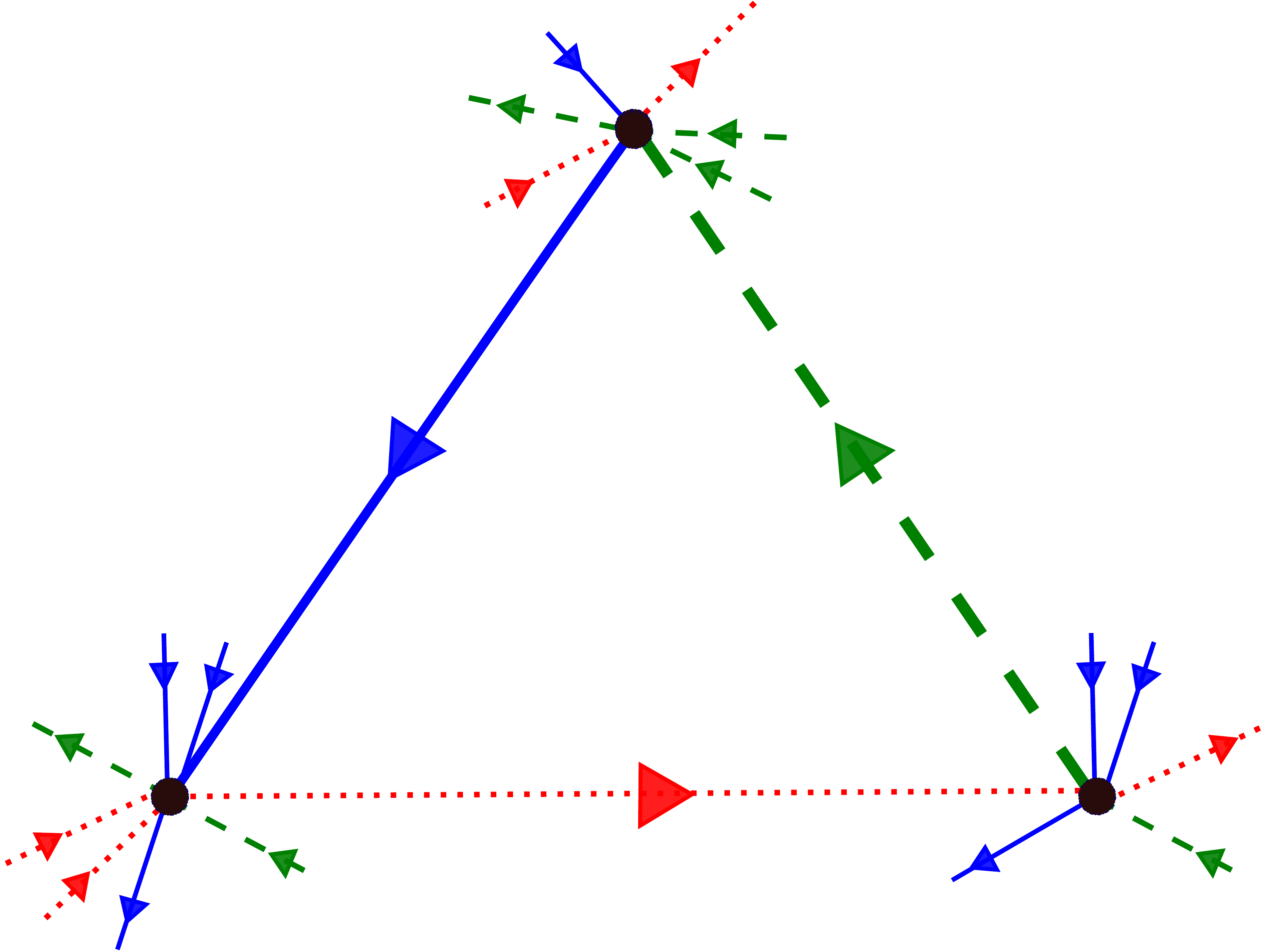}
 \put(49,70){$b$}
 \put(13,5){$a$}
 \put(85,5){$c$}
 \end{overpic}\label{trianglemove2}}
\caption{A move of $\mfix$.  The triangle $\triangle abc$ reverses to $\triangle acb$.}
\label{trianglemove}
\end{figure}

First we consider a Markov chain for sampling the 3-orientations of a given triangulation.  Let $T$ be a planar triangulation with $n$ internal vertices.  Consider the following natural local Markov chain $\mfix$ on the set of all 3-orientations of $T$.  Select a directed 3-cycle at random and reverse its orientation.  We will see that $\mfix$ samples from the uniform distribution, but its efficiency will depend on~$T$.  In Section~\ref{FixedFast} we show that if the maximum degree of internal vertices is at most 6, $\mfix$ is rapidly mixing.  In contrast, in Section~\ref{FixedSlow} we demonstrate a triangulation $T$ (with unbounded degree) for which $\mfix$ takes exponential time to sample from the state space $\stsp(T)$.  Define $\mfix$ as follows (see Figure~\ref{trianglemove}).  

\vspace{.05in} 
\noindent  \underline{{\bf The Markov chain $\mfix$}}

\vspace{.05in}
{
\noindent {\tt Starting at any $\sigma_0 \in \stsp(T)$, iterate the following:} 

 - {\tt Choose a triangle $t$ in $\sigma_i$ u.a.r.\footnote{The abbreviation u.a.r. stands for uniformly at random.} }

 - {\tt If $t$ is a directed cycle, with prob. $\frac{1}{2}$ reverse $t$ to obtain $\sigma_{i+1}.$}

 - {\tt Otherwise, $\sigma_{i+1}=\sigma_{i}$.}
}
\vspace{.05in} 

\noindent Brehm proved $\mfix$ connects the state space $\stsp(T)$~\cite{br}.  Since all valid moves have the same transition probabilities, $\mfix$ converges to the uniform distribution over the state space $\stsp(T)$.

\subsection{ Background on 3-orientations of Planar Triangulations}\label{referencedresults}

In this section we will provide an overview of several results on 3-orientations of planar triangulations and the Markov chain $\mfix$ which we will use in Section \ref{FixedFast} to show that $\mfix$ converges quickly when the maximum degree of the triangulation is at most 6.  Brehm~\cite{br} provides a detailed analysis of the robust structure of 3-orientations of planar graphs.  In particular, he constructs a framework which shows that the set of 3-orientations form a distributive lattice and that for a planar triangular graph $G$, any two 3-orientations of $G$ are connected by a series of moves of the Markov chain $\mfix$.  As part of this effort, Brehm examines a potential function on the faces of the graph, which we will see is useful to upper-bound the number of 3-orientations for a given triangulation and the maximum distance between any two 3-orientations.

To show connectivity of $\mfix$, Brehm considers first the case of 4-connected planar triangulations.  In this case, every triangle of $G$ is a face and he shows that it is possible to get between any two 3-orientations by a sequence of rotations of directed facial triangles.  Suppose now that a planar triangulation $G$ has exactly one triangle $t$ which does not bound a face.  For any non-facial cycle, Brehm shows that in any 3-orientation of $G$, the edges in the region bounded by that cycle that are incident to some vertex $v$ of the cycle must be directed towards $v$.  This implies that no face $f$ contained within $t$ that shares an edge with $t$ can be bounded by a directed triangle; hence such faces can never be rotated.  In fact, this implies (see~\cite{br} for details) that $G$ can be regarded as the cross product of the triangulation $G|_{t}$, the restriction of $G$ to the vertices on the boundary of $t$ and the vertices contained within the region bounded by $t$ and the triangulation $G\setminus t$ obtained by removing all vertices and edges contained within the region bounded by $t$.  Thus by allowing $\mfix$ to rotate arbitrary directed triangles, this amounts to extending $\mfix$ to the triangulation $G\setminus t$ which is now 4-connected.  The same arguments will hold when $G$ has many non-facial triangles.  Thus he obtains the following theorem.

\begin{Theorem}[Brehm]\label{Connectivity}
For any planar triangulation $G$, $\mfix$ connects the set of all 3-orientations of~$G$.
\end{Theorem}

\noindent In our setting, we use the fact that $G|_t$ is independent of $G\setminus t$ to show that the mixing time of $\mfix$ is the maximum of the mixing times of each 4-connected piece of $G$, subject to the delay which results from the fact that $\mfix$ only attempts to update one 4-connected piece at a time.  

Brehm defines a \emph{potential} $X$ of a 4-connected planar triangulation as follows.
\begin{Definition}
A \emph{potential} $X$ of a 4-connected planar triangulation $G$ is a mapping $f\rightarrow x_f$ from the interior faces to the natural numbers such that
\begin{itemize}
\item $x_f=0$ if the boundary of $f$ contains an exterior edge
\item $|x_f-x_g|\leq 1$ holds for any two adjacent faces $f,g$.
\end{itemize}
The \emph{value} of a potential $X$ is defined by $\sum_f x_f$.
\end{Definition}

It turns out that there is a bijection between 3-orientations of $G$ and a subset of the potentials of $G$, called \emph{induced} potentials, and that each move of $\mfix$ changes the potential of a face by $\pm 1$ (see \cite{br} for details).  Every triangulation $G\in \cT_n$ has $2n+1$ faces (not counting the infinite face).  It is easy to see that for any potential of $G$, the maximum value for any face is at most $\lfloor\frac{2n+1}{2}\rfloor$ since each face can only differ from its neighbors by at most 1 and faces adjacent to the boundary have value 0.  Thus, the maximum number of steps of $\mfix$ required to get between two 3-orientations of $G$ is bounded by the maximum value of any potential of $G$.  This implies Lemma~\ref{MaxDist}(a).  Moreover, the number of 3-orientations of a graph $G$ is bounded by the number of induced potentials of $G$.  Since each face in a potential is within 1 from each of its adjacent faces, the number of induced potentials is at most $3^{2n+1}$.

\begin{Lemma}\label{MaxDist}
Let $G$ be a 4-connected planar triangulation. 
\begin{enumerate}
\item[(a)] The maximum distance between two 3-orientations of $G$ is at most ${(2n+1)^2/2}. $
\item[(b)] The number of 3-orientations of $G$ is at most $3^{2n+1}.$
\end{enumerate}
\end{Lemma}

\subsection{$\mfix$ mixes rapidly for $\Delta_I(T)\leq 6$}\label{FixedFast}
\begin{figure} [!htb]
\centering
\includegraphics[scale=.08]{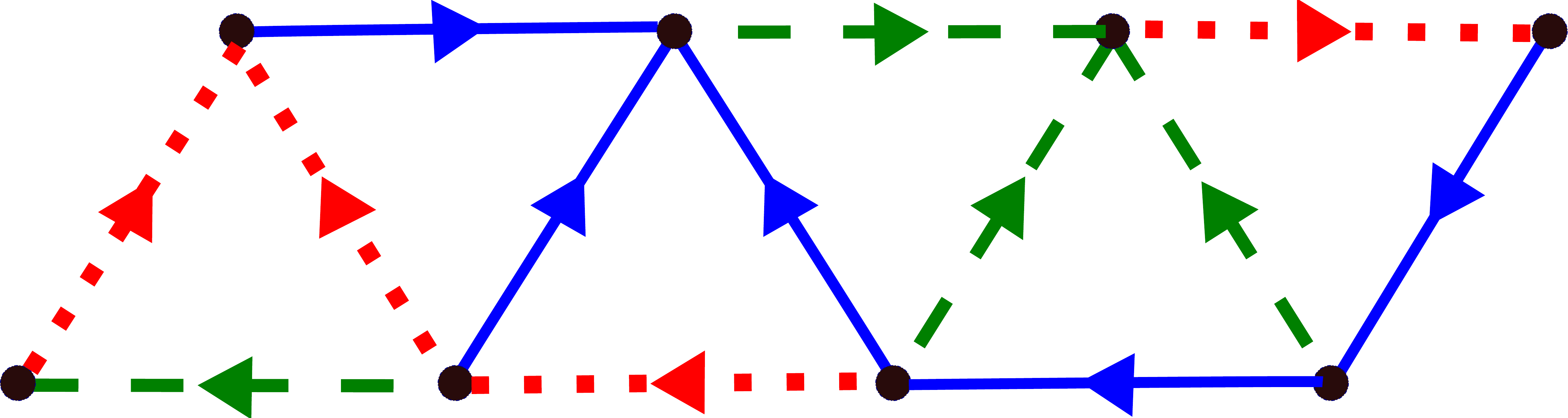}
 \put(-101,10){$f_1$}
 \put(-86,15){$f_2$}
 \put(-70,10){$f_3$}
 \put(-54,15){$f_4$}
 \put(-38,10){$f_5$}
 \put(-20,15){$f_6$}
  \hspace{.2 in}
 \includegraphics[scale=.08]{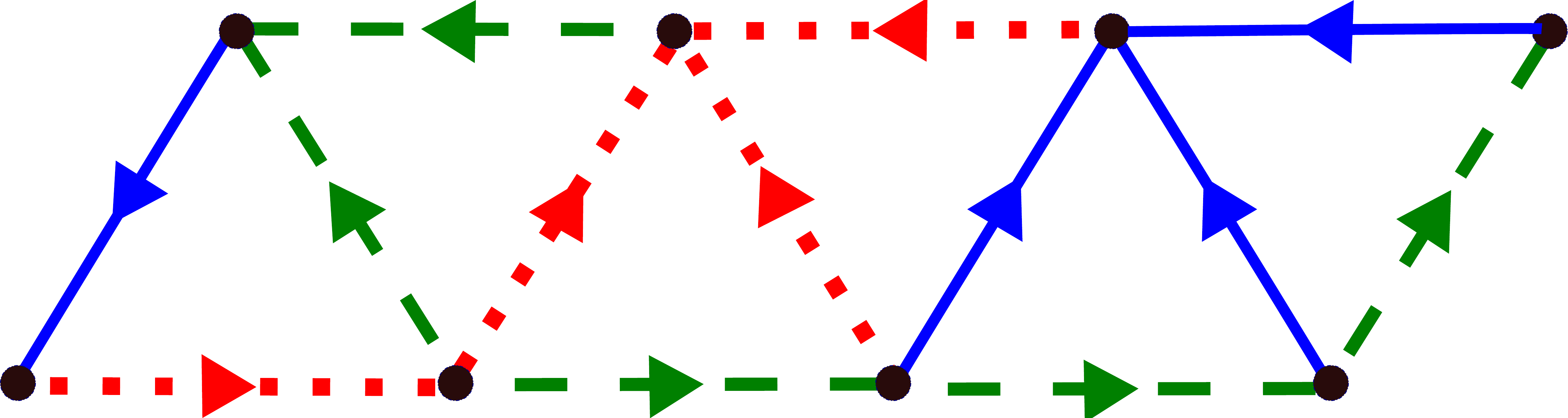}
 \put(-101,10){$f_1$}
 \put(-86,15){$f_2$}
 \put(-70,10){$f_3$}
 \put(-54,15){$f_4$}
 \put(-38,10){$f_5$}
 \put(-20,15){$f_6$}
 \caption{A tower of length 6.}
\label{tower}
\end{figure}
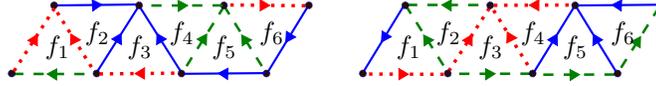

In this section we prove $\mfix$ is rapidly mixing on the state space $\stsp(T)$, if $T$ is a planar triangulation with $\Delta_I(T)\leq 6$.    
First, we introduce an auxilliary chain $\mtower$, which we will then we use to derive a bound on the mixing time of $\mfix$. The Markov chain $\mtower$ involves \emph{towers} of moves of $\mfix$, based on the nonlocal chain introduced in \cite{lrs}. 
Notice that if a face $f$ cannot move then two of its edges have the same orientation and the other edge does not (see, e.g., the face $f_1$ in Figure~\ref{tower}a).  We call this edge the \emph{disagreeing edge} of $f$.  Define a tower of length $k$ to be a path of faces $f_1,f_2,\ldots, f_k$ such that the following three conditions are met: $f_k$ is the only face which is bounded by a directed cycle (i.e. it has a move); for every $1\leq i<k$, the disagreeing edge of $f_i$ is also incident to $f_{i+1}$; and every vertex $v$ is incident to at most three consecutive faces in the path (see Figure~\ref{tower}).  The idea is that once the edges of $f_k$ are reversed, then the edges of $f_{k-1}$ can be reversed, and so on.  We call $f_1$ the beginning of the tower, and $f_k$ the end.  Notice that every face is the beginning of at most one tower (it may be a tower of length 1).  The effect of making these moves is to reverse the edges of the directed cycle surrounding the tower (although the colors on the internal edges also change).

\vspace{.2in}

\noindent  \underline{{\bf The Tower Markov chain $\mtower$}}

\vspace{.05in}{
\noindent {\tt Starting at any $\sigma_0$, iterate the following:} 

$\bullet$ \ {\tt Choose a (finite) face $f$ u.a.r.  }

$\bullet$ \ {\tt If $f$ is the beginning of a tower of length $k$, then \\
\indent w.p.  $\begin{cases}\frac{1}{6k}: & k\geq 2\\ \frac{1}{2}: & k=1\end{cases}$ reverse this tower to obtain $\sigma_{i+1}.$}

$\bullet$ \ {\tt Otherwise, $\sigma_{i+1}=\sigma_{i}$.}

\vspace{.2in}
}
\comment{
\begin{itemize}
\item{\tt Choose a (finite) face $f$ u.a.r.  }

\item{\tt If $f$ is the beginning of a tower of length $k$, then with \\ prob.
  $\begin{cases}\frac{1}{6k}: & k\geq 2\\ \frac{1}{2}: & k=1\end{cases}$ reverse this tower to obtain $\sigma_{i+1}.$}

\item{\tt Otherwise, $\sigma_{i+1}=\sigma_{i}$.}

\end{itemize}}

\comment{The Markov chain $\mtower$ operates as follows.  Starting at any $\sigma_0$, iterate the following: Choose a (finite) face $f$ u.a.r.; if $f$ is the beginning of a tower of length $k$, then with probability
  $\begin{cases}1/(6k) & \text{if } k\geq 2\\ 1/2 & \text{if } k=1\end{cases}$ reverse this tower to obtain $\sigma_{i+1};$ else, $\sigma_{i+1}=\sigma_{i}$.}
The moves of $\mfix$ are a subset of the moves of $\mtower$, so $\mtower$ is connected as well.    

We first consider the case that $T$ is 4-connected.  Notice if $T$ is 4-connected, every 3-cycle is facial, so $\mfix$ selects a face and rotates the edges around that face if possible.  
The bulk of the work to prove Theorems~\ref{AllTriangulationsTower} and~\ref{AllTriangulations} is to show that $\mtower$ is rapidly mixing when $T$ is 4-connected.

\begin{Theorem}\label{4conn}
Let $T$ be a 4-connected planar triangulation with $\Delta_I(T)\leq 6$.  Then the mixing time of $\mtower$ on the state space $\stsp(T)$ satisfies $$\tau(\epsilon) =O(n^5\ln \epsilon^{-1}).$$
\end{Theorem}
First we prove Theorem~\ref{4conn} and then we apply the comparison theorem to prove that $\mfix$ is also rapidly mixing and extend the result to non-4-connected triangulations using a result of Brehm~\cite{br}, proving Theorems~\ref{AllTriangulationsTower} and~\ref{AllTriangulations}. The main tool we will use to prove Theorem~\ref{4conn} is the following path coupling theorem, due to Dyer and Greenhill \cite{dg}.  

\begin{Theorem}[Dyer and Greenhill] \label{path}Let $\phi$ be an integer-valued metric on $\stsp \times \stsp$ which takes values in $\{0,...,B\}$.  Let $U$ be a subset of $\stsp \times \stsp$ such that for all $(x_t,y_t) \in \stsp \times \stsp$ there exists a path $x_t = z_0,z_1,...,z_r = y_t$ between $x_t$ and $y_t$ such that $(z_i,z_{i+1}) \in U$ for $0\leq i < r$ and $\sum_{i=0}^{r-1}\phi(z_i,z_{i+1})=\phi(x_t,y_t).$  Let $\m$ be a Markov chain on state space $\stsp$ with transition matrix $P$.  Consider any random function $f: \stsp \rightarrow \stsp$ such that $P[f(x) = y] = P(x,y)$ for all $x,y \in \stsp$, and define a coupling of the Markov chain by $(x_t,y_t) \rightarrow (x_{t+1},y_{t+1})=(f(x_t),f(y_t))$.  If $E[\phi(x_{t+1},y_{t+1})]\leq \phi(x_t,y_t)$, for all $x_t,y_t \in U$, let $\alpha > 0$ satisfy $\Pr [\phi(x_{t+1},y_{t+1}) \neq \phi(x_t,y_t)] \geq \alpha$ for all $t$ such that $x_t \neq y_t$.  Then the mixing time of $\m$ on the state space $\stsp$ satisfies
$$\tau (\epsilon) \leq 2\left\lceil\frac{eB^2}{\alpha}\right\rceil\lceil \ln \epsilon^{-1}\rceil.$$
\end{Theorem}

\vspace{.1in}
\noindent {\em Proof of Theorem \ref{4conn}.}
Let $T$ be a 4-connected planar triangulation with $\Delta_I(T) \leq 6$.  
Define the distance $d$ between any two 3-orientations in the state space $\stsp(T)$ to be the minimum number of steps of $\mfix$ from one to the other.  Assume $\sigma, \tau\in \stsp(T)$ and $\tau$ is obtained from $\sigma$ by reversing a facial triangle $f$.  We use the trivial coupling, which chooses the same face for $\sigma$ and $\tau$ at every step.  Suppose without loss of generality that the edges of $f$ are clockwise in $\sigma$.  There are two obvious moves that decrease the distance, namely when the $\mtower$ selects the face $f$ and chooses to direct the cycle clockwise or counterclockwise, each of which happens with probability $1/(2(2n+1))$.  Moreover, any move of $\mtower$ that does not involve an edge of $f$ occurs with the same probability in $\sigma$ and $\tau$, and hence is neutral.

We call a tower \emph{bad} if it contains a neighbor $f'$ of $f$ that is not the end of the tower.  In this case, we say this bad tower is \emph{associated with} $f'$.  On the other hand, a tower is \emph{good} if it ends in $f$, or if it ends in a face $f'$ adjacent to $f$ and contains no other faces adjacent to $f$.  We will show that the good towers in $\sigma$ have corresponding good towers in $\tau$, while the bad towers in $\sigma$ fail in $\tau$, and therefore increase the distance.  Any tower that is neither good nor bad does not contain an edge of $f$, so it is neutral with respect to the distance.

\begin{figure} [!htb]
\centering
\subfloat[]{\begin{overpic}[scale=.12]{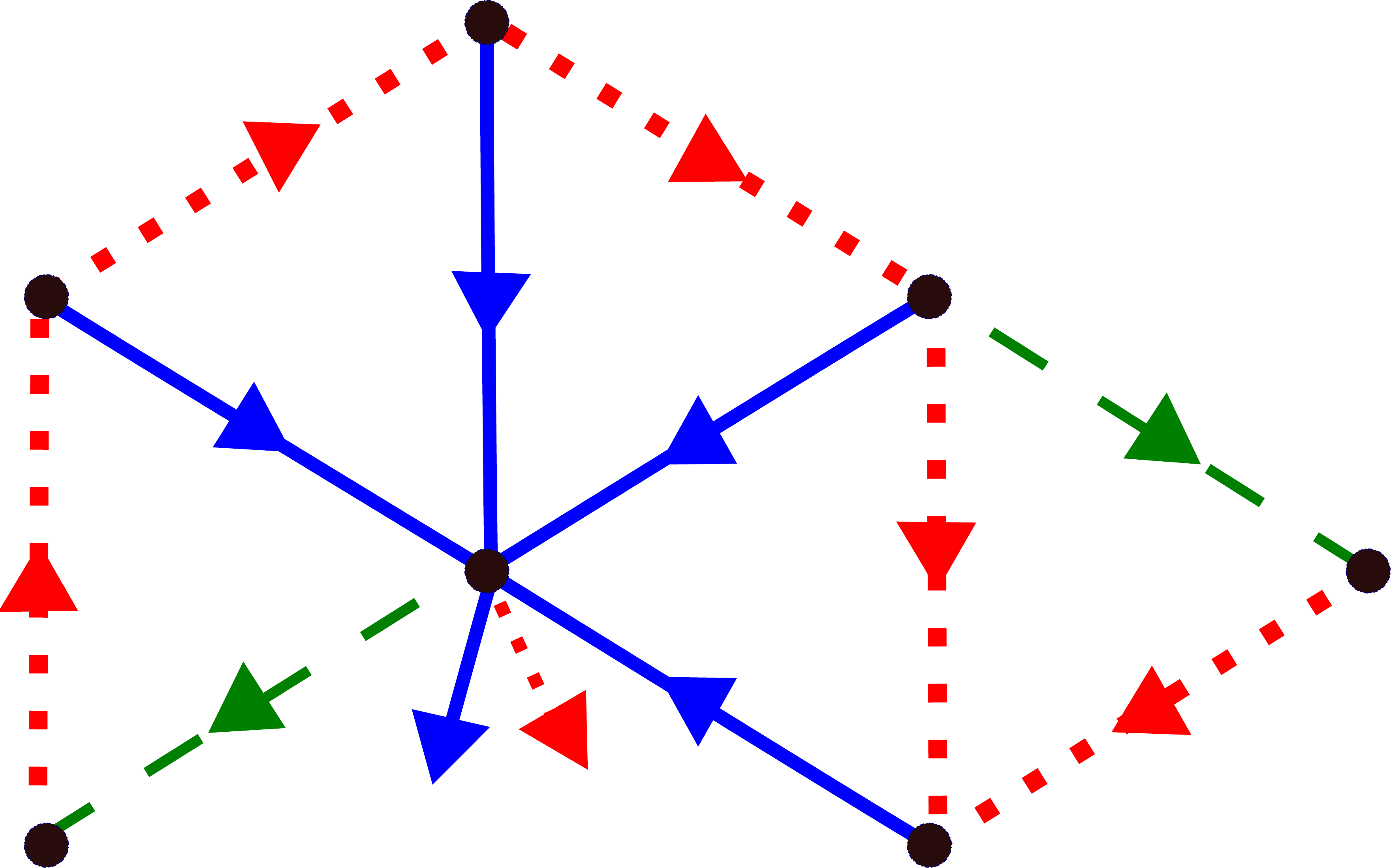}
\small
\put(12,19){$f$}
\put(30,27){$v$}
\put(20,39){$f_{k}$}
\put(40,39){$f_{k-1}$}
\put(48 ,19){$f_{k-2}$}
\put(72,19){$f_{k-3}$}
\end{overpic}}
\hspace{1in}
\subfloat[]{\begin{overpic}[scale=.12]{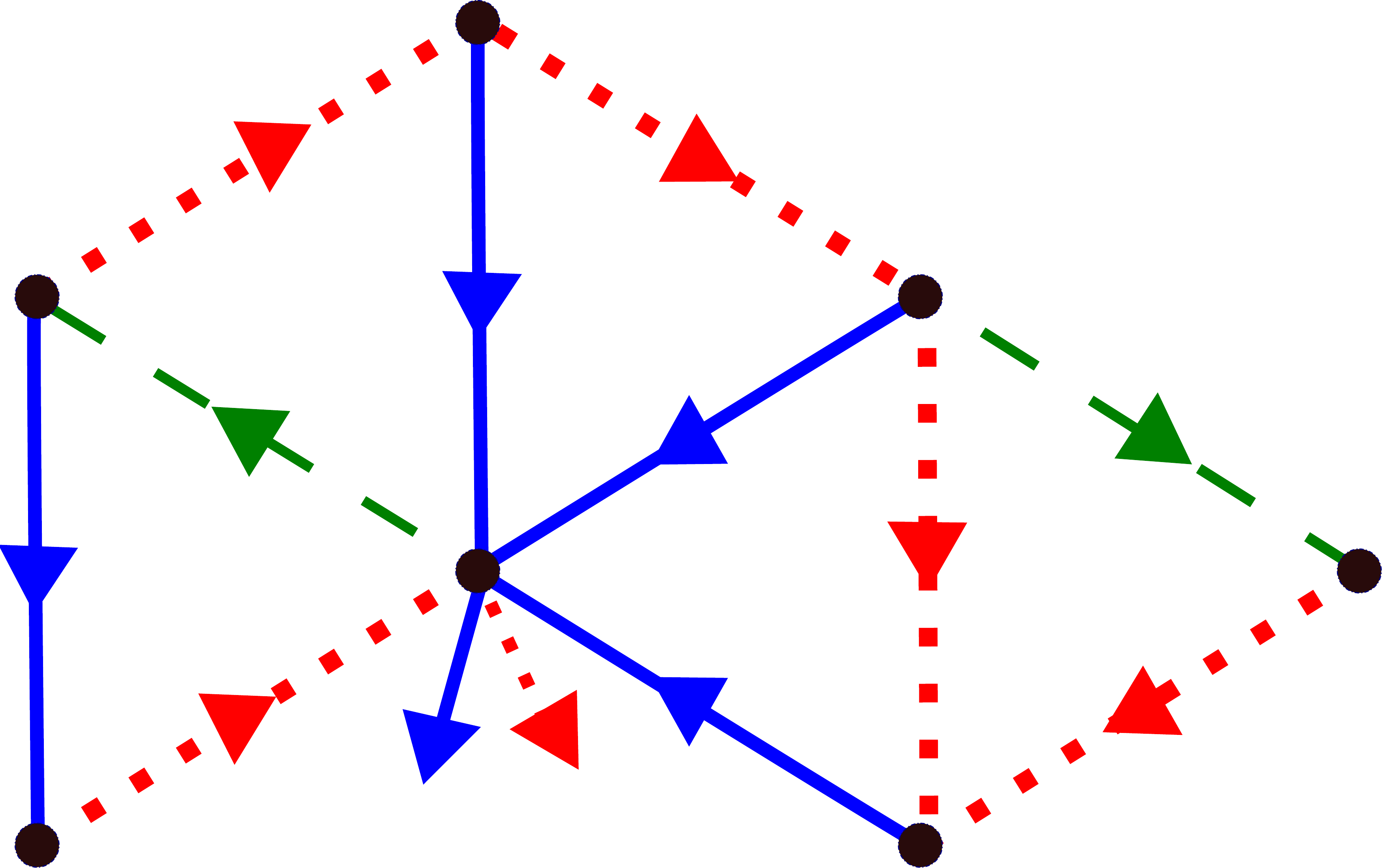}
\small
\put(12,19){$f$}
\put(29,27){$v$}
\put(20,39){$f_{k}$}
\put(40,39){$f_{k-1}$}
\put(48 ,19){$f_{k-2}$}
\put(72,19){$f_{k-3}$}
  \end{overpic}}
\caption{A tower $(f_1,f_2,\ldots,f_k)$ in $\sigma$ for which $(f_1,f_2,\ldots, f_k,f)$ is not a tower in $\tau$.  Notice deg($v)\geq 7$.}
\label{indegree4}
\end{figure}
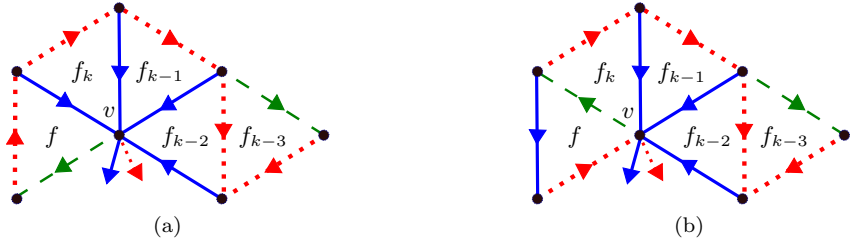

Suppose $k\geq 1$, $(f_1,f_2,\ldots, f_k)$ is a good tower in $\sigma$, and $f_k$ is adjacent to $f$.  We claim that $(f_1,f_2,\ldots, f_k, f)$ is a good tower in $\tau$.  It is clear that in $\tau$, $f$ is the only one of these faces that is bounded by a cycle, and that upon rotating $f$, the tower $(f_1,f_2,\ldots, f_k)$ is possible.  We must check two things: that $(f_1,f_2,\ldots, f_k, f)$ is a path of faces (i.e. does not contain any cycle of faces), and that every vertex is incident to at most three consecutive faces.  The first condition is clear, since $f_k$ is the only neighbor of $f$ in $\{f_1,f_2,\ldots, f_k\}$, and $(f_1,f_2,\ldots, f_k)$ is a path of faces.  Suppose the second condition does not hold.  Then there is a vertex $v$ incident to $f, f_k, f_{k-1}$, and $f_{k-2}$.  The edges between faces $f_{k-2}$ and $f_{k-1}$ and between $f_{k-1}$ and $f_{k}$ are either both incoming to $v$ or both outgoing from $v$ (see Figure~\ref{indegree4}).  Moreover, since the edge between $f_{k-2}$ and $f_{k-1}$ is the disagreeing edge of $f_{k-2}$, the two edges  of $f_{k-2}$ incident to $v$ are either both incoming to $v$ or both outgoing from $v$ (similarly the two edges of $f_{k}$ incident to $v$ are either both incoming to $v$ or both outgoing).  Hence there are four edges incident to $v$ which are all incoming or all outgoing; a contradiction since a vertex of degree at most 6 with exactly three outgoing edges can have at most three incoming edges as well.
Therefore if a good tower of length $k\geq 1$ begins on a face $f_1$ and ends on a neighbor $f_k$ of $f$ in $\sigma$ then there is a corresponding tower of length $k+1$ that begins on $f_1$ and ends on $f$ in $\tau$. 

Thus we have shown that if  $(f_1,f_2,\ldots, f_k)$ is a good tower in $\sigma$, then $(f_1,f_2,\ldots$
$\ldots, f_k, f)$ is a good tower in $\tau$.  
On the other hand, it should be clear that if $(f_1,f_2,\ldots, f_k, f)$ is a good tower of length $k+1\geq 2$ that ends on $f$ in $\sigma$ then $(f_1,f_2,\ldots, f_k)$ is a good tower of length $k$ in $\tau$.  In either case, if $k\geq 2$ then the expected change in distance given the choice of these towers is 
$$\frac{1}{2n+1}\left(-\frac{1}{6(k+1)}+k\left(\frac{1}{6k}-\frac{1}{6(k+1)}\right)\right)=0.$$  If $k=2$ then the expected change in distance is 
$$\frac{1}{2n+1}\left(-\frac{1}{12} + \left(\frac{1}{2}-\frac{1}{12}\right)\right)=\frac{1}{3(2n+1)}.$$ 

We point out that if $\sigma$ and $\tau$ have good towers using a neighbor $f'$ of $f$ then no bad tower in $\sigma$ or $\tau$ is associated with $f'$; that is, if there exists a bad tower containing $f'$ then $f'$ is the end of the tower.  Suppose without loss of generality that the good tower is longer in $\tau$ than in $\sigma$.   
Then the edge between $f$ and $f'$ is the disagreeing edge of $f'$ in $\sigma$ so the only way to tower is towards $f$, so $f'$ is not in a bad tower in $\sigma$.  On the other hand, in $\tau$, $f'$ is bounded by a cycle, so it must be the end of any tower containing it.  
We claim that $\sigma$ and $\tau$ can have at most two bad towers associated with a given face $f'$ adjacent to $f$.  It is clear that $\sigma$ (resp., $\tau$), has at most one bad tower that begins in $f'$, which is defined by the disagreeing edge of $f'$ in $\sigma$.  However, $\sigma$ may have a bad tower that uses $f'$ but does not begin in $f'$.  Let $(f_1,f_2,\ldots, f_k)$ be such a tower.  We will show that $f'=f_2$; suppose not, so that $f'=f_i,$ where $i\geq 3$.  Then as above, there is a vertex $v$ that is incident to $f, f_i, f_{i-1},$ and $f_{i-2}$, and the same proof will show that $v$ must have either in-degree at least 4 or out-degree at least 4, which is a contradiction.  Therefore bad towers associated with $f'$ must either begin in $f'$ or in a neighbor $f_1$ of $f'$.  Moreover, if there is a bad tower in $\sigma$ ($\tau$) beginning in $f_1$ then in both $\sigma$ and $\tau$, the edge between faces $f'$ and $f_1$ is $f_1$'s disagreeing edge, which means that $\tau$ (resp. $\sigma$) cannot have a bad tower beginning in $f'$.  Therefore there are at most two bad towers in $\sigma$ or $\tau$ associated with $f'$.  The expected change in distance given that a bad tower of length $k\geq 2$ is chosen is $2k/(6k(2n+1))=1/(3(2n+1))$.  Therefore
$$\E[\Delta d] \leq \frac{1}{2n+1}\left( -2\left(\frac{1}{2}\right) + 3\left(\frac{1}{3}\right)\right)=0.$$

We will see that for any $(\sigma_t,\tau_t)$ (not necessarily at distance 1), $\text{Pr}( d(\sigma_t,\tau_t) \neq d(\sigma_{t+1},\tau_{t+1})) \geq 1/(4n+2).$  This follows from the connectivity proof of Brehm~\cite{br}.  For any such pair, there exists a path of transitions of $\mfix$ from $\sigma_t$ to $\tau_t$ of length $d(\sigma_t,\tau_t)$, and the first of these transitions occurs with probability at least $1/(4n+2)$, since each face is selected with probability at least $1/(2n+1)$, and the move succeeds with probability $1/2$.  Clearly this move decreases the distance (the number of steps between the configurations) by 1.
By Theorem~\ref{path} and the bound on the distance between any two 3-orientations given in Lemma~\ref{MaxDist}a, we see the mixing time of $\mtower$ over 4-connected triangulations satisfies
$$\tau(\epsilon) \leq \lceil e(2n+1)^5/2\rceil \lceil\ln \epsilon^{-1} \rceil = O(n^5 \ln \epsilon^{-1} ) .$$
\qed

Next, we prove Theorems~\ref{AllTriangulationsTower} and~\ref{AllTriangulations}.  We will use Theorem~\ref{4conn} and the comparison method, Theorem~\ref{Comparison}, introduced in Section \ref{mix_mvar} to bound the mixing time of $\mfix$ in terms of the mixing time of $\mtower$ in the case of 4-connected triangulations. We will also extend the analysis to all planar triangulations by showing that $\mfix$ operates as a product of independent Markov chains, each acting on a 4-connected planar triangulation.  Thus, we will need one final detail, which is the following straightforward theorem, proved in~\cite{bmrs}.

\begin{Theorem}
  \label{crossproductthm}
  Suppose the Markov chain $\m$ is a product of $M$ independent Markov
  chains $\m_1,\m_2,\ldots, \m_M$, where $\m$ updates $\m_i$ with probability
  $p_i$, where $\sum_i p_i=1$. If $\tau_i(\epsilon)$ is the mixing time for
  $\m_i$ on the state space $\stsp$ and $\tau_i(\epsilon)\geq 4\ln\epsilon$ for each $i$, then the mixing time of $\m$ on the state space $\stsp$ satisfies
  \[\tau(\epsilon)\leq \max_{i=1,2,\ldots, M} \ \frac{2}{p_i}\ 
  \tau_i\left(\frac{\epsilon}{2M}\right).\]
\end{Theorem}

\vspace{.2in}
\noindent {\em Proof of Theorems~\ref{AllTriangulationsTower} and~\ref{AllTriangulations}.}

First, we compare $\mtower$ with $\mfix$ using Theorem~\ref{Comparison} to derive a bound on the mixing time of $\mfix$ in the case of 4-connected planar triangulations.
To do so we need to bound the constant $A$ given in that theorem.  For each edge $(\sigma, \tau)$ in $\mfix$ that takes a counterclockwise cycle $f$ and makes it clockwise, let $\Gamma(\sigma, \tau)$ denote the set of edges $(x,y)$ of $\mtower$ such that the tower $f_1,f_2,\ldots, f_k$ that takes $x$ to $y$ contains the face $f$.  Since a tower is uniquely determined by its direction and its length, $|\Gamma(\sigma, \tau)|\leq 3n^2$.  Define $A:=\max_{(\sigma, \tau)} \left(| \Gamma(\sigma,\tau)|\right)\leq 3n^2.$
Since each 3-orientation has the same stationary probability, Lemma~\ref{MaxDist}b implies that the minimum weight of any state is $\pi_* \geq 3^{-(2n+1)}.$  Therefore, by Theorems~\ref{Comparison} and~\ref{4conn}, the mixing time of $\mfix$ over 4-connected planar triangulations satisfies
\begin{align*}
\tau(\epsilon)& \leq 4\frac{\log(\frac{1}{\epsilon \pi_*})}{\log(1/(2\epsilon))}A\tau'\\
& = O\left( \frac{\log(\frac{ 3^{2n+1}}{\epsilon})}{\log(1/(2\epsilon))}\ n^2\cdot n^5\ln \epsilon^{-1}\right)\\
& = O\left( \frac{n - \log{\epsilon}}{\log(1/(2\epsilon))}\ n^7\ln \epsilon^{-1}\right)\\
&= O(n^8\ln \epsilon^{-1}).
\end{align*}

Finally, we can extend this to non-4-connected planar triangulations, where $\mfix$ may select non-facial triangles.  
Brehm~\cite{br} proves that if $T$ has a non-facial triangle $C$, the edges on its interior that are incident to $C$ must point towards $C$.  This implies that for all $\sigma\in \stsp(T)$ every face on the interior of $C$ that contains an edge of $C$ is not bounded by a directed cycle, so cannot rotate, regardless of the orientation of $C$.  Thus $\mfix$ acts completely independently on the interior of $C$ from the exterior of $C$, and its mixing time is independent of the mixing time of the exterior as well.  

Let $T$ be a planar triangulation and assume that $C_1,C_2,\ldots, C_k$ are all the non-facial triangles of $T$.  Let $T_i$ be the triangulation consisting of all faces contained within $C_i$ and not within any other non-facial triangle contained within $C_i$.  Let $\tau_i$ be the mixing time of $\mfix$ on $T_i$,$F_i$ be the number of faces within $C_i$, and $n_i$ be the number of internal vertices to $T_i$.  Then $n=\sum_i n_i$ and the number of faces in $T$ is $\sum_{i} F_i$.  Therefore, by Theorem~\ref{crossproductthm}, the mixing time of $\mfix$ on $T$ will be 
\begin{eqnarray*}\label{anon4-conn}
\tau(\epsilon) &=& \max_{i}\left\{\tau_i\left(\frac{\epsilon}{2k}\right)\frac{2n+1}{F_i}\right\}\\
&=& \max_{i}\left\{O\left(n_i^8\ln\left(\frac{2k}{\epsilon}\right)\frac{2n+1}{F_i}\right)\right\}.
\end{eqnarray*}
This is maximized when $n_1=n, k=1$, and $F_1=2n+1$, so
$$\tau( \epsilon) = O(n^8\ln(\epsilon^{-1})).$$
This proves Theorem \ref{AllTriangulations}.  In fact, this shows that $\mfix$ is rapidly mixing on the state space $\stsp(T)$ for any planar triangulation $T$ whose 4-connected triangulations $T_1,T_2,\ldots, T_k$ each have maximum degree (of any internal vertex) 6.  Moreover, the same proof can be applied to the bound on $\mtower$ for 4-connected triangulations to extend to the case of non-4-connected triangulations, proving Theorem~\ref{AllTriangulationsTower}.

\qed

%%%%%%%%%%%%%%%%%%%%%%%%%%%%%%%%%%%%%%%%%%%%%%%%%%%%%%%%%%%%%%%%%%%%%%%%%%%%%%%%%%%%%%%%%%%%%%%%%%%%%%%%%%%%%%%%%

\subsection{$\mfix$ can mix slowly}\label{FixedSlow}
We now exhibit a triangulation on which $\mfix$ takes exponential time to converge.  A key tool is \emph{conductance}, which for an ergodic Markov chain with distribution $\pi$, is $\Phi = \min_{{S\subseteq\stsp}\atop{ \pi(S) \leq 1/2}}\sum_{s_1\in S, s_2\in \bar{S}}\pi(s_1)\prob(s_1,s_2)/{\pi(S)}.$
The following theorem relates the conductance and mixing time (see \cite{js}).
\begin{Theorem}\label{conductance}
For any Markov chain $\m$ with conductance $\Phi$,  the mixing time of $\m$ on state space $\stsp$ satisfies $$\tau (\epsilon) \geq ({4\Phi})^{-1} - 1/2.$$
\end{Theorem}

\begin{figure}[htb]
\centering
\includegraphics[scale=.08]{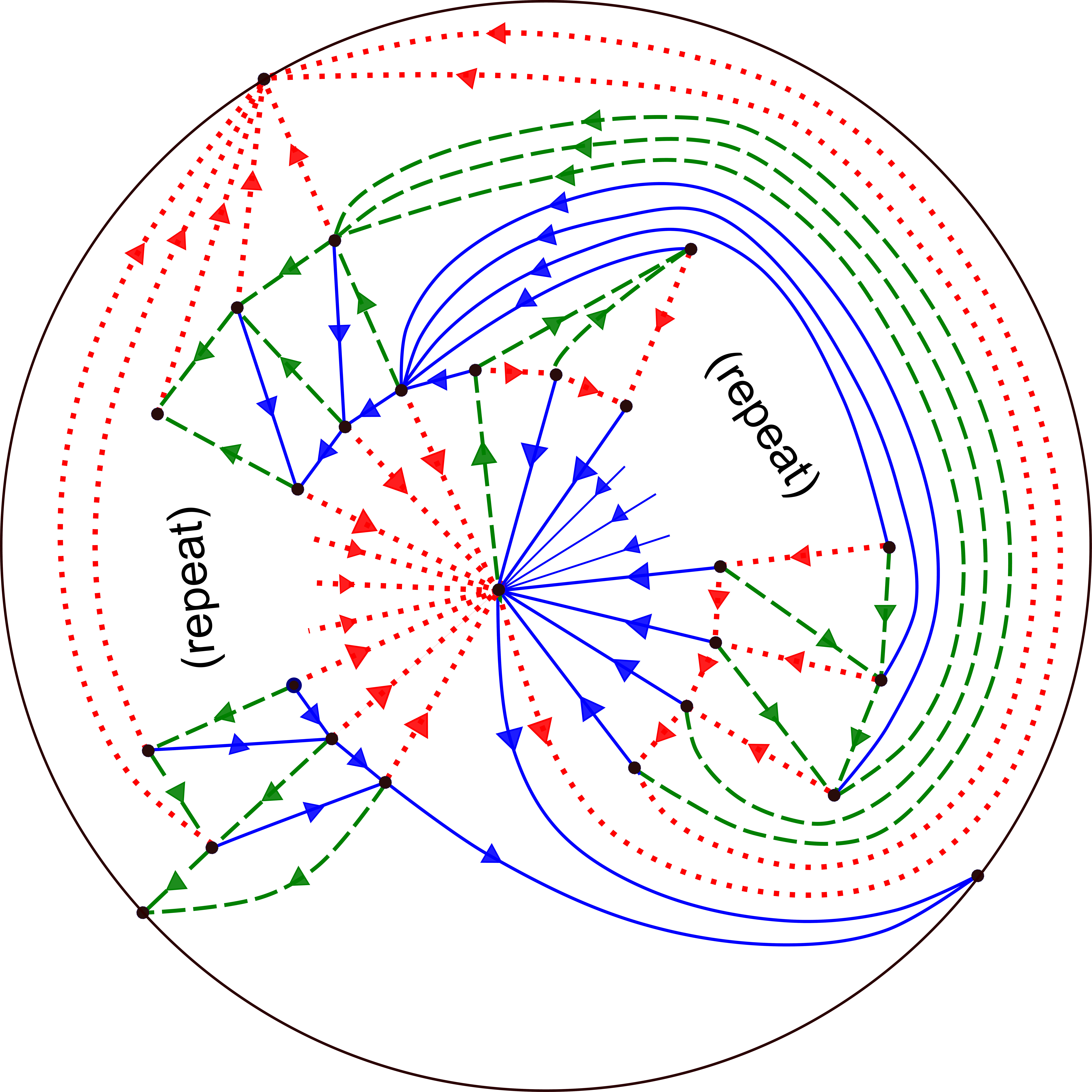}
\put(-225,35){$s_{green}$}\put(-190,210){$s_{red}$}\put(-20,40 ){$s_{blue}$}
\put(-205,50){${\scriptstyle v_{2t+2}}$}\put(-140,60){${\scriptstyle v_{1}}$}\put(-200,80){${\scriptstyle v_{2t+3}}$}
\put(-160,80){${\scriptstyle v_3 }$}\put(-150,70){$\scriptstyle{v_2 }$}
\put(-75,60){${\scriptstyle v_{4t-2} }$}\put(-115,65){$\scriptstyle{v_{2t+1} }$} \put(-66,81){${\scriptstyle v_{4t-3}}$}
\put( -99,79){${\scriptstyle v_{2t}}$}\put( -79,94){${\scriptstyle v_{2t-1}}$}\put(-65,114){${\scriptstyle v_{4t-4}}$}
\put(-82, 110){${\scriptstyle v_{2t-2}}$}
\put(-132,100){${\scriptstyle v_{0}}$}
\put(-150,133){${\scriptstyle v_{t-1}}$} \put(-175,120){${\scriptstyle v_{t-2}}$}
\put(-215,138){${\scriptstyle v_{3t-2}}$} \put(-198,160){${\scriptstyle v_{3t-1}}$}
\put(-169,174){${\scriptstyle v_{3t}}$} \put(-139,141){${\scriptstyle v_{t}}$}
\put(-126,152){${\scriptstyle v_{t+1}}$} \put(-107,150){${\scriptstyle v_{t+2}}$}
\put(-94,140){${\scriptstyle v_{t+3}}$} \put(-90,165){${\scriptstyle v_{3t+1}}$}
\caption{A triangulation for which $\mfix$ mixes slowly.}
\label{triangleslow}
\end{figure}
\comment{
\begin{figure}[htb]
\centering
\includegraphics[scale=.08]{BadCutMiddle.pdf}
\caption{A triangulation for which $\mfix$ mixes slowly.}
\label{triangleslow}
\end{figure}
}
\vspace{.1in}
\noindent \emph{Proof of Theorem \ref{slowmixthm}.}  
For the generalized triangulation $G$ given in Figure~\ref{triangleslow} with $n=4t-2$ internal vertices, $\mfix$ takes exponential time to converge.  Specifically, we show that although there is an exponential number of 3-orientations where the edge $(v_0,v_{t+1})$ is colored blue or red, all paths between these 3-orientations with $(v_0,v_{t+1})$ colored differently must include a 3-orientation where $(v_0,v_{t+1})$ is colored green.  Moreover, we show there is only a single 3-orientation that satisfies this property (namely, the one pictured in Figure~\ref{triangleslow}), which creates a bottleneck in the state space.  

Let $D$ be the set of 3-orientations of $G$ with $(v_0,v_{t+1})$ colored red or green and $\bar{D}$, be the set of 3-orientations with $(v_0,v_{t+1})$ colored blue.  In order to show that both $D$ and $\bar{D}$ are exponentially large we produce a triangulation in each set which contains roughly $t$ directed triangles which do not share any edges and reversing these triangles does not change the colors of the edges adjacent to $v_0$.  Hence each of the $2^t$ choices of the orientations of these triangles gives a distinct 3-orientation with edge $(v_0,v_{t+1})$ colored appropriated.

\begin{figure}[htb]
\centering
\includegraphics[scale=.08]{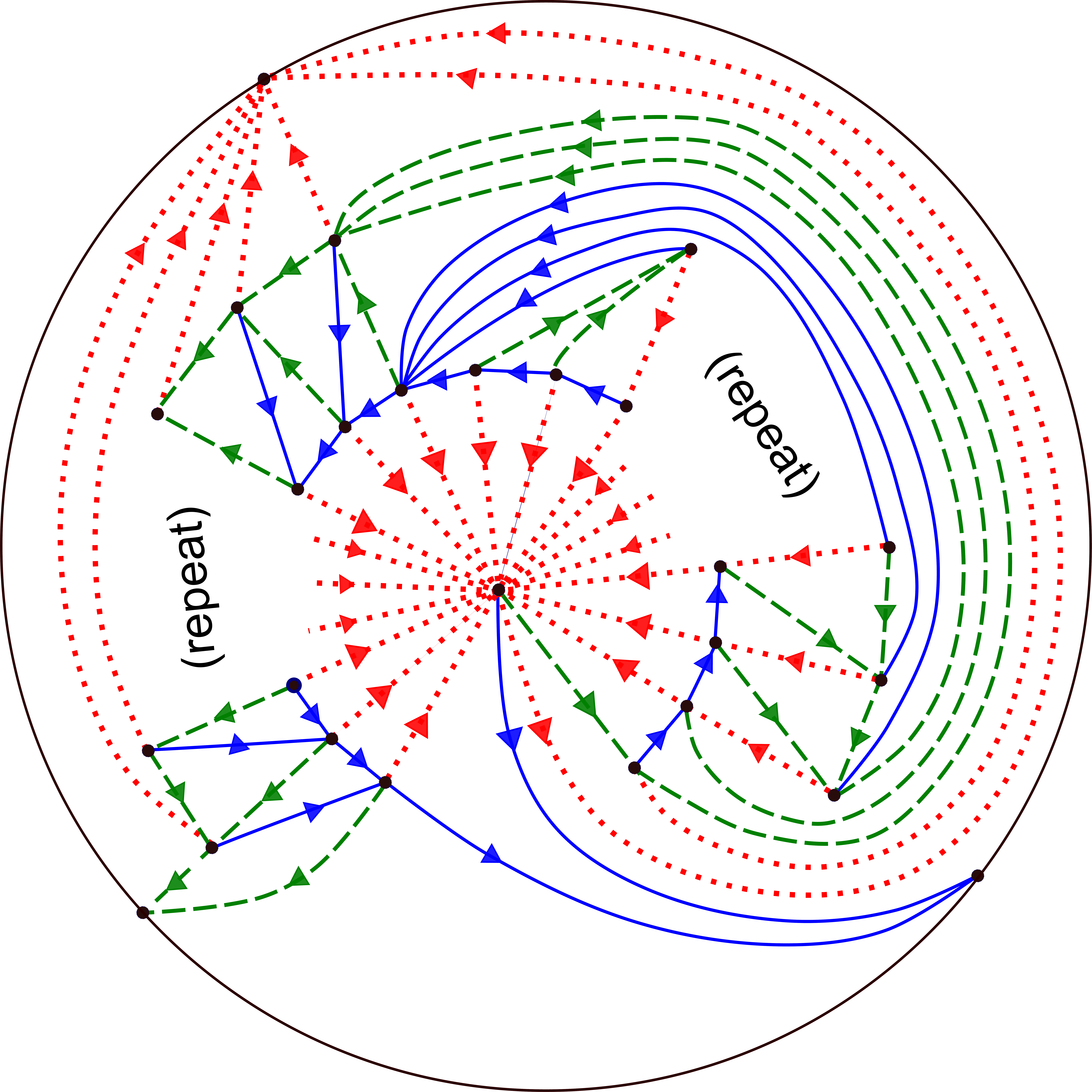}%\put(-110,-15){(a)}
\put(-225,35){$s_{green}$}\put(-190,210){$s_{red}$}\put(-20,40 ){$s_{blue}$}
\put(-132,100){${\scriptstyle v_{0}}$}
\put(-115,65){$\scriptstyle{v_{2t+1} }$}
\put(-126,152){${\scriptstyle v_{t+1}}$}
\put(-102,150){${\scriptstyle T_{t-2}}$}
\put( -74,94){${\scriptstyle T_2}$}
\put(-80,78){${\scriptstyle T_1}$}
\caption{There is an exponential number of 3-orientations with edge $(v_0,v_{t+1})$ colored red corresponding to the different orientations of triangles $T_1,T_2,\ldots,T_{t-2}$.}
\label{abadcutred}
\end{figure}

\begin{figure}[htb]
\centering
\includegraphics[scale=.08]{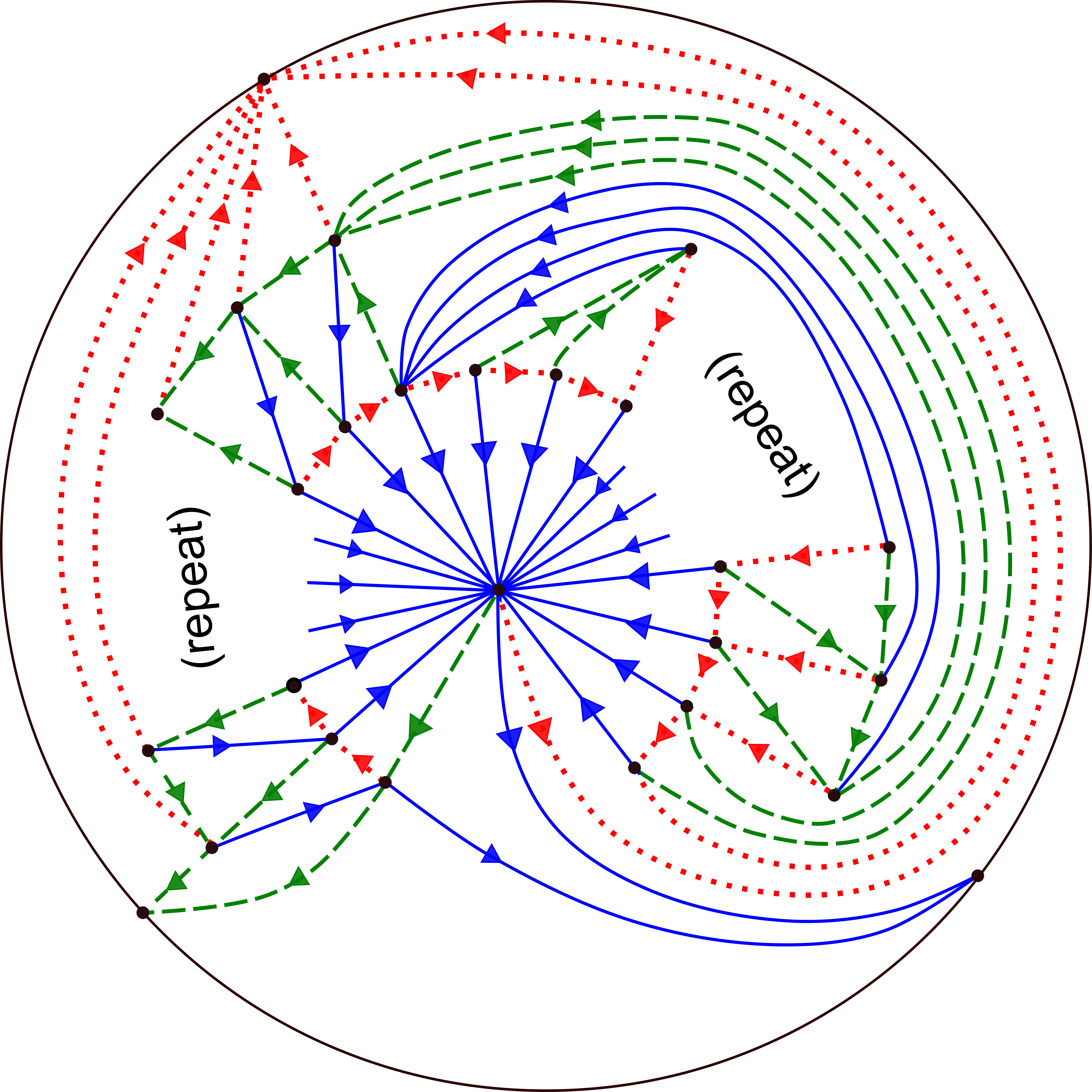}
\put(-225,35){$s_{green}$}\put(-190,210){$s_{red}$}\put(-20,40 ){$s_{blue}$}
\put(-141,63){${\scriptstyle v_{1}}$}\put(-126,152){${\scriptstyle v_{t+1}}$}\put(-142,100){${\scriptstyle v_{0}}$}
\put(-178,74){${\scriptstyle S_{t-2} }$}\put(-167,63){${\scriptstyle S_{t-1} }$}\put(-165,135){${\scriptstyle S_2}$} \put(-152,144){${\scriptstyle S_1}$}
\caption{There is an exponential number of 3-orientations with edge $(v_0,v_{t+1})$ colored blue corresponding to the different orientations of triangles $S_1, S_2,\ldots, S_{t-2}$.}
\label{abadcutblue}
\end{figure}

First consider the 3-orientation in Figure~\ref{abadcutred} which is in $D$.  Notice that triangles $T_1, T_2, .... T_{t-2}$ do not share any edges and reversing these triangles does not change the color of any edges adjacent to $v_0$.  Each of these triangles has 2 possible orientations and each of these $2^{t}$ choices of the orientations of the triangles gives a distinct 3-orientation with edge $(v_0,v_{t+1})$ colored red implying that 

\begin{equation}\label{aboundD}
|D| \geq 2^t = 2^{(n-6)/4}.
\end{equation}  

Next, consider the coloring in Figure~\ref{abadcutblue}.  Notice that triangles $S_1, S_2, .... S_{t-1}$ do not share any edges and reversing these triangles does not change the color of any edges adjacent to $v_0$.  Each of these triangles has 2 possible orientations and each of these $2^{t-1}$ choices of the orientations of the triangles gives a distinct 3-orientation with edge $(v_0,v_{t+1})$ colored blue implying that 
\begin{equation}\label{aboundbarD}
|\bar{D}| \geq 2^{t-1} = 2^{(n-2)/4}.
\end{equation}

Next, we show that there is only one 3-orientation of $G$ with $(v_0,v_{t+1})$ colored green, corresponding to Figure~\ref{triangleslow}.  By the Vertex Condition, if edge $(v_0,v_{t+1})$ is green then edges $(v_0,v_1), (v_0,v_2), ..., (v_0, v_t)$ must all be directed toward $v_0$ and colored red; this is because the edge $(v_0,s_{blue})$ is blue and directed towards $s_{blue}$ in every 3-orientation of $G$.  Similarly, edges $(v_0,v_{t+2}),$  $ (v_0,v_{t+3}), \ldots, (v_0, v_{2t+1})$ must all be blue and directed toward $v_0$.  Since $v_{t+1}$ has degree 4 and has the incoming green edge $(v_0,v_{t+1})$, the other edges incident to $v_{t+1}$ are determined; $(v_t,v_{t+1})$ is blue, $(v_{t+1}, v_{3t+1})$ is green and $(v_{t+1}, v_{t+2})$ is red all directed away from $v_{t+1}$.  Knowing the colors and direction of the edges incident to $v_0$ and $v_{t+1}$ forces the color and direction of the edges incident to $v_{t+2}$.  Similarly $v_{t+1}$ and $v_{t+2}$ force $v_{3t+1}$ and $v_0$, $v_{t+2}$ and $v_{3t+1}$ force $v_{t+3}$ and so on until the color and direction of all edges incident to vertices $v_{t+1},v_{t+2}\ldots, v_{2t+1}$ and $v_{3t+1}, v_{3t+2},\ldots, v_{4t-2}$ are forced.  Next consider $v_t$; we know the edges $(v_t,v_{t+1}), (v_t,v_{3t+1}), (v_t, v_{3t+2}), \ldots, (v_t, v_{4t-2})$ are all blue and directed toward $v_t$ which implies $(v_t,v_{3t})$ and $(v_t, v_{t-1})$ must be directed outward and green and blue respectively.  Now consider $v_{3t}$; we have already shown that all edges incident to $v_{3t}$ except for $(v_{3t},v_{t-1}), (v_{3t},v_{3t-1}), (v_{3t},v_{s_1})$ have a forced color and are directed inward.  Thus, these 3 edges must all be directed outwards with colors blue, green and red respectively.  Similarly, knowing the colors and directions of all edges incident to $v_{3t},v_t$ and $v_0$ forces the colors and directions of edges incident to $v_{t-1}$ and $v_{3t}, v_{t-1}$ forces $v_{3t-1}$ and so on for the remaining vertices.  Since all the edge colors and orientations are fixed, this implies there is a unique 3-orientation with $(v_0,v_{t+1})$ colored green.  To go from a configuration where edge $(v_0,v_{t+1})$ has color red (blue) to blue (resp., red) one must go through a coloring where the edge is green.  This is because the only choices for edge $(v_0,v_{t+1})$ are red directed toward $v_0$, blue directed toward $v_0$, and green directed away, and any move that changes the color must also change the direction.

Finally, given the bound on $|D|$ and $|\bar{D}|$, we derive a bound on the mixing time of $\mfix$.  Let $g$ be the single 3-orientation which has edge $(v_0,v_{t+1})$ colored green.  If $\pi(D) \leq 1/2$ then combining the definition of conductance with the bound on $|D|$ yields

$$
\begin{array}{rl}
\Phi_{\mfix} &\ \leq\ \frac{1}{\pi(D)} \sum_{d_1 \in D, d_2 \in \bar{D}}\pi(d_1)\prob(d_1,d_2)\\
&\ =\ \frac{1}{\pi(D)} \sum_{d_2 \in \bar{D}}\pi(g)\prob(g,d_2) \leq \frac{\pi(g)}{\pi(D)}  \leq \frac{1/Z}{2^{(n-6)/4}/Z} = \frac{1}{2^{(n-6)/4}}.\\
\end{array}
$$

\noindent If $\pi(D) > 1/2$ then $\pi(\bar{D}) \leq 1/2$ and so by detailed balance and the bound on $|\bar{D}|$,
$$
\begin{array}{rll}
\Phi_{\mfix} &\ \leq\ \frac{1}{\pi(\bar{D})} \sum_{d_1 \in \bar{D}, d_2 \in D}\pi(d_1)\prob(d_1,d_2) &\ = \ \frac{1}{\pi(\bar{D})} \sum_{d_1 \in \bar{D}}\pi(d_1)\prob(d_1,g) \\
&\ =\ \frac{1}{\pi(\bar{D})} \sum_{d_1 \in \bar{D}}\pi(g)\prob(g,d_1) &\ \leq \    \frac{\pi(g)}{\pi(\bar{D})} \leq \frac{1/Z}{2^{(n-2)/4}/Z} = \frac{1}{2^{(n-2)/4}}.\\
\end{array}
$$
In both cases, $\Phi_{\mfix} \leq 2^{-(n-6)/4}$.  Applying Theorem \ref{conductance} proves that the mixing time of $\mfix$ satisfies
$$\tau(\epsilon) \geq 2^{(n-14)/4)} - \frac{1}{2}.$$\qed
\begin{Remark}Combining this result with the comparison argument in Section \ref{FixedFast} shows that $\mtower$ can also take exponential time to converge. 
\end{Remark}

\comment{
Given that the local sampling algorithm is not always rapidly mixing, the obvious question is whether there exists some other algorithm for sampling 3-orientations of a fixed triangulation.  The obvious next attempt might be to allow the Markov chain to select any directed cycle and change its direction.  Indeed, a simple coupling argument implies that such a Markov chain will be rapidly mixing, as long as the Markov chain chooses each directed cycle of length $k$ with probability proportional to $\frac{1}{k}$.  However, this approach seems fatally flawed, as there is no known algorithm for sampling directed cycles with this distribution.
}

%%%%%%%%%%%%%%%%%%%%%%%%%%%%%%%%%%%%%%%%%%%%%%%%%%%%%%%%%%%%%%%%%%%%%%%%%%%%%%%%%%%%%%%%%%%%%%%%%%%%%%%%%%%%%%%%%%%%%%%%%%%%%%%%%%%%%

\section{Sampling the 3-orientations of triangulations on $n$ internal vertices}\label{vary}
\begin{figure}[htb]
\centering
\comment{
\includegraphics[scale=.12]{c2c1swap1.pdf}
\put(-97,42){$x$}
\put(-39,0){$z$}
\put(2,42){$y$}
\put(-39,80){$w$}
\put(-51,-15){(a)}
\hspace{1in}\includegraphics[scale=.12]{c2c1swap2.pdf}
\put(-97,42){$x$}
\put(-39,0){$z$}
\put(2,42){$y$}
\put(-39,80){$w$}
\put(-51,-15){(b)}}
\includegraphics[scale=.07]{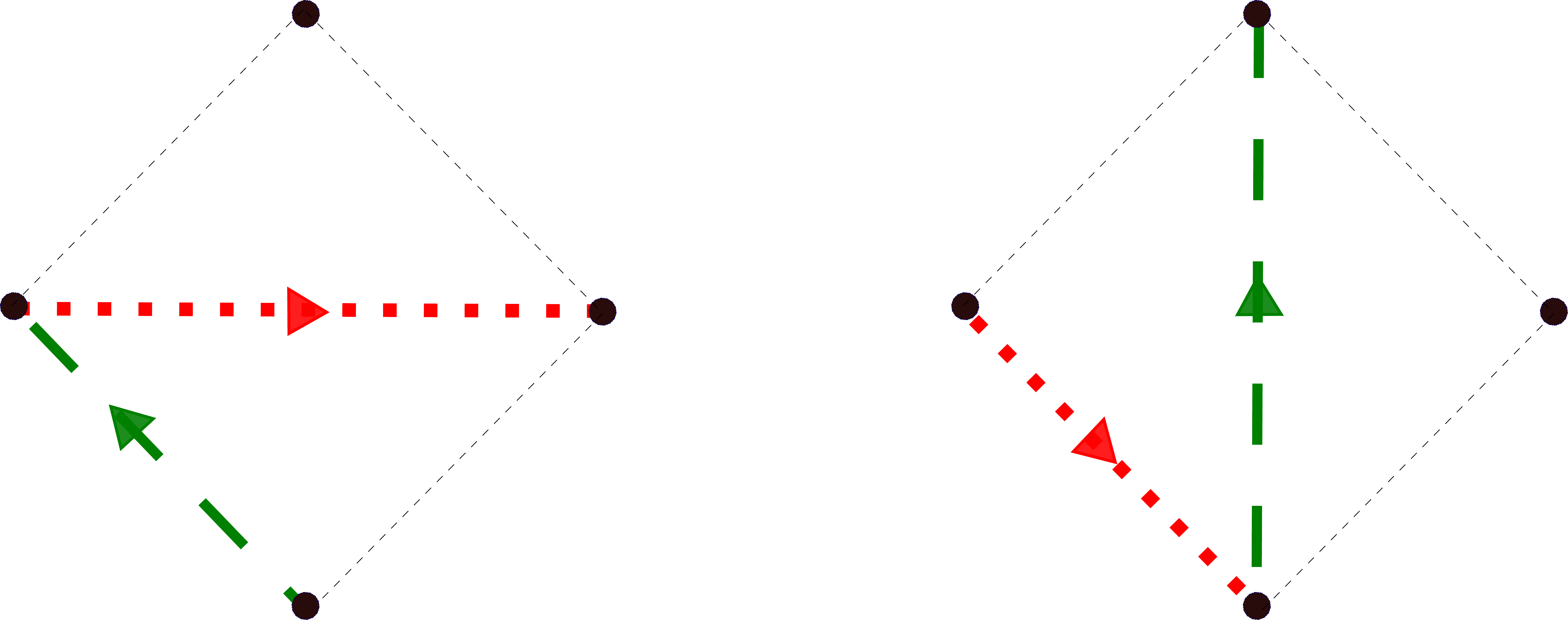}
\put(-136,20){$x$}
\put(-100,0){$z$}
\put(-78,20){$y$}
\put(-100,48){$w$}
\put(-55,20){$x$}
\put(-20,0){$z$}
\put(0,20){$y$}
\put(-21,48){$w$}
\caption{A red/green swap.}
\label{adjuppic}\label{c2c1swap}
\end{figure}

In this section, we define the local Markov chain $\mvar$ for sampling uniformly from $\stsp_n$ and show that $\mvar$ is always rapidly mixing.  Our argument relies on a bijection with pairs of Dyck paths to relate the mixing time of a local Markov chain on Dyck paths to that of $\mvar$. 
Define $\mvar$ as follows (see Figure~\ref{c2c1swap}b).

\vspace{.05in}
\noindent  \underline{{\bf The Markov chain $\mvar$}}

\vspace{.05in}
{
\noindent {\tt Starting at any $\sigma_0\in\stsp_n$, iterate the following:} 

 - \ { \tt Choose facial triangles $T_1$ and $T_2$ with shared edge $\overrightarrow{xy}$ u.a.r.}

-  \  { \tt Choose an edge $\overrightarrow{zx}$ from $T_1\cup T_2$ u.a.r., if one exists.  With\\
\indent  prob. $1/2$ replace the path $\{\overrightarrow{zx}, \overrightarrow{xy}\}$ by the path $\{\overrightarrow{xz},\overrightarrow{zw}\}$ where $w$ \\
\indent is the remaining vertex of $T_1\cup T_2$.}

 - \ { \tt Otherwise, $\sigma_{i+1}=\sigma_{i}$.}

}
\vspace{.05in}

\noindent If the edge $\arc{zx}$ with color $c_i$ is replaced by the edge $\arc{xz}$ with color $c_j$, we call this a {\bf $\bf c_j/c_i$ swap}.  Bonichon, Le Sa\"ec and Mosbah showed in~\cite{blm} that $\mvar$ connects the state space $\stsp_n$.
Since all valid moves have the same transition probabilities, this implies that $\mvar$ converges to the uniform distribution over state space~$\stsp_n$.

\subsection{The bijection between $\stsp_n$ and pairs of Dyck paths}\label{dyck}
\begin{figure}[htb]
\centering
\includegraphics[scale=.06]{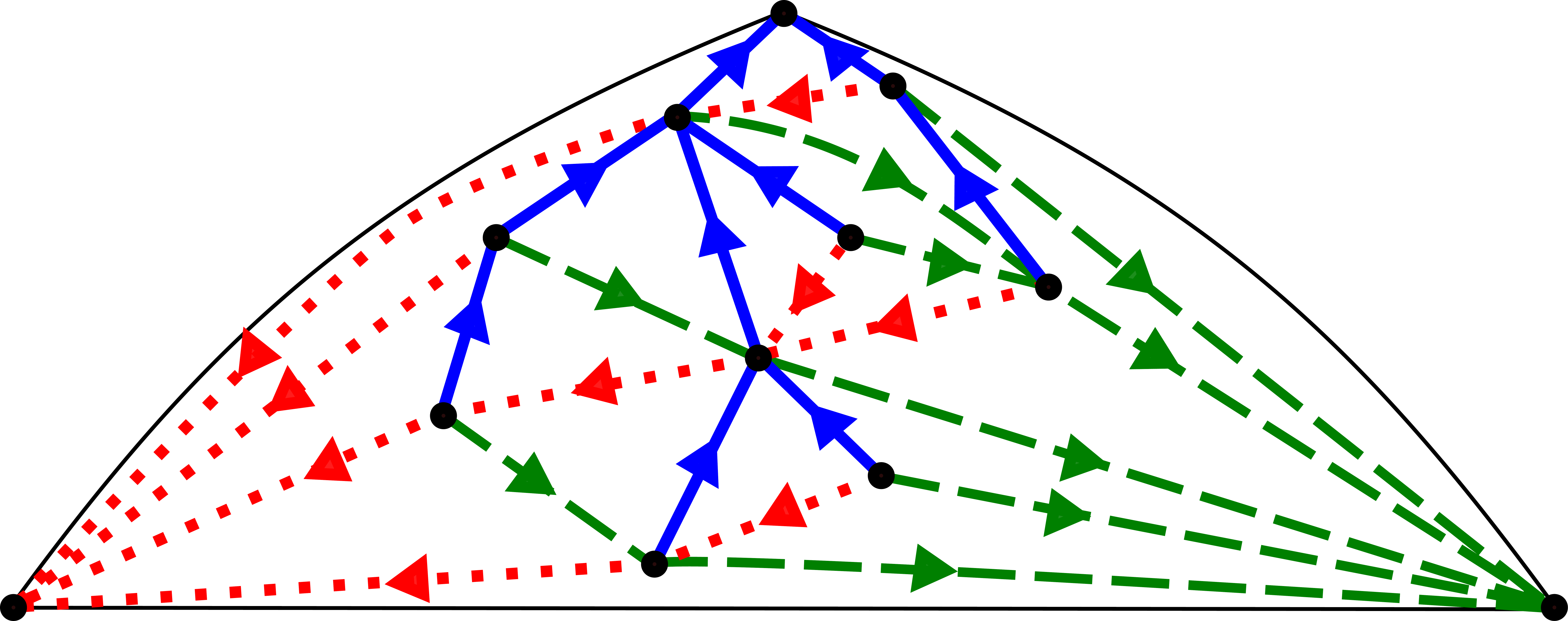}\put(0,0){$s_{green}$} \put(-215,0){$s_{red}$} \put(-95,75){$s_{blue}$}
\hspace{1in}\includegraphics[scale=.07]{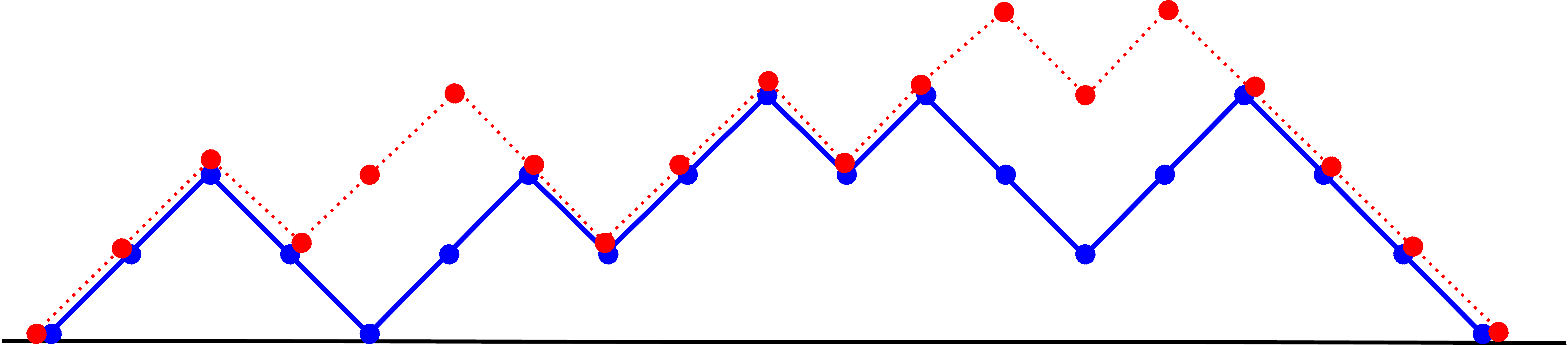}
\caption{The bijection between 3-orientations and Dyck paths.}
\label{bijection}
\end{figure}

The key to bounding the mixing time of $\mvar$ is a bijection between $\stsp_n$ and pairs of nonoverlapping Dyck paths of length $2n$, introduced by Bonichon \cite{bo}.  Dyck paths can be thought of as strings $a_1a_2\cdots a_{2n}$ containing an equal number of $1$'s and $-1$'s, where for any $1\leq k\leq 2n$, $\sum_{i=1}^k a_i \geq 0.$  Recall that a 3-orientation of a triangulation can be viewed as a union of three trees, one in each color.  In the bijection, the bottom Dyck path corresponds to the blue tree\comment{in $\sigma\in \stsp_n$}, and the top Dyck path indicates the degree of each vertex in the red tree.  The green tree is determined uniquely by the blue and red trees.  More specifically, given $\sigma \in \stsp_n$, to determine the bottom Dyck path, start at the root of the blue tree and trace along the border of the tree in a clockwise direction until you end at the root.  The first time you encounter a vertex, insert a $1$ in the Dyck path, the second time you encounter the vertex insert a $-1$.  Let $v_1,v_2,\ldots, v_n$ be the order of the vertices as they are encountered by performing this DFS traversal of the blue tree in a clockwise direction and define $L$ to be the resulting linear order on the vertices. Let $d_i$ be number of incoming red edges incident to $v_i$.  Let $r$ be the number of incoming red edges incident to $s_{red}$.  The top Dyck path is as follows $1(-1)^{d_2}1(-1)^{d_3}1(-1)^{d_4}\ldots1(-1)^{d_n}1(-1)^{r}$.  The structure of the 3-orientation guarantees that the top path will never cross below the bottom path.  See Figure~\ref{bijection}, and $\cite{bo}$ for details.

We will deduce that $\mvar$ is rapidly mixing by comparing it to $\mdyck$, an efficient Markov chain on (pairs of) Dyck paths which was introduced by Luby, Randall and Sinclair \cite{lrs}.  The algorithm proceeds as follows.  At each step select a point on one of the two Dyck paths uniformly at random.  If the point is a local maximum (or minimum) then push it down (or up) with probability 1/2 as shown in Figure~\ref{dyckpaths}(a-b).  If this move is blocked by a local maximum (or minimum) in the bottom (or top) Dyck path as shown in Figure~\ref{dyckpaths}c then push both Dyck paths down (or up) with probability 1/2 as shown in Figure~\ref{dyckpaths}(c-d).  The following theorem due to Wilson \cite{wil} bounds the mixing time of $\mdyck$:

\begin{figure}[htb]
\centering
\includegraphics[scale=.05]{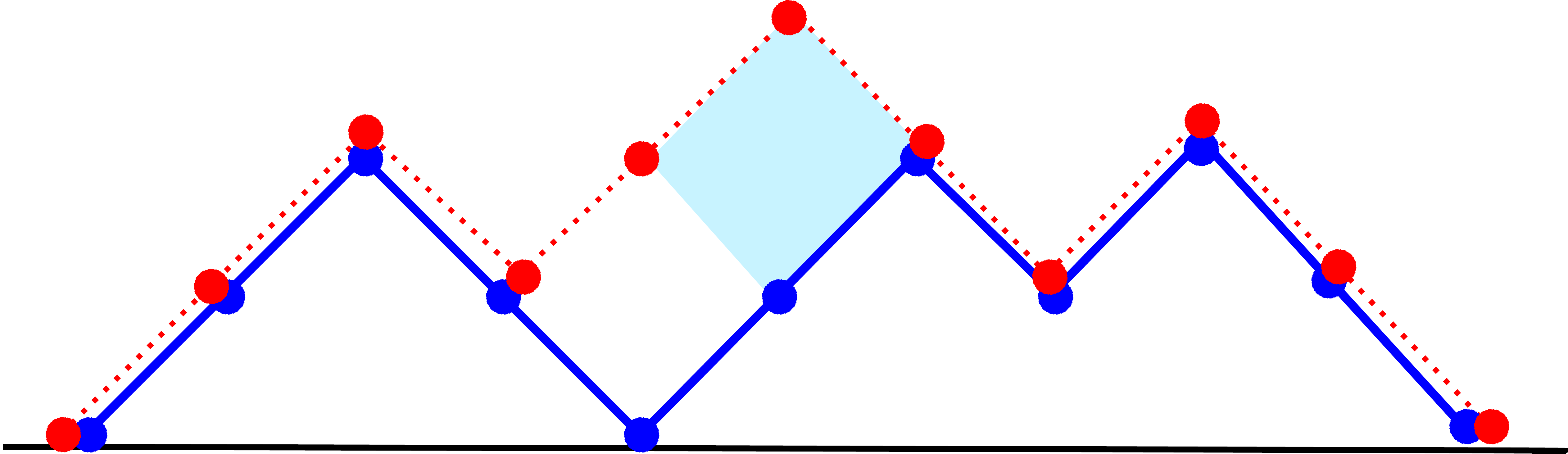}\put(-45,-15){(a)}
\put(-5,10){${\displaystyle \leftrightarrow}$}
\hspace{.03in}\includegraphics[scale=.05]{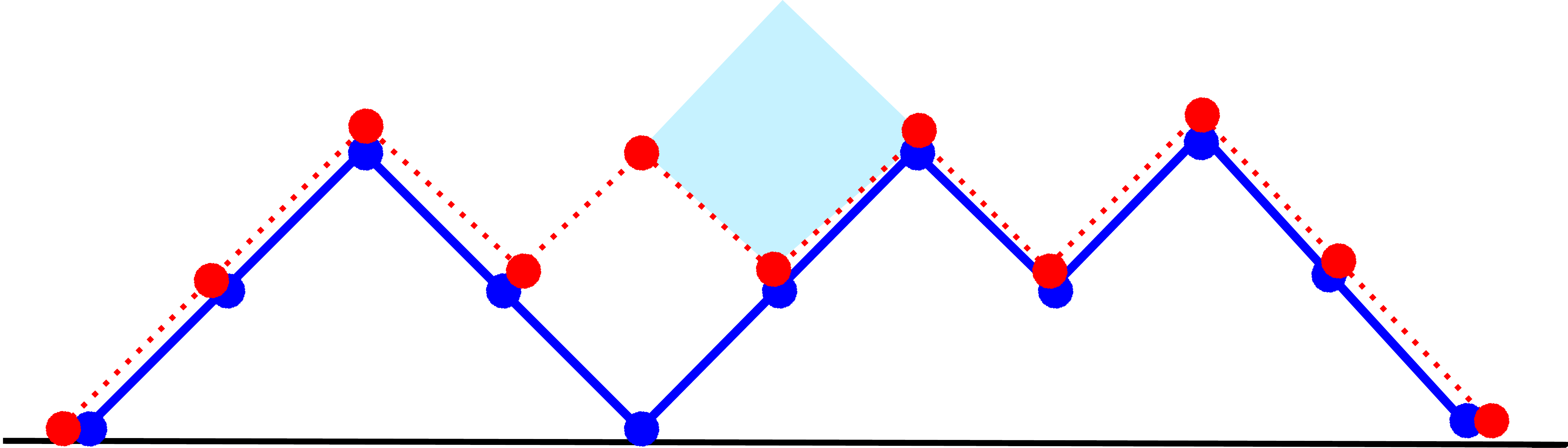}\put(-45,-15){(b)}
\hspace{.1in}
\includegraphics[scale=.05]{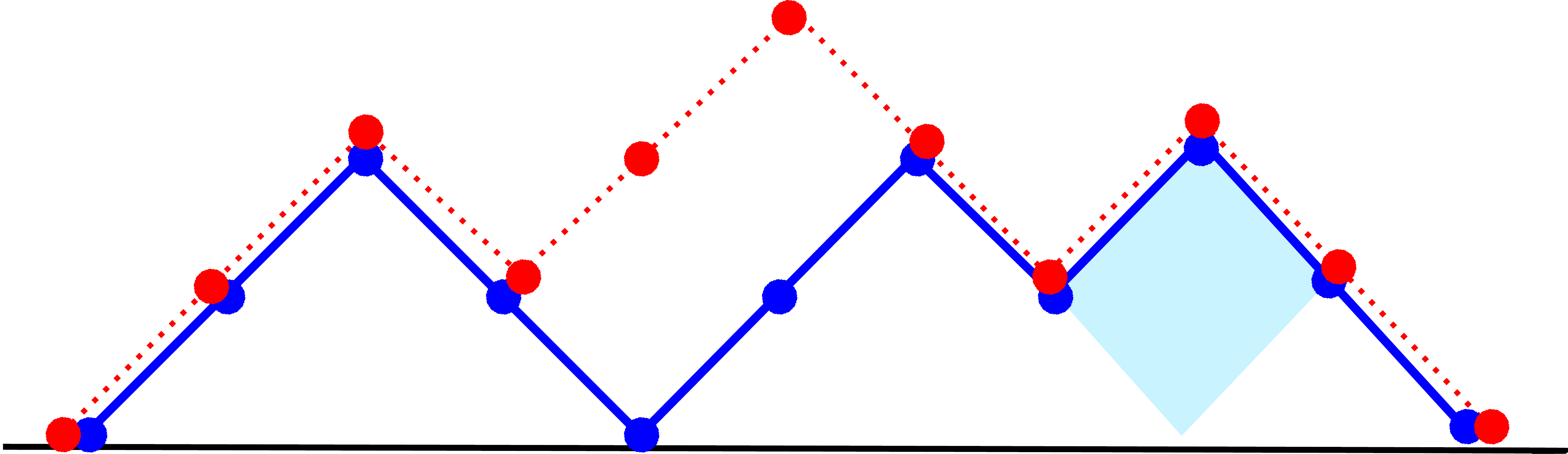}\put(-45,-15){(c)}
\put(-5,10){${\displaystyle \leftrightarrow}$}
\hspace{.03in}\includegraphics[scale=.05]{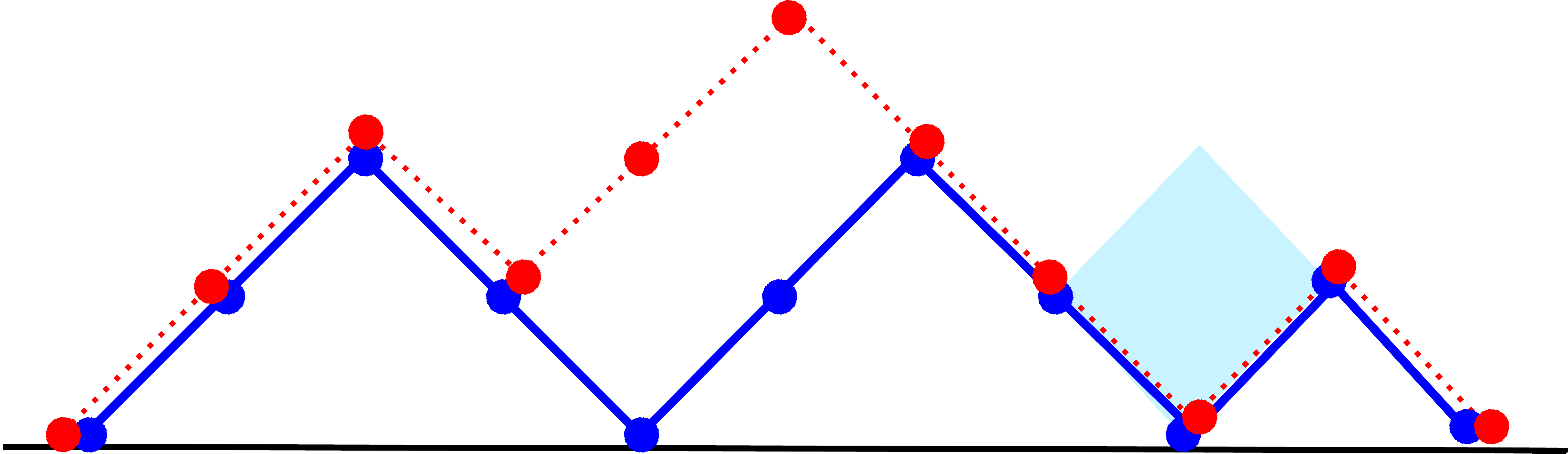}\put(-45,-15){(d)}
\caption{Two moves of the Markov chain $\mdyck$.}
\label{dyckpaths}
\end{figure}

\begin{Theorem}[Wilson]\label{mixdyck} The mixing time of $\mdyck$ on state space $\stsp_n$ satisfies $$\tau(\epsilon) = \Theta(n^3\log (n/\epsilon)).$$\end{Theorem}

\noindent Using the above bijection, the Markov chain $\mdyck$ on Dyck paths can be translated into a Markov chain on 3-orientations of triangulations, but its moves are quite unnatural in that setting.  We obtain a bound on the mixing time of $\mvar$ using Theorem~\ref{mixdyck} together with a careful comparison argument.

\subsection{$\mvar$ mixes rapidly}\label{mix_mvar}
Next we show that $\mvar$ is an efficient algorithm for sampling from the state space $\stsp_n$ by comparing the moves of $\mvar$ to the moves of $\mdyck$.  The comparison theorem of Diaconis and Saloff-Coste \cite{dsc} allows us to compare the mixing times of two reversible Markov chains $\m$ and $\m '$ on the same state space $\stsp$.  Assume they have the same stationary distribution $\pi$ and mixing times $\tau$ and $\tau'$ respectively.  
Let $E(\m) = \{(X,Y): \m(X,Y) > 0\}$ and $E(\m') = \{(X,Y): \m'(X,Y) > 0\}$ denote the transitions of the two Markov chains, viewed as directed graphs.  For each $(X,Y)\in E(\m')$, define a canonical path $\gamma_{XY}=(X=X_0,X_1,\cdots,X_k = Y)$ with $(X_i,X_{i+1})\in \m$, and let $|\gamma_{XY}|$ denote the length of the path.  Let $\Gamma(Z,W) = \{(X,Y) \in E(\m'): (Z,W) \in \gamma_{XY}\}$ be the set of canonical paths that use the transition $(Z,W)$ of $\m$.  Let $\pi_* = \min_{X \in \stsp}\pi(X)$.  Finally, define 
$$A = \max_{(Z,W) \in E(\m)}\sum_{\Gamma(Z,W)}|\gamma_{XY}|\pi(X)\m'(X,Y) /(\pi(Z)\m(Z,W)).$$
We will use the following version of comparison, due to Randall and Tetali \cite{rt}.

\begin{Theorem}[Randall and Tetali]\label{Comparison} With the above notation, the mixing time of $\m$ on the state space $\stsp$ satisfies
$$\tau (\epsilon) \leq 4\frac{\log(\frac{1}{\epsilon \pi_*})}{\log(1/(2\epsilon))}A\tau'(\epsilon).$$
\end{Theorem}

\begin{figure} [!htb]
\centering
\includegraphics[scale=.1]{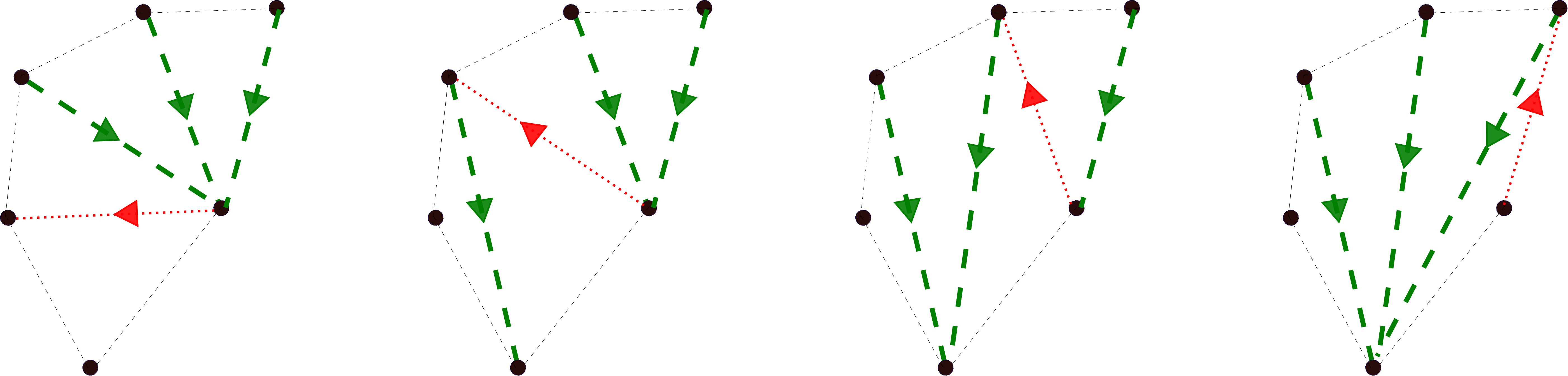}
\put(-330,30){$x$}
\put(-275,30){$v$}
\put(-265,75){$y$}
\put(-242,30){$x$}
\put(-187,30){$v$}
\put(-175,75){$y$}
\put(-154,30){$x$}
\put(-99,30){$v$}
\put(-87,75){$y$}
\put(-66,30){$x$}
\put(-10,30){$v$}
\put(2,75){$y$}
\caption{A sequence of red/green swaps.}
\label{seqofswaps}
\end{figure}

Next, we introduce some notation.  Let $c_1$ be blue, $c_2$ be red, and $c_3$ be green.  Given a vertex $v$ and $i\in \{1,2,3\}$, the unique outgoing edge of $v$ with color $c_i$ is called \emph{$v$'s $c_i$ edge}.  We also define the \emph{first} (\emph{last}) incoming $c_i$-edge of $v$ to be the incoming $c_i$-edge of $v$ that is in a facial triangle with $v$'s $c_{i-1}$ edge (respectively, $v'$s $c_{i+1}$ edge, where the subscripts are taken modulo 3).   In our canonical paths, we will often need to move a $c_j$ edge, say $\arc{vx}$, from some neighbor $x$ of $v$ to another neighbor $y$ of $v$ across several $c_i$ edges.  This is achieved through a sequence of $c_j/c_i$ swaps as in Figure~\ref{seqofswaps}.

Finally, we can prove Theorem~\ref{mixing_mvar}.
\vspace{.1in}

\noindent \emph{Proof of Theorem \ref{mixing_mvar}.}  In order to apply Theorem \ref{Comparison} to relate the mixing time of $\mvar$ with the mixing time of $\mdyck$ we need to define, for each transition of $\mdyck$, a canonical path using transitions of $\mvar$.  Then we will show that each transition of $\mvar$ is not used by too many canonical paths.   We consider four cases, depending on whether the move of $\mdyck$ takes a peak to a valley and which Dyck path it involves.  If the move $e=(X,Y)$ affects both paths, we view the move as two separate moves $(X,Z)$ and $(Z,Y)$, one on each path, and we concatenate the canonical paths as follows: $\gamma_{X,Y}=(\gamma_{X,Z},\gamma_{Z,Y})$.  Hence in the following, we assume that the transitions of $\mdyck$ affect only the top or the bottom Dyck path.

\vspace{.2in}
\underline{\textbf{A Peak to Valley Move on the Top Dyck Path:}}\ \ \
Let $e=(X,Y)$ be a transition of $\mdyck$ that affects the top path only and moves from a peak to a valley.  
Suppose $e$ moves the $i^{\text{th}}$ 1 (where $i>1$) on the top path to the right one position (i.e. the Dyck path move swaps the $i^{\text{th}}$ 1 with a $-1$ on its right, changing a peak to a valley).  From the bijection, we know this move does not affect the blue tree and corresponds to, in the red tree, increasing the incoming degree of $v_i$ by one and decreasing the incoming degree of $v_{i+1}$ by one.  If $v_i$ and $v_{i+1}$ are adjacent in the blue tree (there is a blue edge $\overrightarrow{v_{i+1}v_i})$ then this implies that there is a red/green swap involving $v_{i}$'s green edge and $v_{i+1}$'s first incoming red edge.  This swap is exactly the peak to valley move, so $\gamma_{XY}=e$.   Otherwise, 
we define two stages in the canonical path $\gamma_{XY}$.  

To assist in defining the canonical paths, let $g_i$ be the parent of $v_i$ in the green tree.  First, we will show that $g_i$ is not $s_{green}$; that is, $v_i$ cannot have its green edge point to the green source $s_{green}$ as long as $e=(X,Y)$ is a valid transition of $\mdyck$.  For the sake of contradiction, suppose $\arc{v_is_{green}}$ is an edge in $X$.
 Let $P$ be the path from $s_{blue}$ to $v_i$ in the blue tree combined with the edge $\overrightarrow{v_{i}s_{green}}$.  This path partitions the vertices of $X$ into two sets according to which side of the path they are on.  Let $S$ be the set of vertices on the same side of $P$ as $v_{i+1}$ (not including the vertices along $P$) and let $\bar{S}$ be the remaining vertices.  Let $x\in V(X)$ and define $x$'s red edge by $\arc{xr_x}$.  We claim that $L(x)<L(r_x)$; this is a direct result of a bijection between Dyck paths and 3-orientations given in \cite{bo} (this is a different bijection from the one given in Section \ref{prelim}).  Now follow the blue path from $x$ to $s_{blue}$ until you reach the first vertex $z'$ that has an incoming blue edge from a vertex $z$ such that $L(z) > L(x)$.  Define $z_x = z$.  We claim that in all 3-orientations with this same blue tree, $L(r_x) \geq L(z_x)$.  To prove this, we need some facts from~\cite{bo}.  Specifically, they show that any 3-orientation can be obtained by starting from a fixed 3-orientation, called a \emph{star realizer}, which is unique to each blue tree, and then applying a series of transformations.  The star realizer satisfies the property that $L(r_x)\geq L(z_x)$ for any vertex $x$, and each transformation maintains this property.

This implies that any red edge in $X$ directed toward vertices in $S$ must be directed toward vertices in $S$ for all 3-orientations with the same blue tree, which implies that there is no configuration $Y$ which has the same blue tree as $X$ and the same incoming red degree sequence except for increasing $v_{i}$ by one and decreasing $v_{i+1}$ by one.  This is a contradiction since we have assumed $(X,Y)$ is a valid move.  Hence we have shown that $g_i$ is not $s_{green}$.

\begin{figure}[htb]
\centering
\includegraphics[scale=.12]{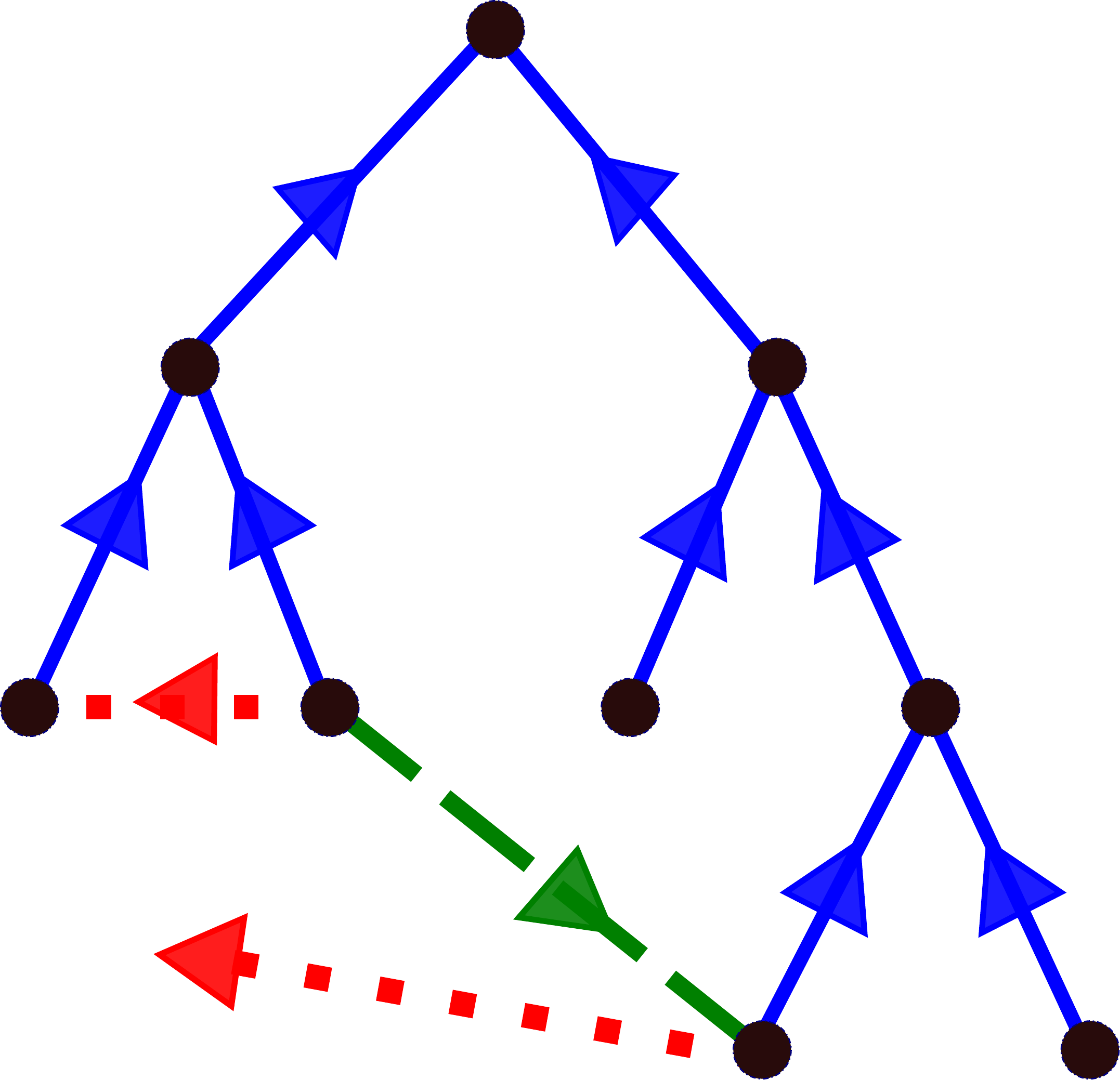}
\put(-21,0){${\scriptstyle g_i}$}\put(-51,27){${\scriptstyle v_i}$} 
\put(-73,10){${\scriptstyle v_j}$} 
\caption{The vertex $v_i$'s red and green edges prevent $v_j$ from satisfying $L(v_{i+1}) \geq L(v_j) \geq L(g_i)$. }
\label{block}
\end{figure}

\begin{figure}[htb]
\centering
\includegraphics[scale=.07]{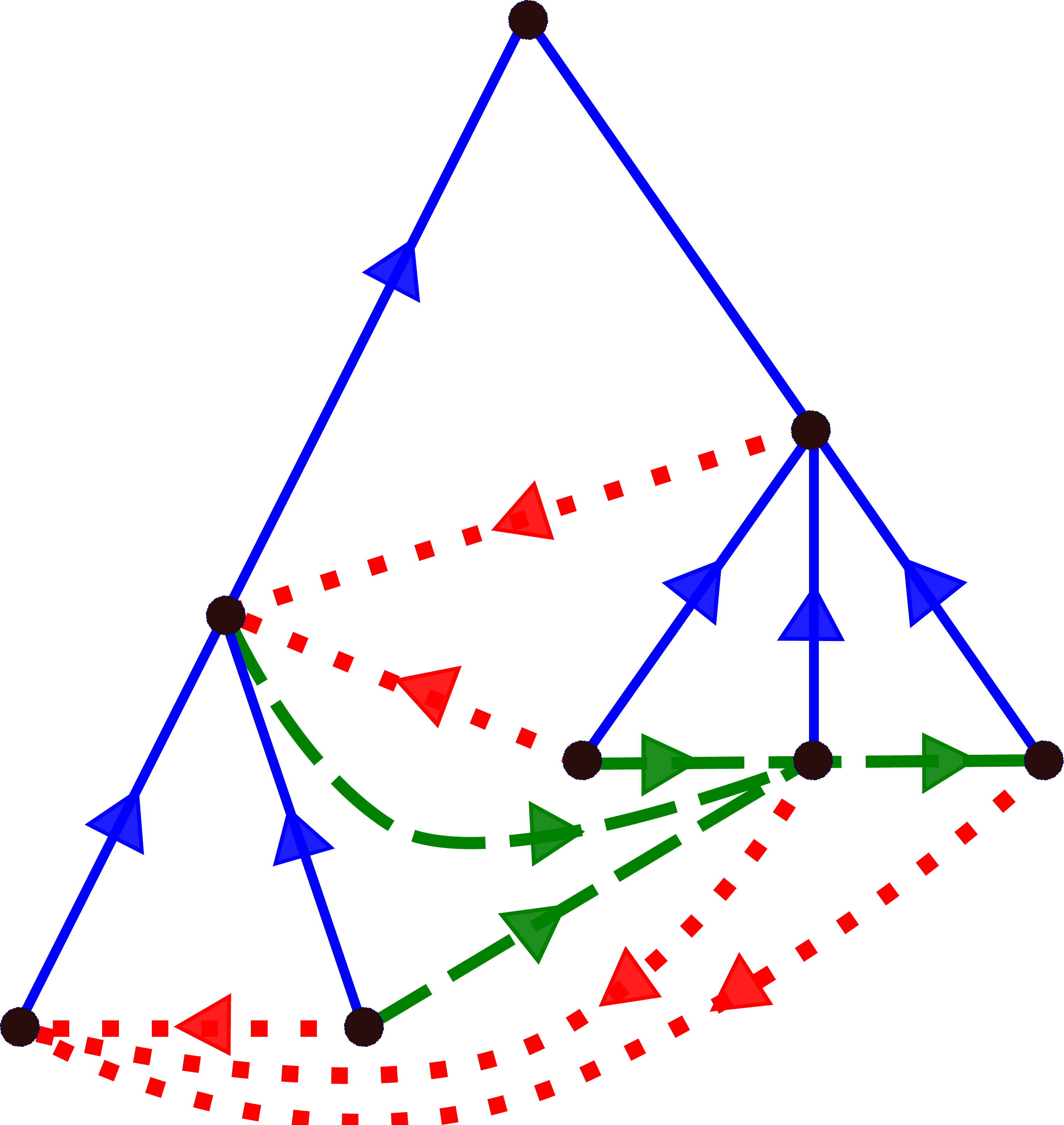}
\put(-64,31){${\scriptstyle v_{i+1}}$}
\put(-33,25){${\scriptstyle v_i}$}
\put(-70,5){${\scriptstyle v_j}$}
\put(-14,16){${\scriptstyle g_i}$}
\put(2,40){$\rightarrow$}
\hspace{.2 cm}
\includegraphics[scale=.07]{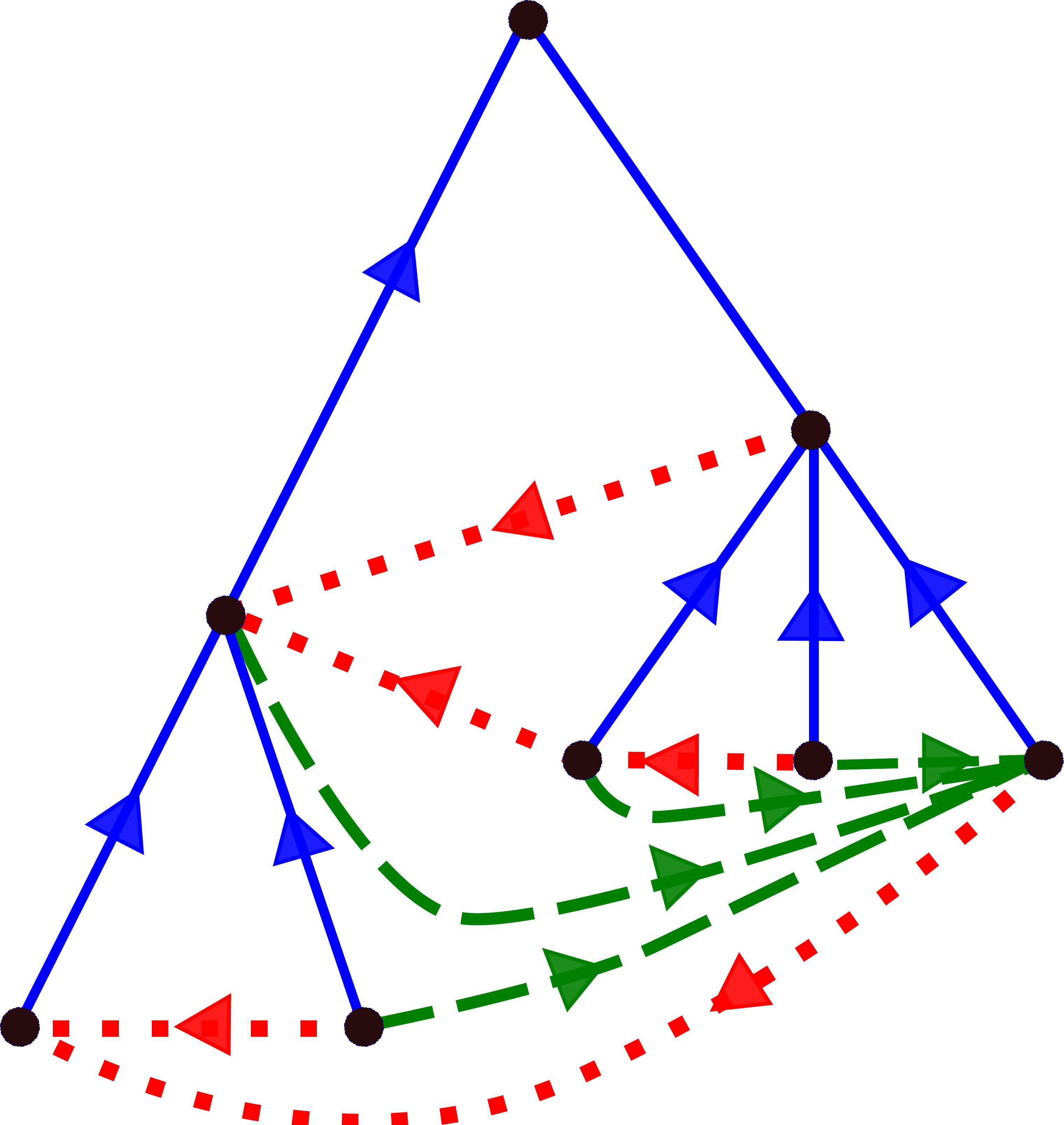}
\put(-64,31){${\scriptstyle v_{i+1}}$}
\put(-33,25){${\scriptstyle v_i}$}
\put(-70,5){${\scriptstyle v_j}$}
\put(2,40){$\rightarrow$}
\hspace{.2 cm}
\includegraphics[scale=.07]{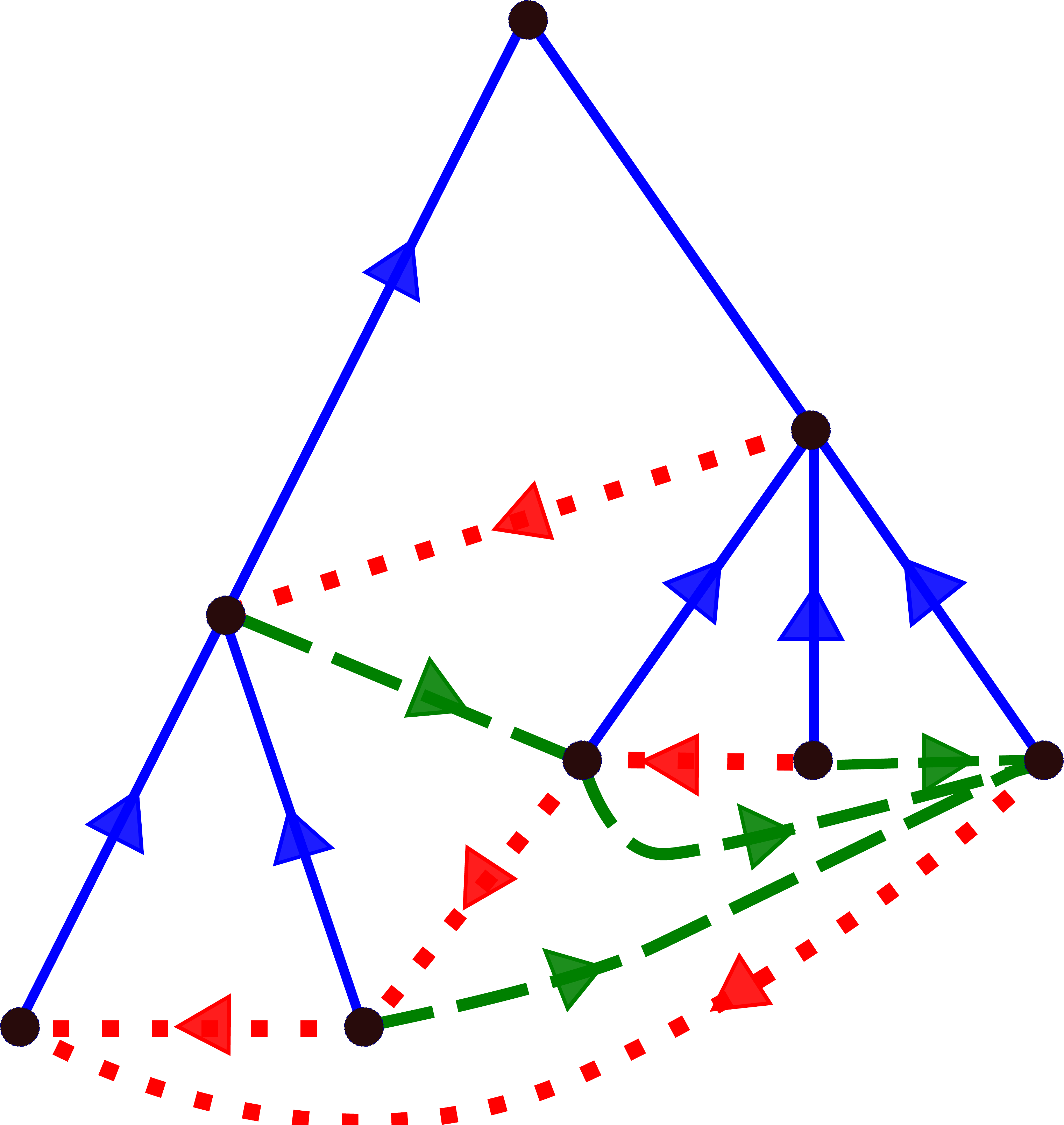}
\put(-64,31){${\scriptstyle v_{i+1}}$}
\put(-33,25){${\scriptstyle v_i}$}
\put(-70,5){${\scriptstyle v_j}$}
\put(2,40){$\rightarrow$}
\hspace{.2 cm}
\includegraphics[scale=.07]{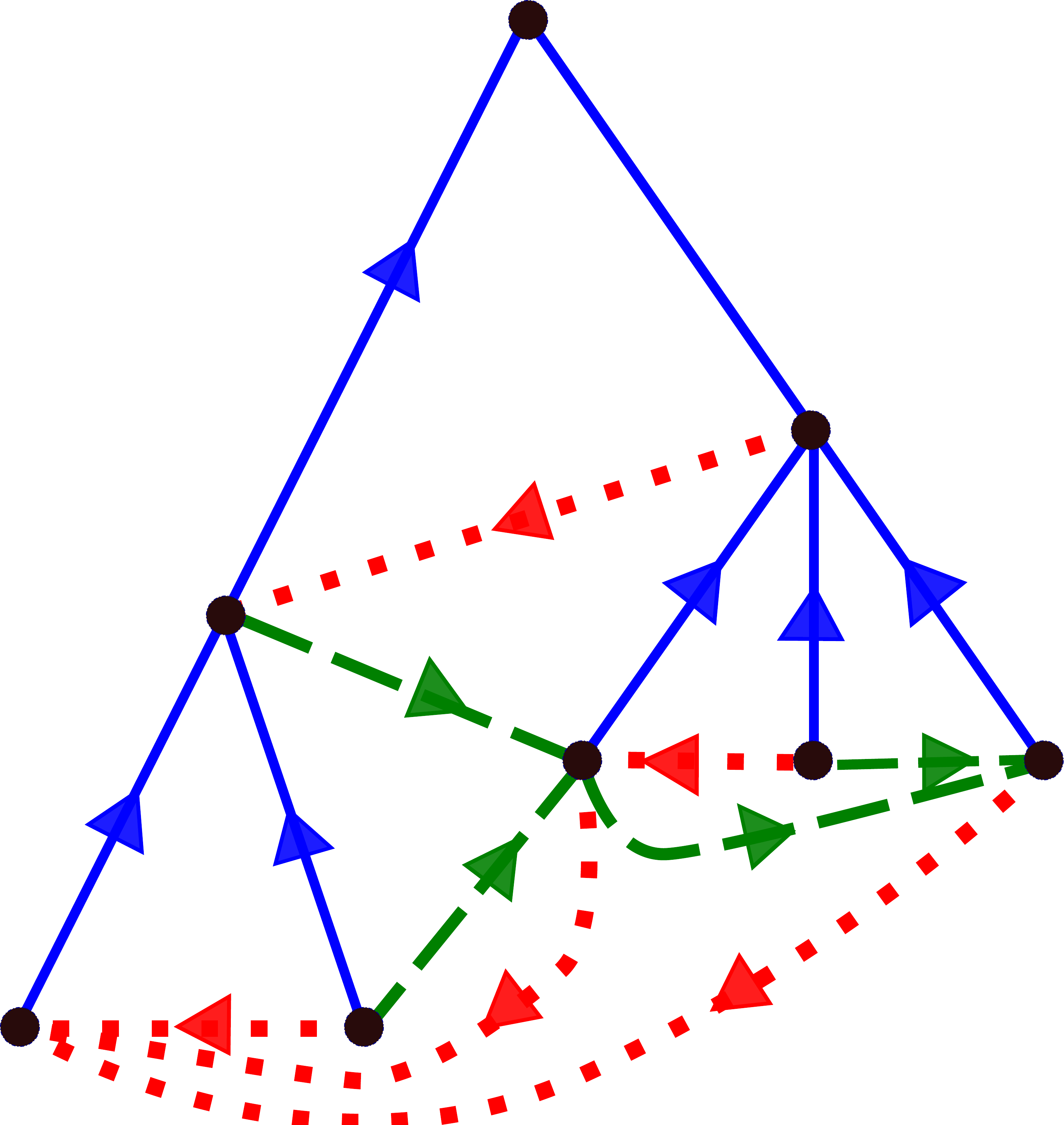}
\put(-64,31){${\scriptstyle v_{i+1}}$}
\put(-33,25){${\scriptstyle v_i}$}
\put(-70,5){${\scriptstyle v_j}$}
\caption{The canonical path to move a peak down to a valley, top Dyck path.}
\label{toppath}
\end{figure}

Next, let $v_j$ be the parent of $g_i$ in the red tree.  Notice that $L(v_i)<L(v_{i+1})\leq L(v_j)$, since $L(v_j) > L(g_i)$ and $v_i$'s red and green edges prevent $v_j$ from satisfying $L(v_{i+1}) \geq L(v_j) \geq L(g_i)$ as shown in Figure~\ref{block}.  In the first stage of the path $\gamma_{XY}$ we make the sequence of red/green swaps centered at $g_i$ that move the red edge $\overrightarrow{g_iv_j}$ to $\overrightarrow{g_iv_i}$ without affecting any other red edges as shown in Figure~\ref{toppath}b, step 1 (see Figure~\ref{seqofswaps} for detail on the sequence of swaps).  In the second stage we transfer an incoming red edge from $v_{i+1}$ to $v_j$, completing $\gamma_{X,Y}$.  Recall that $L(v_j) > L(v_{i+1})$, so we do this iteratively by moving an incoming red edge $\overrightarrow{yx}$ either to one of $x$'s neighbors in the blue tree that is larger in $L$ or to $x$'s parent in the red tree, which is also larger in $L$, if it is a leaf and has no neighbors as shown in Figure~\ref{toppath}b, step 2 and 3.   We claim it is always possible to make one of these moves.  If $x$ has a neighbor $y$ in the blue tree such that $L(y) = L(x) + 1$ then there must be a green/red swap centered at $x$ and involving $x$'s green edge that moves an incoming red edge from $x$ to $y$ as desired.  Next, if $x$ is a leaf with red edge $\overrightarrow{xr_x}$ then again there is a green/red swap centered at $x$ involving $x$'s green edges that moves an incoming red edge from $x$ to $r_x$ as desired.  Finally notice that, using this canonical path, we never bypass $v_j$ because the original red edge $\overrightarrow{g_iv_j}$ blocked any blue leaves between $v_{i+1}$ and $g_i$ from having red parents higher in $L$ than $v_j$.

Given a transition $(Z, W)$ of $\mvar$ we must bound the number of canonical paths $\gamma_{XY}$ using this edge.  To do so, we analyze the amount of information needed in addition to $(Z, W)$ to determine $X$ and $Y$ uniquely.  We record the vertex $v_{i}$ and the vertex $v_j$. If $v_i$ and $v_{i+1}$ are adjacent we record $v_{i+1}$ instead of $v_j$.  Notice in this case the canonical path only involves red/green and green/red swaps.  If we are moving a red edge to a higher vertex in $L$ then we are in stage 2 and otherwise we are in stage 1.  Given this information we can uniquely recover $X$ and $Y$.  We only need to record two vertices, so in this case there are at most $n^2$ canonical paths which use any edge $(Z, W)$. 

\vspace{.2in}
\underline{\textbf{A Valley to Peak Move on the Top Dyck Path:}}\ \ \ Next consider the case where $e=(X,Y)$ changes a valley to a peak.  Assume that $e$ moves the $i$th 1 on the top Dyck path to the left one position.  From the bijection we know that this move corresponds to, in the red tree, decreasing the incoming degree of $v_i$ by one and increasing the incoming degree of $v_{i+1}$ by one.  If $v_i$ and $v_{i+1}$ are physically adjacent in the blue tree then there is a green/red swap centered at $v_i$ and involving $v_i$'s green edge that performs exactly these changes so we will define $\gamma_{XY}$ to be this single move.   If $v_i$ and $v_{i+1}$ are not adjacent then we will define two stages in $\gamma_{XY}$.  Let $r_i$ be $v_i$'s parent in the red tree.  In the first stage we transfer an incoming red edge from $v_i$ to $r_i$ using the green/red swap centered at $v_i$ and involving $v_i$'s green edge.  In the second stage we will repeatedly use the canonical path defined for peak to valley moves to transfer an incoming red edge from $r_i$ to $v_{i+1}$.  

Again given a transition $(Z, W)$ of $\mvar$ we must upper bound the number of canonical paths $\gamma_{XY}$ that use this edge in this case.  First, we will record a bit to determine which stage we are in.  For both stages we will record the vertex $v_{i}$ and the vertex $r_i$. If $v_i$ and $v_{i+1}$ are adjacent we will record $v_{i+1}$ instead of $r_i$.  For the second stage if we are in the process of making a peak to valley move which affects the vertices $v_p$ and $v_{p+1}$, we will record the vertex $v_p$.  In addition we will record the extra vertex $v_j$ that is needed when we are in the middle of a peak to valley move, as explained in the peak to valley case above.  Given this information we can uniquely recover $X$ and $Y$.  This implies that in this case there are at most $2n^4$ canonical paths which use any edge $(Z, W)$.  

\begin{figure} [!htb]
\centering
\subfloat[]{\begin{overpic}[scale=.1]{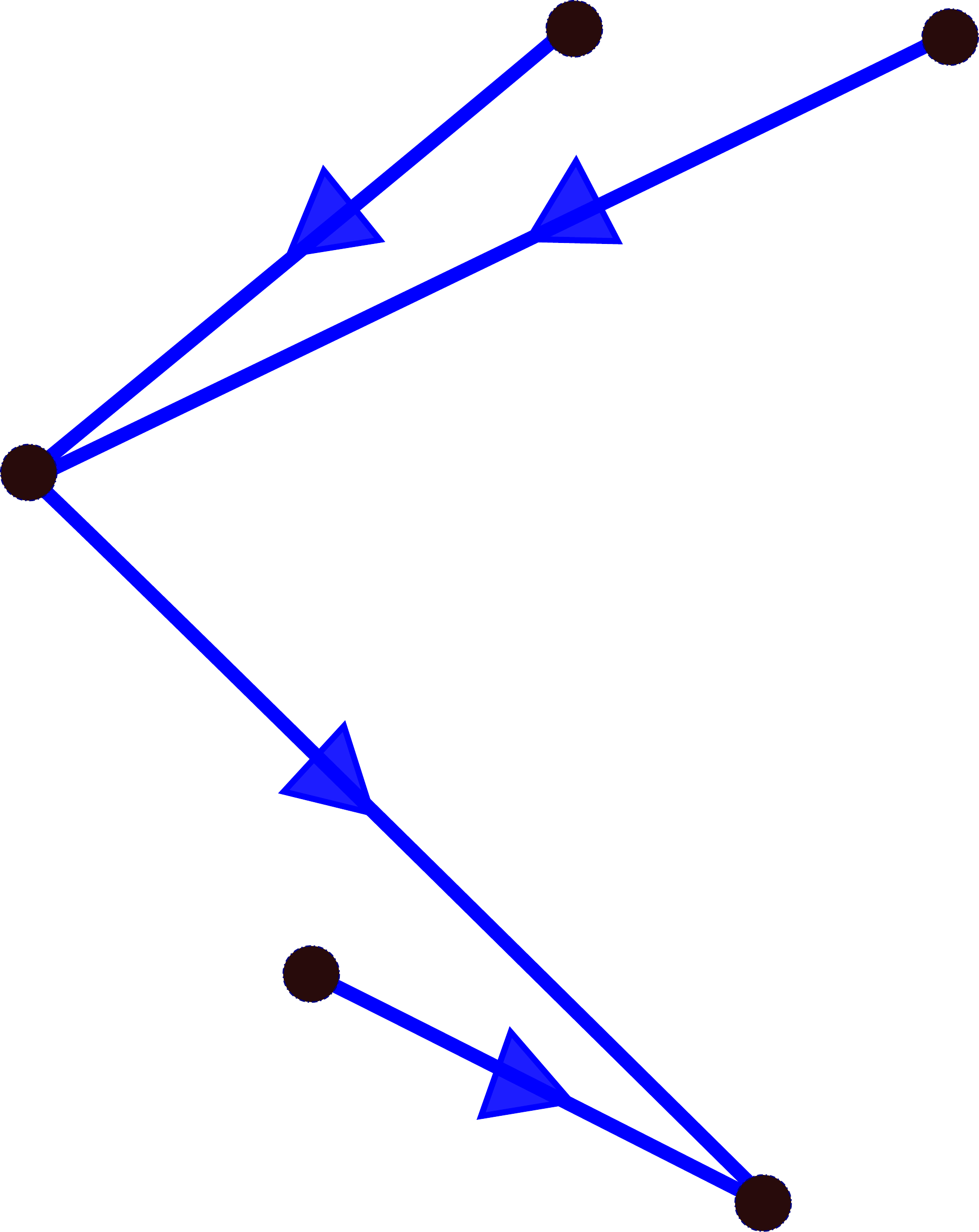}
 \put(15,20){$b$}
 \put(-5,50){$a$}
 \put(67,0){$c$}
\end{overpic}\label{avtop1}}
\hspace{1 in}
\subfloat[]{\begin{overpic}[scale=.1]{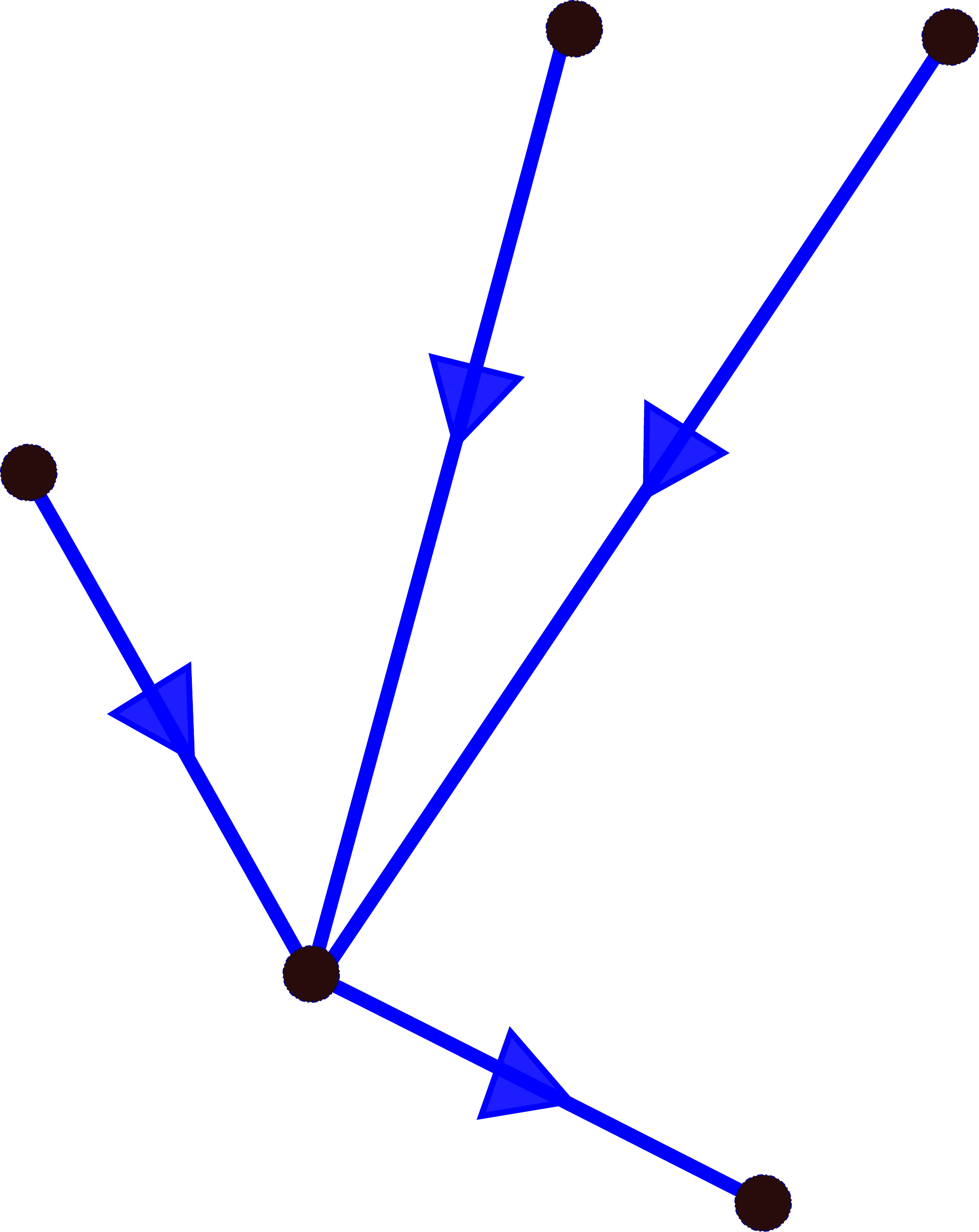}
 \put(15,20){$b$}
 \put(-5,50){$a$}
 \put(67,0){$c$}
 \end{overpic}\label{avtop2}}
\caption{A move of $\mdyck$ that moves the bottom Dyck path from a valley to a peak takes the blue tree from (a) to (b).}
\label{avtop}
\end{figure}

\begin{figure} [!htb]
\centering
\subfloat[]{\begin{overpic}[scale=.09]{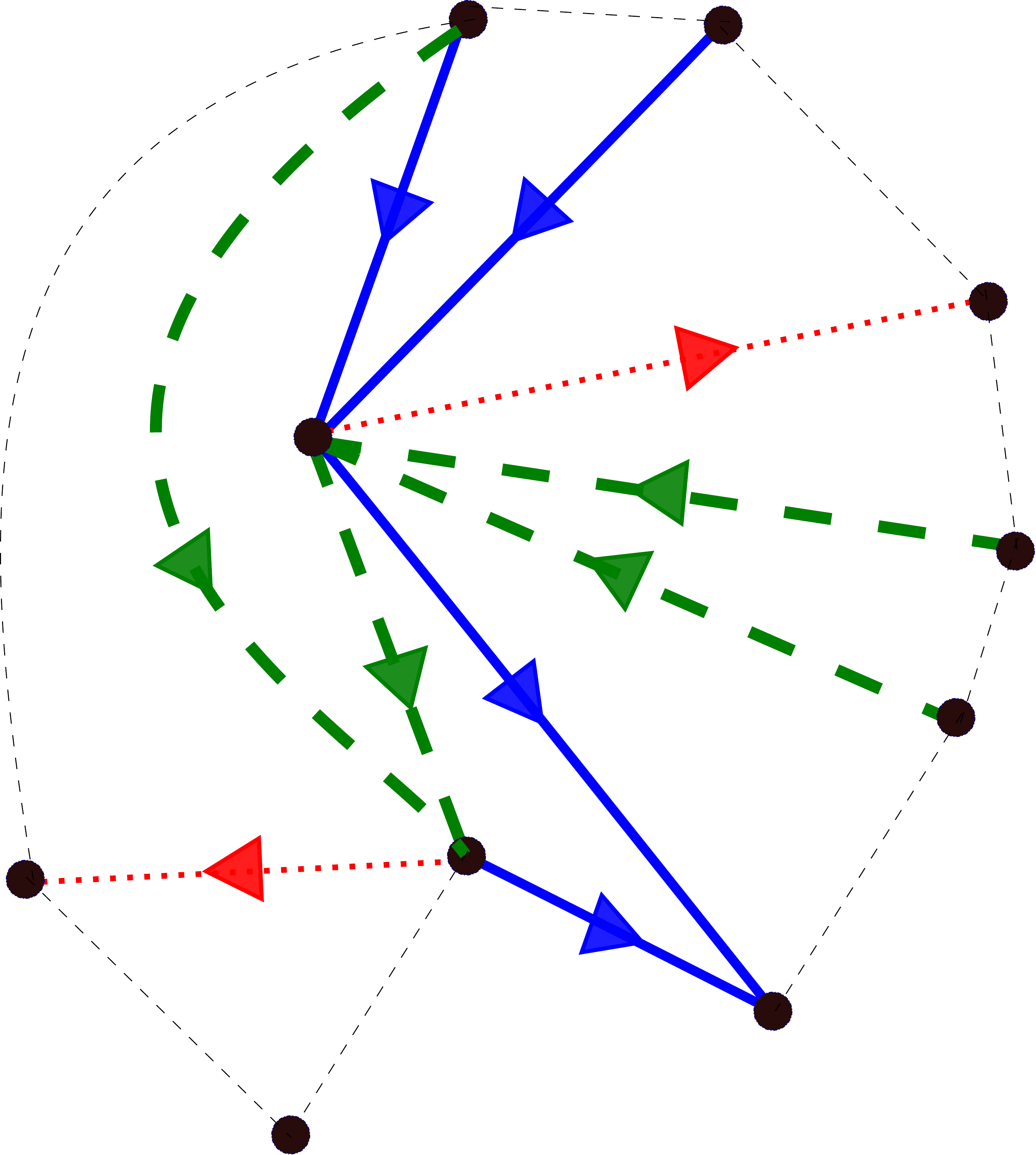}
\put(18,60){$a$}
\put(35,17){$b$}
\put(70,10){$c$}
\put(80,27){$d$}
\put(-8,18){$\head_b$}
\put(83,78){$\head_a$}
\end{overpic}\label{avalleybluetreeA}}
\hspace{.7 cm}
\subfloat[]{\begin{overpic}[scale=.09]{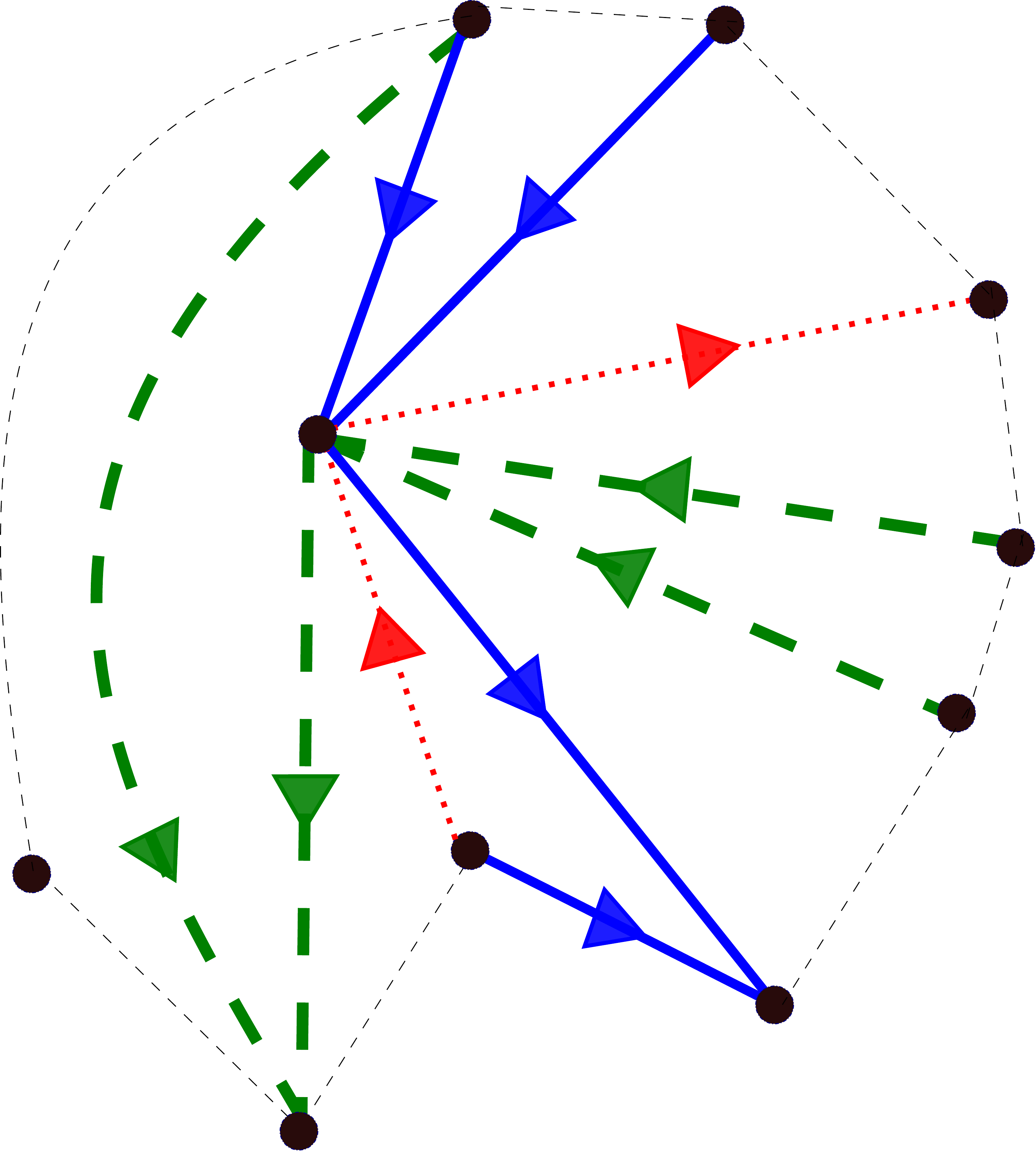}
\put(18,60){$a$}
\put(35,17){$b$}
\put(70,10){$c$}
\put(80,27){$d$}
\put(-8,18){$\head_b$}
\put(83,78){$\head_a$}
\end{overpic}\label{avalleybluetreeB}}
\hspace{.7 cm}
\subfloat[]{\begin{overpic}[scale=.09]{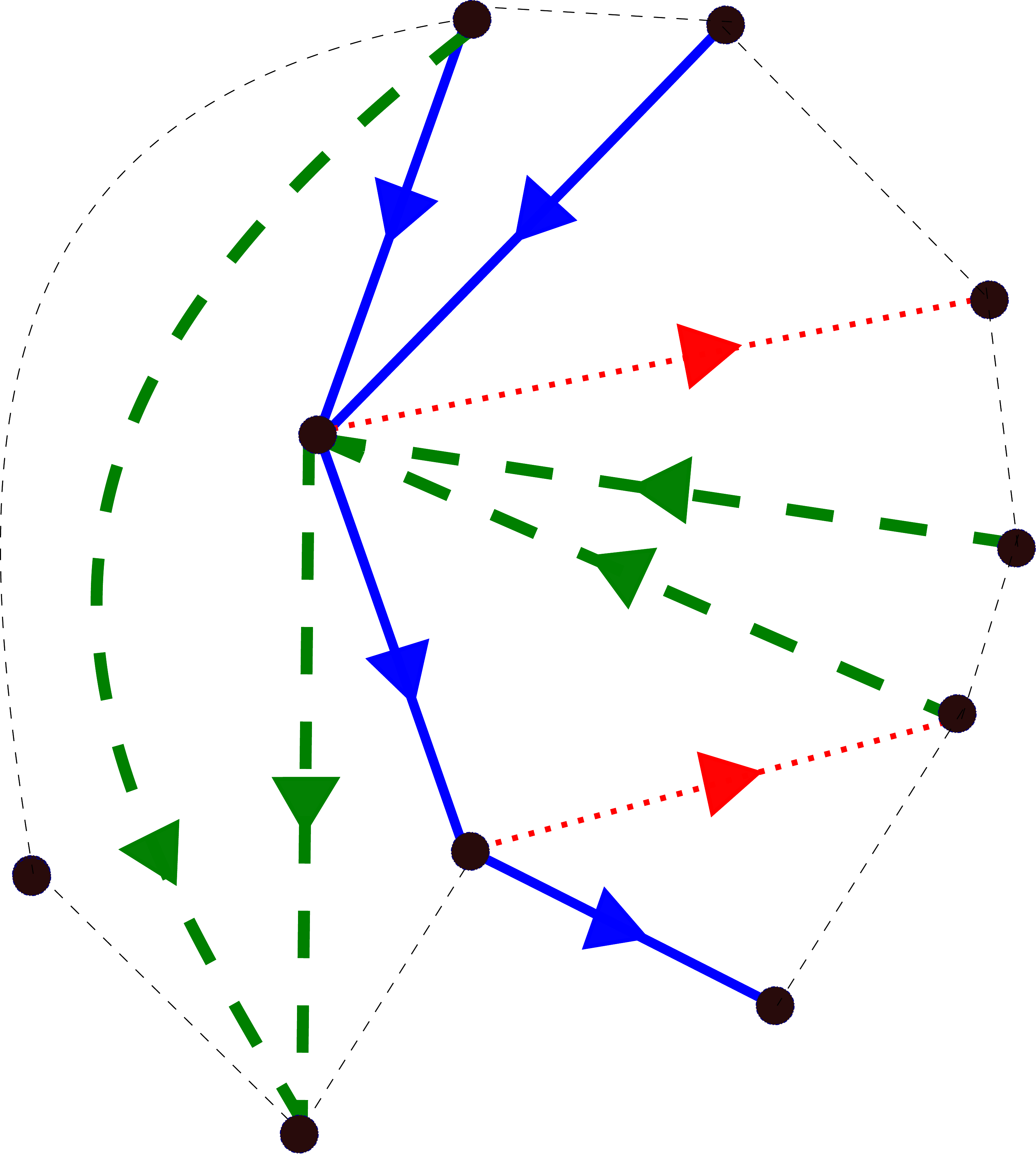}
\put(18,60){$a$}
\put(35,17){$b$}
\put(70,10){$c$}
\put(80,27){$d$}
\put(-8,18){$\head_b$}
\put(83,78){$\head_a$}
\end{overpic}\label{avalleybluetreeC}}
\hspace{.7 cm}
\subfloat[]{\begin{overpic}[scale=.09]{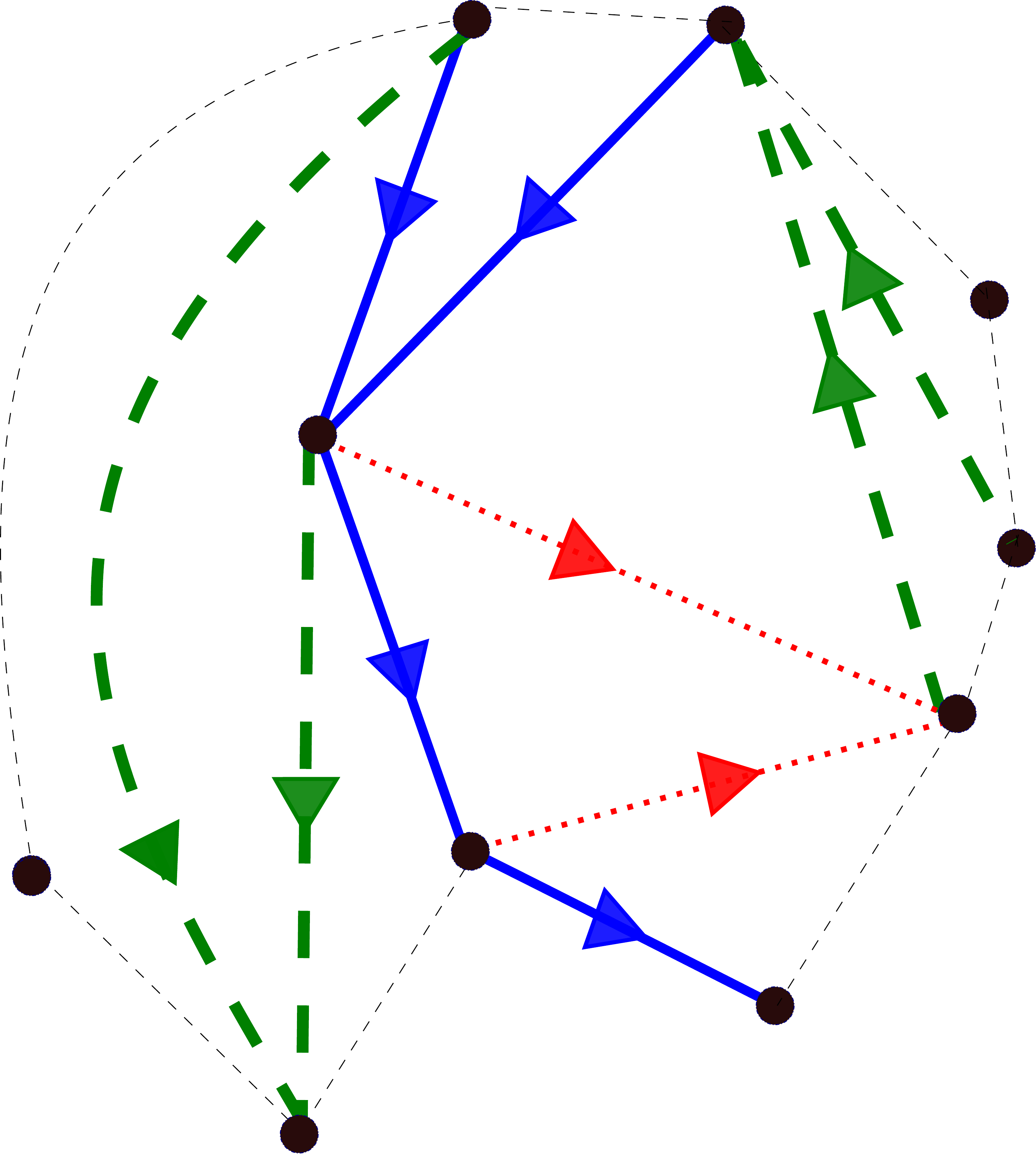}
\put(18,60){$a$}
\put(35,17){$b$}
\put(70,10){$c$}
\put(80,27){$d$}
\put(-8,18){$\head_b$}
\put(83,78){$\head_a$}
\end{overpic}\label{avalleybluetreeD}}
\hspace{.7 cm}
\subfloat[]{\begin{overpic}[scale=.09]{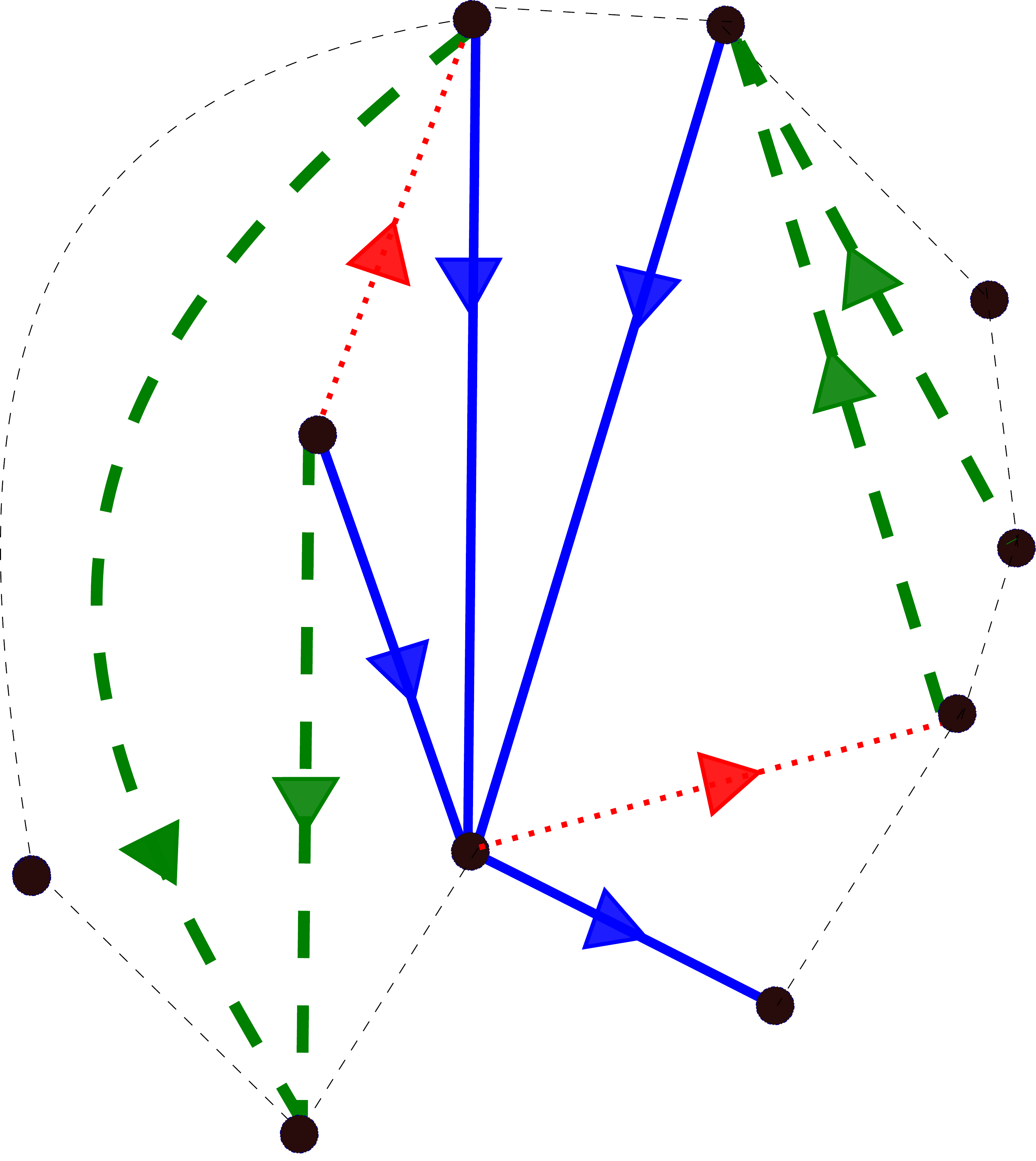}
\put(18,60){$a$}
\put(35,17){$b$}
\put(70,10){$c$}
\put(80,27){$d$}
\put(-8,18){$\head_b$}
\put(83,78){$\head_a$}
\end{overpic}\label{avalleybluetreeE}}
\caption{Canonical path to move a valley up to a peak in the blue tree.}
\label{avalleybluetree}
\end{figure}

\underline{\textbf{A Valley to Peak Move on the Bottom Dyck Path:}}\ \ \ Let $e=(X,Y)$ be a transition of the tower chain $\mtower$ on the bottom Dyck path, and suppose the edge $e$ is a move that takes a valley of the bottom Dyck path and moves it to a peak.  This affects the blue tree as follows (see Figure~\ref{avtop}): $a$'s blue edge moves from $c$ to $b$, and all the (blue) children of $\overrightarrow{ac}$ (if any exist) become children of $\overrightarrow{bc}$.  However, to define the canonical path between these two configurations, it is necessary to also know what the top Dyck path looks like, as it determines the red (and therefore the green) tree.  Figure~\ref{avalleybluetree} shows how the red and green tree might look.  Our path will first update the blue tree from $X$ to match the blue tree of $Y$, and then update the red tree (and therefore the green tree) to match the red tree of $Y$ using the steps outlined in the previous two cases.  For a vertex $v\in T$ let $\head_v$ denote the head of $v'$s red edge in $X$. We will go through 4 distinct stages in the canonical path.  In stage 1 the blue edge of $a$ moves from $c$ to $b$.  Then in stage 2, $a$'s red edge moves into position for stage 3, where all incoming blue edges to $a$ move down to point to $b$.  Finally, in stage 4 we repair the red tree.

\noindent \textbf{Stage 1.} Given the vertex condition (Figure~\ref{vertexconditionpic}a) for $b$ and the bijection between the bottom Dyck path and the blue tree, there is no edge in the angle $\angle acb$ and thus $a,b, $ and $c$ form a triangle in our triangulation.  Vertex $b$ may have some green edges coming in between its blue edge $\arc{bc}$ and its red edge $\arc{b\head_b}$.  If so, then $\arc{ab}$ is an edge, as in Figure~\ref{avalleybluetree}a.  The first step along the canonical path from $X$ to $Y$ is to rotate $b$'s red edge to point to $a$ (see Figure~\ref{avalleybluetree}(a-b)).  This is accomplished through a sequence of red/green swaps, one for each green edge coming into $b$, as in Figure~\ref{seqofswaps}.  Now we are ready to move the blue edge $\arc{ac}$ to $\arc{ab}$ by swapping with $b$'s red edge, as in Figure~\ref{avalleybluetree}(b-c).  
  
\noindent \textbf{Stage 2.} Consider vertex $a$.  Counterclockwise from $a$'s blue edge, there may be some green edges coming into $a$, followed by the red edge $\arc{a\head_a}$.  If there are green edges, let $d$ be the tail of the last incoming green edge to $a$; however, if there are no green edges, then $\head_a=d$.    In this stage of our canonical path, $a$'s red edge moves from $\head_a$ to $d$ (if $\head_a\neq d$) by a series of red/green swaps.  See Figure~\ref{seqofswaps}(c-d).

\noindent \textbf{Stage 3.} Next, we use $a$'s red edge to move all the blue children of $\arc{ab}$ to point to $b$, one at a time, in a clockwise manner.  See Figure~\ref{seqofswaps}(d-e).  Now the blue tree is completely fixed.

\noindent \textbf{Stage 4.} Finally, we must repair the red tree.  Notice that the red edges of $a$ and $b$ are the only red edges that we moved; we must move them to their proper place.  We must increase the degree of $\head_b$ and $\head_a$ back to match their indegree in $X$.  To do this without affecting the blue tree, we first make $a$'s red edge point to $\head_b$ (we'll call this {\bf Stage 4a}) and then make $b$'s red edge point to $\head_a$ ({\bf Stage 4b}).  These moves can each be accomplished by a sequence of red/green swaps without affecting the blue tree.

Given a transition $(Z, W)$ of $\mvar$, we must upper bound the number of canonical paths $\gamma_{X,Y}$ that use this edge.  If $(Z, W)$ is in stage 1, we need to remember vertices $a$ and $ \head_b$, and a bit to tell us whether or not we have moved $\arc{ac}$ yet.  Given this information, we can recover $b$ and $c$ and can undo all red/green swaps in order to get back to $X$.  Given $X$ we can find $Y$ since we know which valley to flip up.  If $(Z, W)$ is in stage 2, then we only need to record $\head_b$, since $(Z,W)$ moves $a$'s red edge, so we know $a$.  To get back to the last configuration in stage 1, we just need to move $a$'s red edge counter-clockwise until it can't make any more red/green swaps.  Thus we can get back to the last configuration in stage 1, and using $a$ and $\head_b$ we can recover $X$.  If $(Z, W)$ is in stage 3, we need to record $\head_b$.  Each move in stage 3 takes a child of $\arc{ab}$ and moves it to point to $b$.  Hence, we know $a$.  Notice that since $\triangle abc$ was facial in $X$, all blue edges coming into $b$ in $\sigma_1$ before $\arc{ab}$ (in the counterclockwise direction) were children of $\arc{ac}$ in $X$.  Thus, given $a$, we know that we must use $a$'s red edge to move each of these children back up to $a$.  This brings us back to the last configuration in stage 2; using $a$ and $\head_b,$ we can recover $X$.  If $(Z, W)$ is in stage 4a, then we know that the blue tree agrees with $Y$, and the red edges of $a$ and $b$ are the only red edges in a different position in $Z$ than in $Y$.  Since $(Z,W)$ moves $a$'s red edge, we know $a$, and $b$ is the vertex that shares $a$'s blue edge.  Given $b$, it is easy to recover $r_a$, since to find $r_a$, just move $b$'s red edge counter-clockwise until it can't make any more red/green swaps.  We need to record $\head_b$ and then it is easy to get to $Y$.  If $(Z, W)$ is in stage 4b, then we know that the blue tree agrees with $Y$ and $b$'s red edge is the only red edge that is in a different position in $Z$ than in $Y$.  Since $(Z,W)$ moves $b$'s red edge, we know $b$.  As described in stage 4a, it is easy to then find $r_a$; move $b$'s red edge all the way there to get to $Y$.  In each of the four stages we need to record a maximum of 2 vertices and a single bit.  This implies that in this case there are $O(n^2)$ canonical paths which use any edge $(Z, W)$. 

\begin{figure} [!htb]
\centering
\subfloat[]{\begin{overpic}[scale=.09]{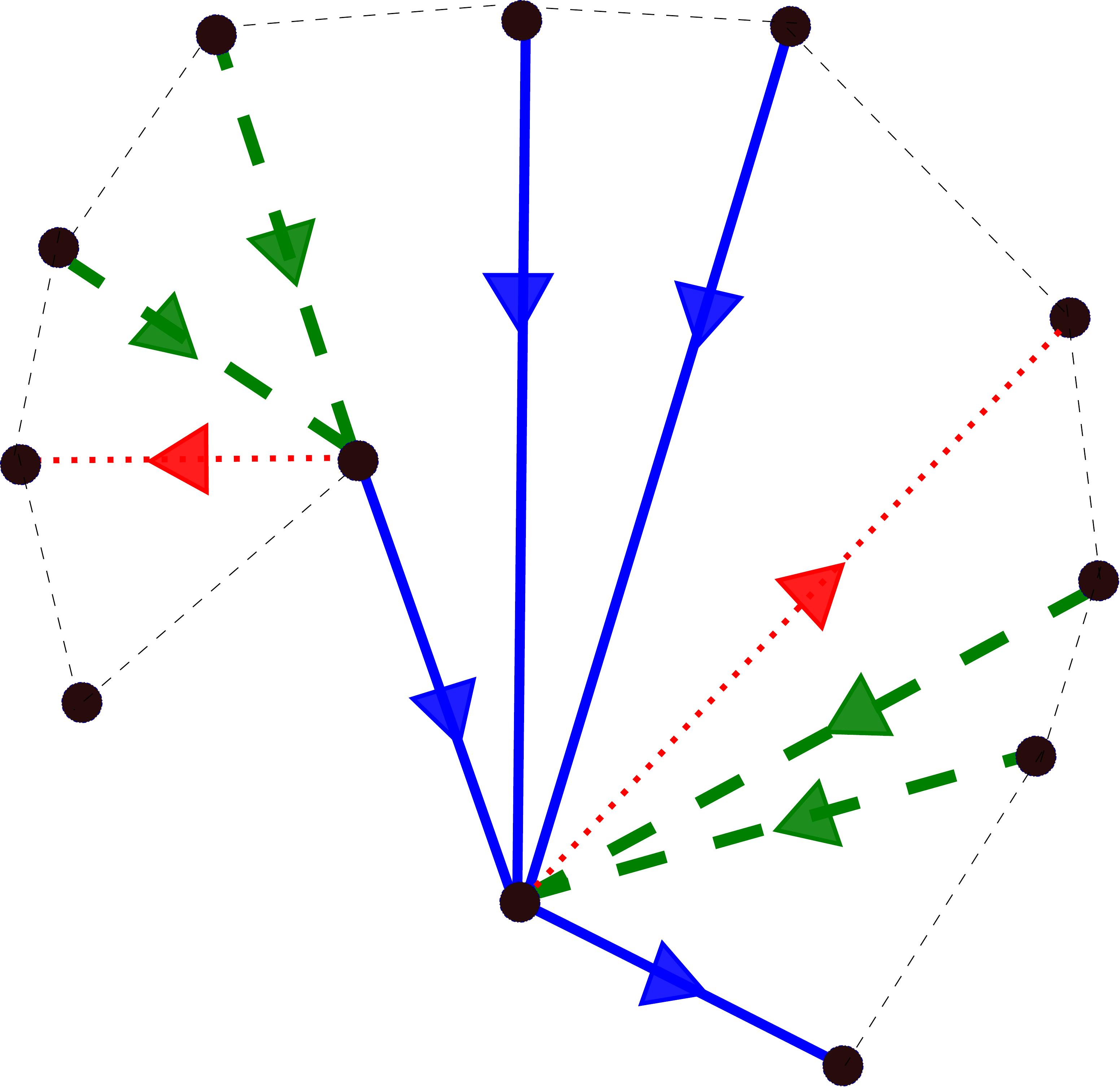}
\put(23,48){$a$}
\put(35,17){$b$}
\put(75,5){$c$}
\put(96,27){$d$}
\put(-2,80){$\head_a$}
\put(90,75){$\head_b$}
\put(73,96){$f$}
\put(50,98){$g$}
\end{overpic}\label{apeakbluetreeA}}
\hspace{.6 cm}
\subfloat[]{\begin{overpic}[scale=.09]{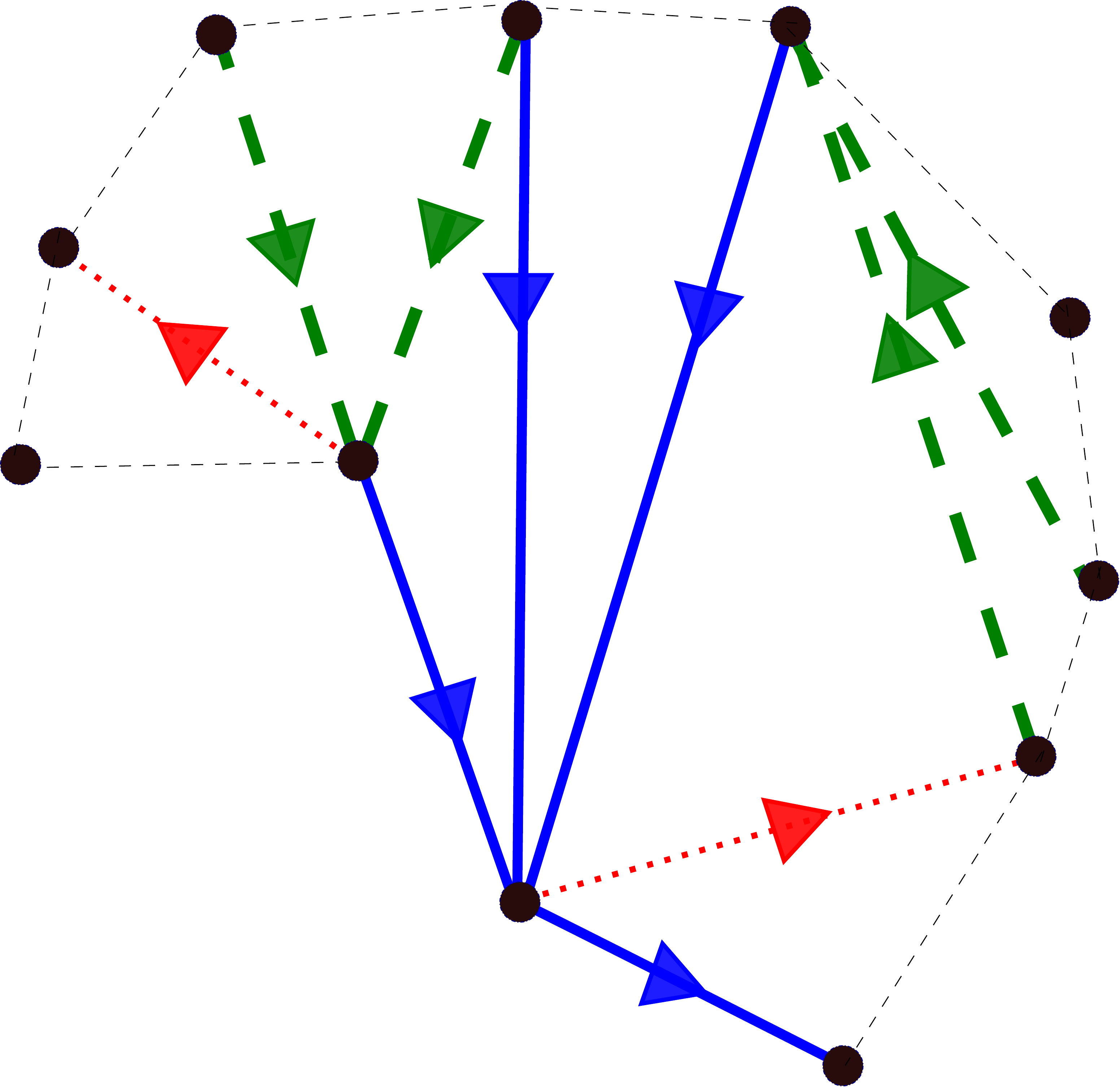}
\put(23,48){$a$}
\put(35,17){$b$}
\put(75,5){$c$}
\put(96,27){$d$}
\put(-2,80){$\head_a$}
\put(90,75){$\head_b$}
\put(73,96){$f$}
\put(50,98){$g$}
\end{overpic}\label{apeakbluetreeB}}
\hspace{.6 cm}
\subfloat[]{\begin{overpic}[scale=.09]{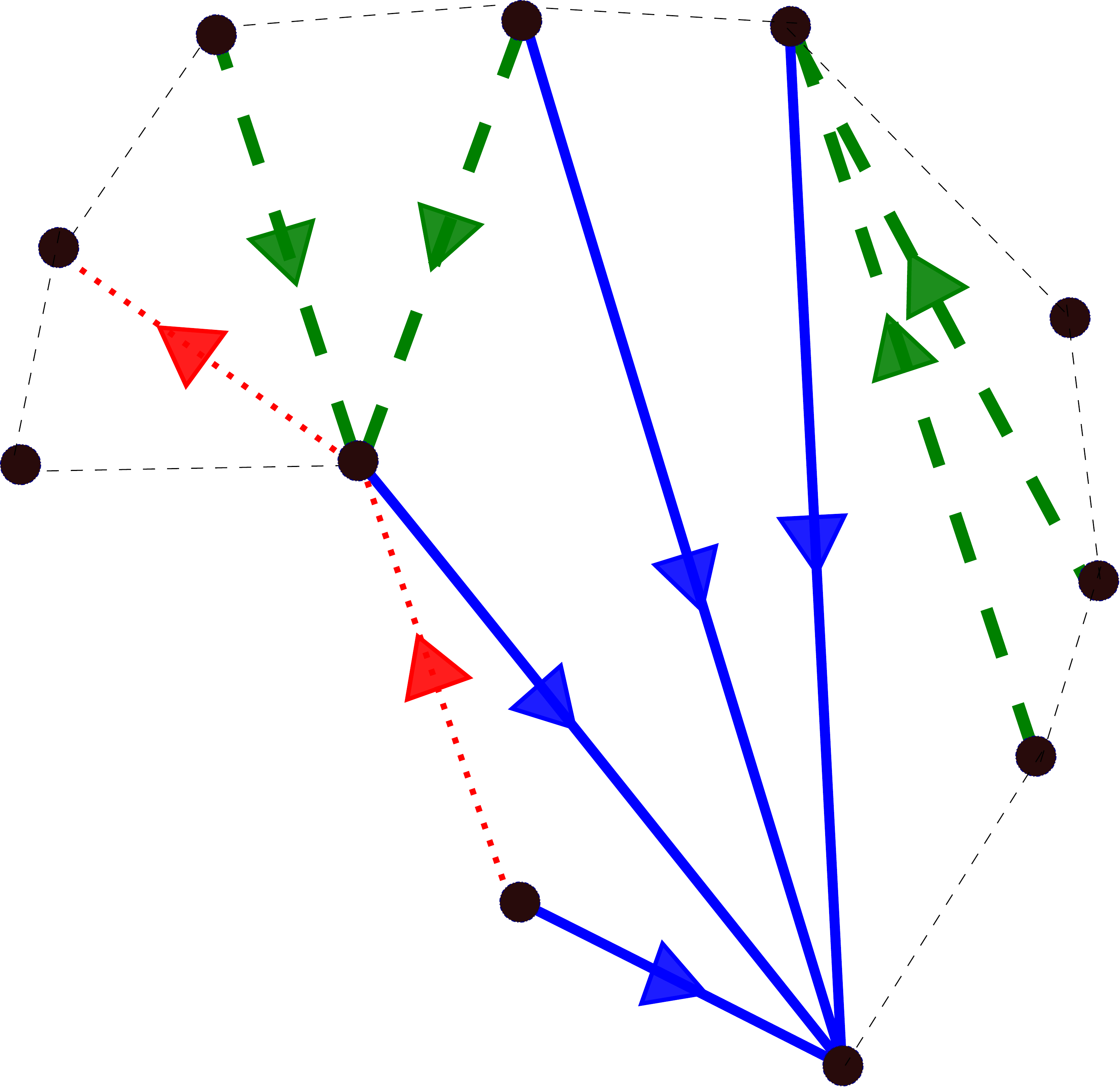}
\put(23,48){$a$}
\put(35,17){$b$}
\put(75,5){$c$}
\put(96,27){$d$}
\put(-2,80){$\head_a$}
\put(90,75){$\head_b$}
\put(73,96){$f$}
\put(50,98){$g$}
\end{overpic}\label{apeakbluetreeC}}
\hspace{.6 cm}
\subfloat[]{\begin{overpic}[scale=.09]{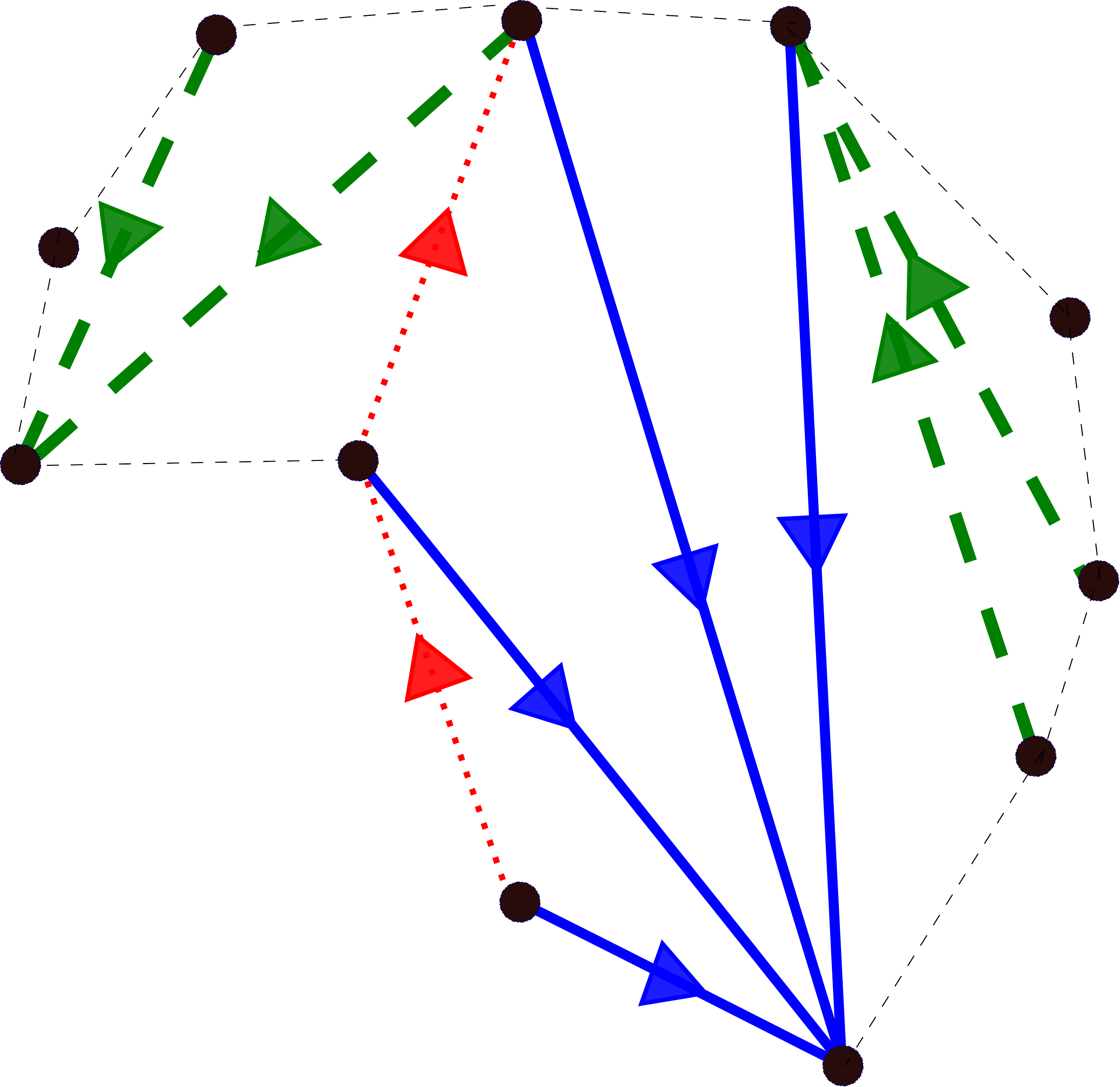}
\put(23,48){$a$}
\put(35,17){$b$}
\put(75,5){$c$}
\put(96,27){$d$}
\put(-2,80){$\head_a$}
\put(90,75){$\head_b$}
\put(73,96){$f$}
\put(50,98){$g$}
\end{overpic}\label{apeakbluetreeD}}
\hspace{.6 cm}
\subfloat[]{\begin{overpic}[scale=.09]{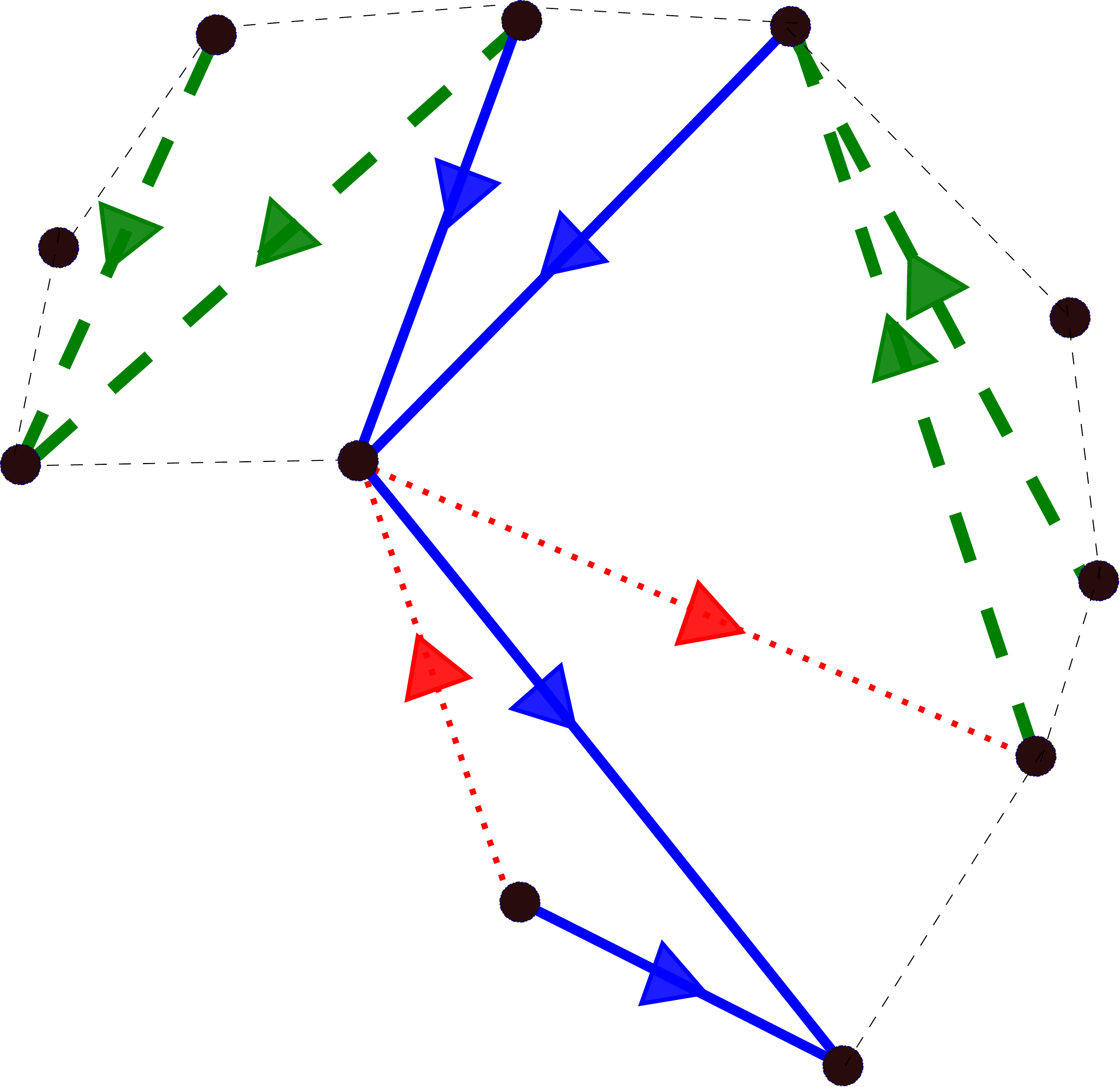}
\put(23,48){$a$}
\put(35,17){$b$}
\put(75,5){$c$}
\put(96,27){$d$}
\put(-2,80){$\head_a$}
\put(90,75){$\head_b$}
\put(73,96){$f$}
\put(50,98){$g$}
\end{overpic}\label{apeakbluetreeE}}
\hspace{.6 cm}
\caption{Canonical path to move a peak down to a valley in the blue tree.   }
\label{apeakbluetree}
\end{figure}

\vspace{.2in}
\underline{\textbf{A Peak to Valley Move on the Bottom Dyck Path:}}\ \ \ Next suppose $e=(X,Y)$ is a move that takes a peak of the bottom Dyck path and moves it to a valley.  This moves the blue tree in Figure~\ref{avtop}b to Figure~\ref{avtop}a.  In general, $a$'s blue edge moves from $b$ to $c$, and all the (blue) children of $\overrightarrow{bc}$ clockwise from $\arc{ab}$ become children of $\overrightarrow{ac}$.  It is important to note that in order for the peak to valley move to be possible, $a$ must be a leaf of the blue tree.  Figure~\ref{apeakbluetree} shows how the red and green trees might look.  Using the vertex condition (Figure~\ref{vertexconditionpic}a) for $b$, going clockwise around $b$ there are $k_1\geq 0$ blue edges coming into $b$ after $\arc{ab}$, followed by $b$'s red edge, then there are $k_2\geq 0$ green edges coming in, and finally $b$ has its blue edge $\arc{bc}$.   As above, $\head_v$ is the head of $v'$s red edge.  We will go through 5 stages in the canonical path.  

\noindent \textbf{Stage 1.} If there are green edges coming into $b$, let $d$ be the tail of the last incoming green edge; if $b$ has no green edges coming in, then set $d=\head_b$.  During the first stage of the canonical path, $b$'s red edge must move from $\head_b$ to $d$ (if $\head_b\neq d$) by making $k_2$ red/green swaps.

\noindent \textbf{Stage 2.} Let $f$ be the last blue incoming edge to $b$.  Now we can move $k_1+1$ blue edges pointing to $b$ (namely, all blue edges between $\arc{ab}$ and $\arc{fb}$, inclusive) to point to $c$.  Do this sequentially and counterclockwise, using $b$'s red edge to swap.  See Figure~\ref{apeakbluetree}(b-c).

\noindent \textbf{Stage 3.} Vertex $a$ may have green edges entering between the edges $\arc{a\head_a}$ and $\arc{ac}.$   Let $g$ be the tail of $a$'s last incoming green edge, if it exists; otherwise, $g=\head_a$.  If $g\neq \head_a$, we move $a$'s red edge from $\head_a$ to $g$.  

\noindent \textbf{Stage 4.} Using $a$'s red edge, we can move the $k_1$ blue edges from $c$ back to $a$, correcting the blue tree.

\noindent \textbf{Stage 5.} To repair the red tree, first we need to move $a$'s red edge to point to $\head_b$ ({\bf Stage 5a}).  Notice that $b$'s red edge now points toward $a$. Thus we need to pass a red edge from $a$ to $\head_a$ ({\bf Stage 5b}).  Since $a$ is a leaf of the blue tree, this can be done using a sequence of red/green swaps.

Given a transition $(Z, W)$ of $\mvar$ we must upper bound the number of canonical paths $\gamma_{X,Y}$ that may use this edge.  If $(Z, W)$ is in stage 1, then it moves $b$'s red edge, so we know $b$.  To get back to $X$, move $b$'s red edge counter-clockwise until it can't make any more red/green swaps.  Hence we don't need to remember any vertices.  Given $X$ and $b$, we can find $Y$ since we know which peak to flip down.  If $(Z, W)$ is in stage 2, then it moves some blue edge from $b$ to $c$, so we know both $b$ and $c$.  We must record $f$, but then we can get back to the last configuration in stage 1, and using $b$ we can get back to $X$.  If $(Z, W)$ is in stage 3, then it moves $a$'s red edge, so we know $a$.  We need to remember $\head_a$ to get back to stage 2, and remember $f$ to get back to stage 1.  Given $a$ we know $b$, so we can then get back to $X$.  If $(Z, W)$ is in stage 4, then it moves $a$'s red edge, so we know $a$.  To get back to stage 3, just move $a$'s red edge as far counter-clockwise as possible.  Hence to get back to $X$ we need to record $\head_a$ and $f$.  If $(Z, W)$ is in stage 5a, then we are moving $a$'s red edge, so we know $a$.  Rotate it as far counter-clockwise as possible to find $r_b$.  To recover $Y$, we just need to know $\head_a$ so we can pass an edge from $a$ to $\head_a$.  If $(Z, W)$ is in stage 5b, then we need to remember $a$ and $\head_a$.  Move $a$'s red edge all the way to $\head_a$ to obtain $Y$.  In each of the five stages we need to record a maximum of 2 vertices.  This implies that in this case there are $O(n^2)$ canonical paths which use any edge $(Z, W)$.

We have shown that in each of the four cases above there is a maximum of $O(n^4)$ canonical paths which use any edge $(Z, W)$.  If the move of $\mdyck$ affects both the top and bottom paths, we can think of this move as two moves, each of which affects only the top or bottom path; hence, we concatenate the paths for each of those moves.  Therefore, if we record a bit to decide if the move of $\mdyck$ affects both the top and bottom paths, as well as a bit to decide which of the stages we are in, this implies that across all cases there is a maximum of $O(n^4)$ canonical paths which use any edge $(Z, W)$.  Notice that the maximum length of any path $\gamma_{XY}$ is $O(n^2)$.  We can now upper bound the quantity $A$ which is needed to apply Theorem \ref{Comparison} as follows: 
\begin{eqnarray*}
A &=& \max_{(Z,W) \in E(P)} \left \{\frac{1}{\pi(Z)P(Z,W)}\sum_{\Gamma(Z,W)}|\gamma_{XY}|\pi(X)P'(X,Y) \right \}\\
&\leq& \max_{(Z,W) \in E(P)} \left \{4n\sum_{\Gamma(Z,W)}\frac{O(n^2)}{2n} \right \} \ \leq  \ O(n^6). 
\end{eqnarray*}
Moreover, we can bound $\pi_*$ as follows:
$$\pi_*=\min_{X\in\stsp} \pi(X)=\frac{1}{C_{n+2}C_n - C_{n+1}^2}\geq \frac{1}{30^n}.$$ 
Applying Theorems~\ref{mixdyck} and~\ref{Comparison} , we get the following
\begin{eqnarray*}
\tau(\epsilon)&= & O\left(\frac{\log(\frac{1}{\epsilon \pi_*})}{\log(1/(2\epsilon))}n^6 \cdot n^3\log(n/\epsilon)\right)\\
&= & O\left(\frac{n\log{30} - \log{\epsilon}}{-\log(2\epsilon)}n^9\log(n/\epsilon)\right)\\
&= & O\left({n^{10}\log(n/\epsilon)}\right).
\end{eqnarray*}

\noindent Therefore $\mvar$ is an efficient sampling algorithm for sampling from the set of all 3-orientations over any triangulation on $n$ internal vertices.

\section{Concluding Remarks}\label{conclude}
Several questions remain open. The complexity of enumerating Eulerian orientations in planar graphs of bounded degree is one of the foremost, as raised by \cite{fz}.  
Extending our fast mixing result to triangulations with larger degrees is a natural open problem; perhaps there is an alternate local chain which can sample efficiently from the set of 3-orientations corresponding to any fixed triangulation, without recourse to the bipartite perfect matching sampler of \cite{jsv}.
Finally, our slow mixing example involves vertices of degree $\Omega(n)$, and it would be of interest to find other constructions with constant maximum degree.  \\

\noindent {\bf Acknowledgments}.
The last author thanks Stefan Felsner for introducing him to the problem of sampling 3-orientations and providing useful links to literature.

\end{document}